\documentclass[12pt]{article}
\usepackage[dvipdfmx]{graphicx}
\usepackage{amsfonts,amsmath,amssymb,epsf}
\usepackage{amsmath,braket}
\usepackage{graphicx,hyperref,color}
\usepackage[toc,page]{appendix}
\usepackage{tikz}
\hypersetup{
    bookmarks=true,         
    unicode=false,          
    pdftoolbar=true,        
    pdfmenubar=true,        
    pdffitwindow=false,     
    pdfstartview={FitH},    
    pdfnewwindow=true,      
    colorlinks=true,       
    linkcolor=blue,          
    citecolor=blue,        
    filecolor=blue,      
    urlcolor=blue           
}
\newcommand{\de}{\partial}
\newcommand{\be}{\begin{equation}}
\newcommand{\ba}{\begin{eqnarray}}
\newcommand{\ea}{\end{eqnarray}}
\newcommand{\ee}{\end{equation}}

\newcommand{\f}{\frac}
\newcommand{\s}{\sqrt}
\newcommand{\vp}{\varphi}

\newcommand{\ti}{\tilde}
\newcommand{\ap}{\alpha}

\newcommand{\ddd}{\cdot\cdot\cdot}
\newcommand{\no}{\nonumber \\}
\newcommand{\la}{\langle}
\newcommand{\lb}{\rangle}
\newcommand{\bea}{\begin{eqnarray}}
\newcommand{\eea}{\end{eqnarray}}
\newcommand{\bes}{\begin{equation*}}
\newcommand{\beas}{\begin{eqnarray*}}
\newcommand{\eeas}{\end{eqnarray*}}
\newcommand{\bas}{\begin{array*}}
\newcommand{\eas}{\end{array*}}
\newcommand{\ees}{\end{equation*}}

\newcommand{\ep}{\epsilon}

\usepackage{comment}

\usepackage{mathtools}

\numberwithin{equation}{section}

\begin{document}

\begin{titlepage}
\thispagestyle{empty}

\begin{flushright}
KEK-TH-2399
\\
YITP-22-21
\\
IPMU22-0005
\\
\end{flushright}

\bigskip

\begin{center}
\noindent{\large \bf Holographic Local Operator Quenches in BCFTs}\\
\vspace{2cm}

Taishi Kawamoto$^{1}$,
Takato Mori$^{2,3}$,
Yu-ki Suzuki$^1$,\\
Tadashi Takayanagi$^{1,4,5}$,
and
Tomonori Ugajin$^{1,6}$
\vspace{1cm}\\

{\it $^1$Center for Gravitational Physics,
Yukawa Institute for Theoretical Physics,\\
Kyoto University, 
Kitashirakawa Oiwakecho, Sakyo-ku, Kyoto 606-8502, Japan}\\

{\it $^2$ 
KEK Theory Center, Institute of Particle and Nuclear Studies, 
High Energy Accelerator Research Organization (KEK),\\
1-1 Oho, Tsukuba, Ibaraki 305-0801, Japan}\\

{\it $^3$ 
Department of Particle and Nuclear Physics, School of High Energy Accelerator Science, 
The Graduate University for Advanced Studies (SOKENDAI),\\
Oho 1-1, Tsukuba, Ibaraki 305-0801, Japan}\\

{\it $^4$Inamori Research Institute for Science,\\
620 Suiginya-cho, Shimogyo-ku,
Kyoto 600-8411, Japan}\\

{\it $^{5}$Kavli Institute for the Physics and Mathematics
 of the Universe (WPI),\\
University of Tokyo, Kashiwa, Chiba 277-8582, Japan}\\

{\it $^6$The Hakubi Center for Advanced Research, Kyoto University,\\ 
Yoshida Ushinomiyacho, Sakyo-ku,
Kyoto 606-8501, Japan}

\end{center}

\begin{abstract}
We present a gravity dual of local operator quench in a two-dimensional CFT with conformal boundaries.
This is given by a massive excitation in a three-dimensional AdS space with the end of the world brane (EOW brane). Due to the gravitational backreaction, 
the EOW brane gets deformed in a nontrivial way. We show that the energy-momentum tensor and entanglement entropy computed from the gravity dual and from the BCFT in the large $c$ limit match perfectly. Interestingly, this comparison avoids the folding of the EOW brane in an elegant way. 

\end{abstract}

\end{titlepage}

\newpage

\tableofcontents

\section{Introduction}

In recent progress of holographic dualities, especially the AdS/CFT correspondence \cite{Maldacena:1997re}, the end of world branes (EOW branes) play an important role. One reason for this is that the presence of such objects in AdS backgrounds provides a bottom-up model for a gravity dual of a boundary conformal field theory (BCFTs) \cite{Karch:2000gx,Takayanagi2011,Fujita:2011fp}, called the AdS/BCFT. Another reason is that the brane-world  holography predicts that the dynamics of EOW branes is dual to that of quantum gravity \cite{Randall:1999ee,Randall:1999vf,Gubser:1999vj,Karch:2000ct}. These two different interpretations of the EOW branes: the AdS/BCFT and the brane world, are expected to be equivalent. This is manifest in the calculation of entanglement entropy, i.e. the holographic entanglement entropy formula for AdS/BCFT in the presence of EOW branes \cite{Takayanagi2011,Fujita:2011fp} takes the identical form as the island formula \cite{Penington:2019npb,Almheiri:2019psf,Almheiri:2019hni,Almheiri:2019qdq,Penington:2019kki} which computes the entanglement entropy in the presence of gravity. This correspondence can be regarded as an equivalence between a BCFT and a CFT coupled to gravity, as systematically studied recently in \cite{Suzuki:2022xwv} (see also \cite{Numasawa:2022cni,Kusuki:2021gpt} for related studies from conformal field theoretic approaches). In this way, these ideas of EOW branes are deeply connected at the bottom and their further understandings are expected to be a key ingredient to fully understand quantum gravity. 

So far studies of EOW branes have mainly been limited to holographic setups at zero or at finite temperature. This raises a basic question whether the holographic duality of AdS/BCFT works successfully  in time-dependent backgrounds.
Motivated by this, in this paper we would like to study the dynamics of  EOW branes in the presence of a large and inhomogeneous excitation. In particular, we focus on an analytical model where the excitation is created by a massive particle in  three-dimensional AdS geometry. Via the AdS/BCFT duality, this is dual to a local operator excitation in the holographic two-dimensional BCFT, which may also be called a local operator quench in the presence of a boundary. The local operator quench is defined by the time evolution of a locally excited state, acting a primary operator $O(x)$ on a CFT vacuum $|0\lb$ with a suitable regularization\footnote{
In this paper, our two-dimensional BCFT has either a single boundary at $x=0$ or two boundaries (one at $x=0$ and the other at a specific time-dependent location $x=Z(t)$). Although in the former case, in which $H|0\lb=0$ holds, the definition (\ref{LOS}) is equivalent to the usual definition of the local quench $|\Psi(t)\lb=e^{-itH}e^{-\ap H}O(x=x_a)|0\lb$, we need to modify this for the latter case, in which $H|0\lb\neq 0$, as the time evolution is not unitary but isometry as in \cite{Cotler:2022weg}.
As we will discuss in the paper, \eqref{LOS} is the correct definition for the local quench dual to what is discussed in the paper.
}
\ba
|\Psi(t)\lb=e^{-itH}O(x=x_a,t_E=-\alpha)|0\lb,  \label{LOS}
\ea
where $H$ is the CFT Hamiltonian, $t_E=it$, and $\ap$ is a regularization parameter \cite{Nozaki:2014hna,He:2014mwa,Caputa:2014vaa,Asplund:2014coa} (refer also to \cite{Guo:2015uwa} for an earlier analysis of 
local operator quench in BCFTs). Our gravity model is obtained by introducing an EOW brane in the holographic local quench model \cite{Nozaki2013}. The shape of the EOW brane is deformed by the local excitation via the gravitational backreaction, which gives a novel dynamics of the local quench in the BCFT. Via the double holography, this model is also closely related to the local quench in a two-dimensional gravity studied in \cite{Goto:2020wnk} as a model of black hole evaporation.

Our local quench model is analytically tractable both from the gravity side and the BCFT side. The former can be found by finding the correct asymptotically AdS spacetime with an EOW brane and a massive particle. This gravity dual geometry allows us to calculate the holographic energy-momentum tensor \cite{Balasubramanian:1999re} and holographic entanglement entropy \cite{Ryu:2006bv,Ryu:2006ef,Hubeny:2007xt}.  
On the other hand, the latter can be analyzed via  direct conformal field theoretic computations by employing a suitable conformal map. We will note that a careful choice of the coordinate system in the gravity dual is important when we compare the results of the former with those of the latter. Eventually we will find complete matching between the gravity dual results and the BCFT ones in the large central charge limit.

This paper is organized as follows:
In section 2, we present our gravity dual of the local operator quench by introducing a localized excitation in the AdS/BCFT. We  calculate its holographic energy-momentum tensor in this model. We also examine a coordinate transformation which helps us to identify its BCFT dual. In section 3, we compute the holographic entanglement entropy in our holographic local quench model. In section 4, we provide the BCFT description of our local operator quench and compute the energy-momentum tensor. We compare the results in the gravity dual with those in the BCFT. In section 5, we compute entanglement entropy in the BCFT and observe a consistent result with the identified gravity dual. In section 6, we summarize our conclusions and discuss future problems.

\section{Holographic Local Operator Quench in BCFTs}

In this paper, we  would like to study a local operator quench (\ref{LOS}) in a two-dimensional conformal field theory (CFT) in the presence of boundaries, namely, a two-dimensional BCFT. We analyze the system this both from the gravity dual calculations and from the direct computations in BCFT. In the absence of a boundary, a gravity dual of local operator quench was presented in \cite{Nozaki2013} by employing an asymptotically AdS  spacetime with backreaction of a massive particle infalling in the bulk \cite{Horowitz1999}.

Writing the two-dimensional coordinates as $(t,x)$ of the CFT, we define our BCFT by restricting the spacetime in the right half plane $x>0$. In this paper, for simplicity, we will focus on the local operator quench (\ref{LOS}) with the choice $x_a=0$, i.e. the excitation is localized on the boundary. 
Its gravity dual can be found by inserting an end of the world brane (EOW brane) appropriately in the 
three-dimensional asymptotically AdS with a falling massive particle with backreaction taken into account, following the AdS/BCFT prescription \cite{Takayanagi2011}.

\subsection{Coordinate Change between Poincare and Global AdS} 

In general, the backreaction of a bulk infalling particle in the Poincare AdS can be obtained by pulling back a spacetime with a massive static particle in global AdS \cite{Horowitz1999}.
Consider the AdS$_3$ in the Poincare metric with the radius $R$, 
\ba
ds^2=R^2\left(\frac{dz^2-dt^2+dx^2}{z^2}\right).   \label{Pmet}
\ea
This is transformed into the global AdS$_3$  with the metric
\ba
ds^2=-(r^2+R^2)d\tau^2+\frac{R^2}{r^2+R^2}dr^2+r^2d\theta^2,   \label{Gmet}
\ea
via the following map:
\ba
\s{R^2+r^2}\cos\tau &=& \frac{R\ap^2+R(z^2+x^2-t^2)}{2\ap z},\no
\s{R^2+r^2}\sin\tau &=& \frac{Rt}{z},\no
r\sin\theta &=& \frac{Rx}{z},  \no
-r\cos\theta &=& \frac{-R\ap^2+R(z^2+x^2-t^2)}{2\ap z}.  \label{mappo}
\ea
We chose the range $-\pi \le \theta<\pi$ for the spatial coordinate of the global AdS. The parameter $ \alpha$ is an arbitrary real number, corresponding to a particular isometry of AdS. In this map, 
 the line $x=0$ at the boundary $z=0$ in Poincare AdS is mapped into $\theta=0$ and 
 the time slice $t=0$ is mapped into $\tau=0$.

\begin{figure}
  \centering
  \includegraphics[width=8cm]{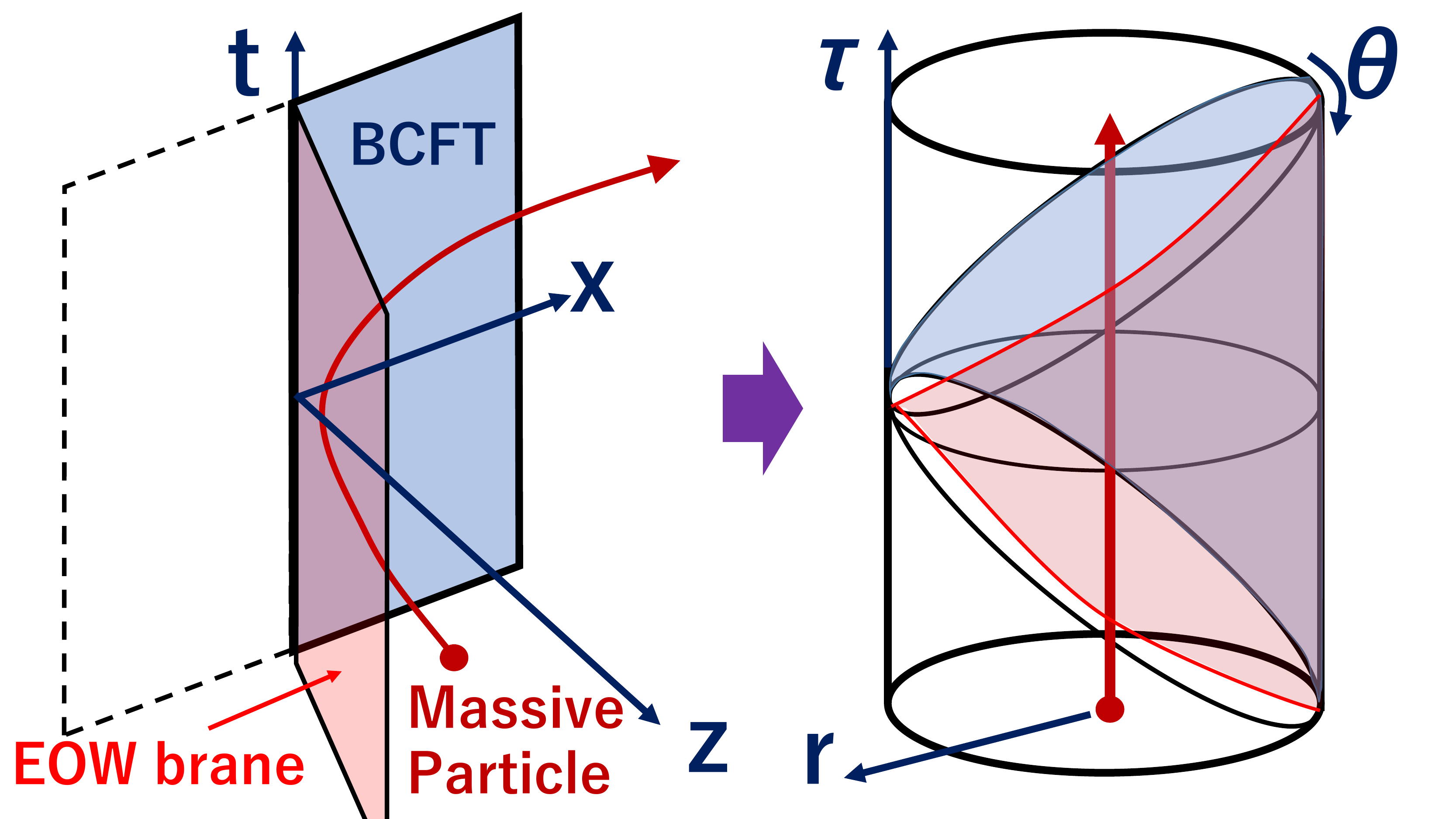}
  \caption{A sketch of the coordinate transformation from the Poincare AdS into a global AdS in the 
presence of a massive particle (the red arrow) and an EOW brane (the red surface).}
\label{mapfig}
\end{figure}

We place a massive particle with a mass $m$ at $r=0$ of the global AdS (\ref{Gmet}).  
In the Poincare AdS (\ref{Pmet}), 
its trajectory is 
\ba
z^2-t^2=\ap^2,\ \ \ x=0,
\ea
as depicted in Fig.\ref{mapfig}. By performing a Wick rotation $\tau=it$, the above trajectory 
becomes a semi-circle $z^2+\tau^2=\ap^2$. Since at the boundary $z=0$, the excitation is at $\tau=\pm \ap$, 
we find that this massive particle is dual to a local operator excitation given by (\ref{LOS}) with $x_a=0$, where the mass $m$ is related to the conformal dimension $\Delta$ of the primary operator $O(x)$ via 
$\Delta\simeq mR$.

Moreover, as can be found by taking the AdS boundary limit $r\to \infty$ and $z\to 0$, the above transformation (\ref{mappo}) corresponds to the following  conformal transformation in the CFT dual
\ba
t \pm x=\ap\tan\left(\frac{\tau\pm \theta}{2}\right).  \label{confmap}
\ea

\subsection{Gravity Dual of an Infalling Particle in AdS} 

If we take into account the gravitational backreaction of this massive particle, then the global AdS metric 
is replaced with 
\ba
ds^2=-(r^2+R^2-M)d\tau^2+\frac{R^2}{r^2+R^2-M}dr^2+r^2d\theta^2,   \label{GBmet}
\ea
where $\theta$ has the periodicity $2\pi$ as $-\pi\leq \theta<\pi$. The mass parameter $M$ is related to the mass $m$ via
\ba
M=8G_N R^2 m.  \label{relam}
\ea
The map  \eqref{mappo} transforms this metric into that of an asymptotically Poincare AdS background \cite{Horowitz1999}.
As identified in \cite{Nozaki2013}, this is a time-dependent background with a local operator insertion at 
$t=x=0$ i.e. given by the local operator quench state (\ref{LOS}) with $x_a=0$. 
The parameter $\ap$ in the map \eqref{mappo} precisely coincides with the regularization parameter in the CFT state (\ref{LOS}).

For $0<M<R^2$, the geometry (\ref{GBmet}) can be transformed into the metric
\ba
ds^2=-(\ti{r}^2+R^2)d\ti{\tau}^2+\frac{R^2}{\ti{r}^2+R^2}d\ti{r}^2+\ti{r}^2d\ti{\theta}^2,
\label{deficitg}
\ea
via the map
\ba
\ti{\tau}=\chi \tau,\ \ \ti{\theta}=\chi \theta,\ \  \ti{r}= \frac{r}{\chi},
\quad\text{where}\quad
\chi=\s{\frac{R^2-M}{R^2}}.
\label{teha}
\ea
Even though this looks like a global AdS$_3$, there is a deficit angle at $r=0$ since the periodicity of the new spatial coordinate is  $\ti{\theta}$ is 
\ba
-\chi \pi \leq \ti{\theta}<\chi \pi, \quad \chi<1.
\label{tildtheta}
\ea

For $M>R^2$, the geometry (\ref{GBmet}) describes a BTZ black hole, where the horizon is situated as $r=\s{M-R^2}$.

\subsection{EOW Branes in AdS/BCFT}\label{sec:m-zero}

We are interested in effects of the presence of a boundary in the above mentioned local quench. Such a boundary in a boundary conformal field theory (BCFT) corresponds to 
 an end of the world brane (EOW brane) in the holographic dual setup, namely the AdS/BCFT \cite{Karch:2000gx,Takayanagi2011,Fujita:2011fp}. In general, the location of the brane is the bulk is fixed by solving the following Neumann boundary condition on the EOW brane \cite{Takayanagi2011}:
  \be
 K_{ab} -K h_{ab} = -8\pi G_{N} {\cal T} h_{ab}. \label{eq:EOW}
 \ee
 
 In the above equations, $K_{ab}$, denotes the extrinsic curvature of the brane profile,  $h_{ab}$ is its induced metric, and $K=h^{ab} K_{ab}$. The extrinsic curvature can be computed from the outward pointing normal vector $n^a$  on the brane as the covariant derivative $\nabla_a n_b$, projected on the brane. Also the parameter ${\cal T}$ is the tension of the EOW brane. 
 In this paper we will focus on the case ${\cal T}\neq 0$ so that the falling particle is off the EOW brane.
 
 In the absence of the massive particle where the metric takes of the form \eqref{Pmet},  
 the following profile of the EOW brane
\ba
x-a=-\lambda z,  \label{eomb}
\ea
 solves \eqref{eq:EOW}. The physical bulk region dual to the BCFT is given by $x-a\geq -\lambda z$.
$\lambda$  is a constant  related to the tension ${\cal T}$ in \eqref{eq:EOW} as
\ba
{\cal T}R=\frac{\lambda}{\s{1+\lambda^2}}.
\ea
The parameter $a$ describes the location of the boundary in the BCFT so that the BCFT is defined for $x>a$.
In this paper, we will set $a=0$ for simplicity. The local operator is inserted at the boundary of BCFT in this case. 
 The EOW brane (\ref{eomb}) corresponds to the following profile in the global AdS$_3$ (\ref{Gmet}):
\ba
r\sin\theta=-\lambda R,
\ea
which is depicted in Fig.\ref{mapfig}.

\subsection{Profiles of EOW Branes}

In the presence of the massive particle, the metric is deformed into the solution (\ref{GBmet}) which is further mapped to the Poincare AdS by the map (\ref{mappo}).  
We can specify the brane profiles in the time-dependent geometry by applying the chain of diffeomorphisms. We begin with the profile of the EOW brane in the transformed geometry (\ref{deficitg}),
\ba
\ti{r}\sin \ti{\theta}=-\lambda R.  \label{profti}
\ea
Thus for $0\leq M<R^2$, the EOW brane profile is found in the original coordinates as 
\ba
r\sin\left(\chi\theta\right)=-\lambda \s{R^2-M}.
\label{beoms}
\ea

\begin{figure}
  \centering
  \includegraphics[width=12cm]{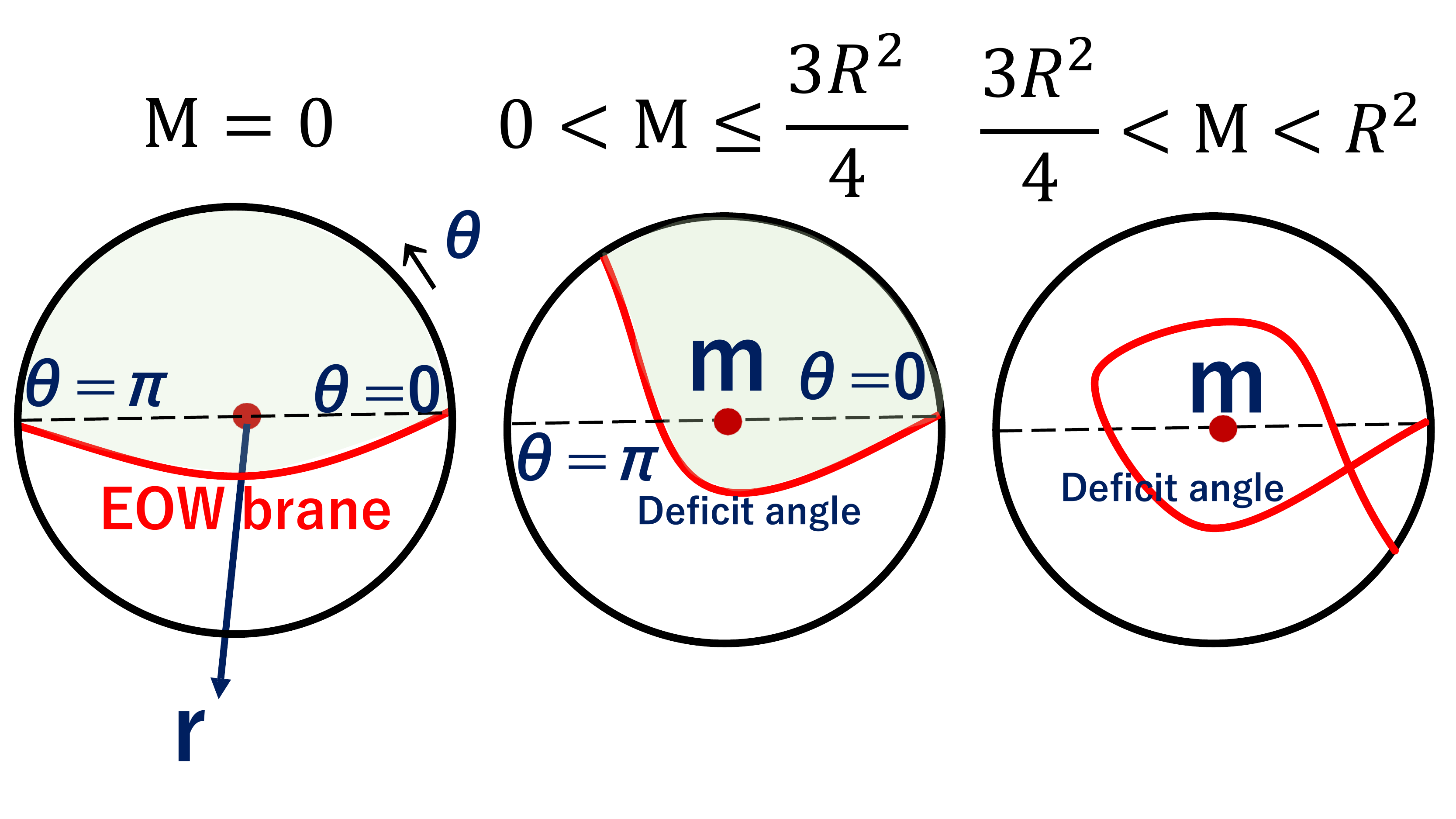}
  \caption{Cross sections at constant $\tau$ for the backreacted geometry with a mass and a positive tension $\lambda>0$. We depicted the EOW brane as the red curved. The light green regions are the gravity duals in the AdS/BCFT. Though for $M>\frac{3}{4}R^2$, the EOW brane gets folded and the gravity dual does not make sense, we do not need this range of the mass when we consider the BCFT dual as we explain around (\ref{relasd}).}
\label{deformationfig}
\end{figure}

For $\lambda>0$, the EOW brane extends from $\theta=0$ to $\theta=-\frac{\pi}{\chi} \simeq 2\pi-\frac{\pi}{\chi}$ as
 depicted in the middle panel of Fig.\ref{deformationfig}. When $M=0$, the brane intersects with the asymptotic boundary at $\theta=0$ and $\theta =\pi$. As we increase the mass  of the bulk particle, the brane profile is eventually  bent, and the coordinate distance between the two end points gets closer.
 
\begin{figure}
  \centering
  \includegraphics[width=12cm]{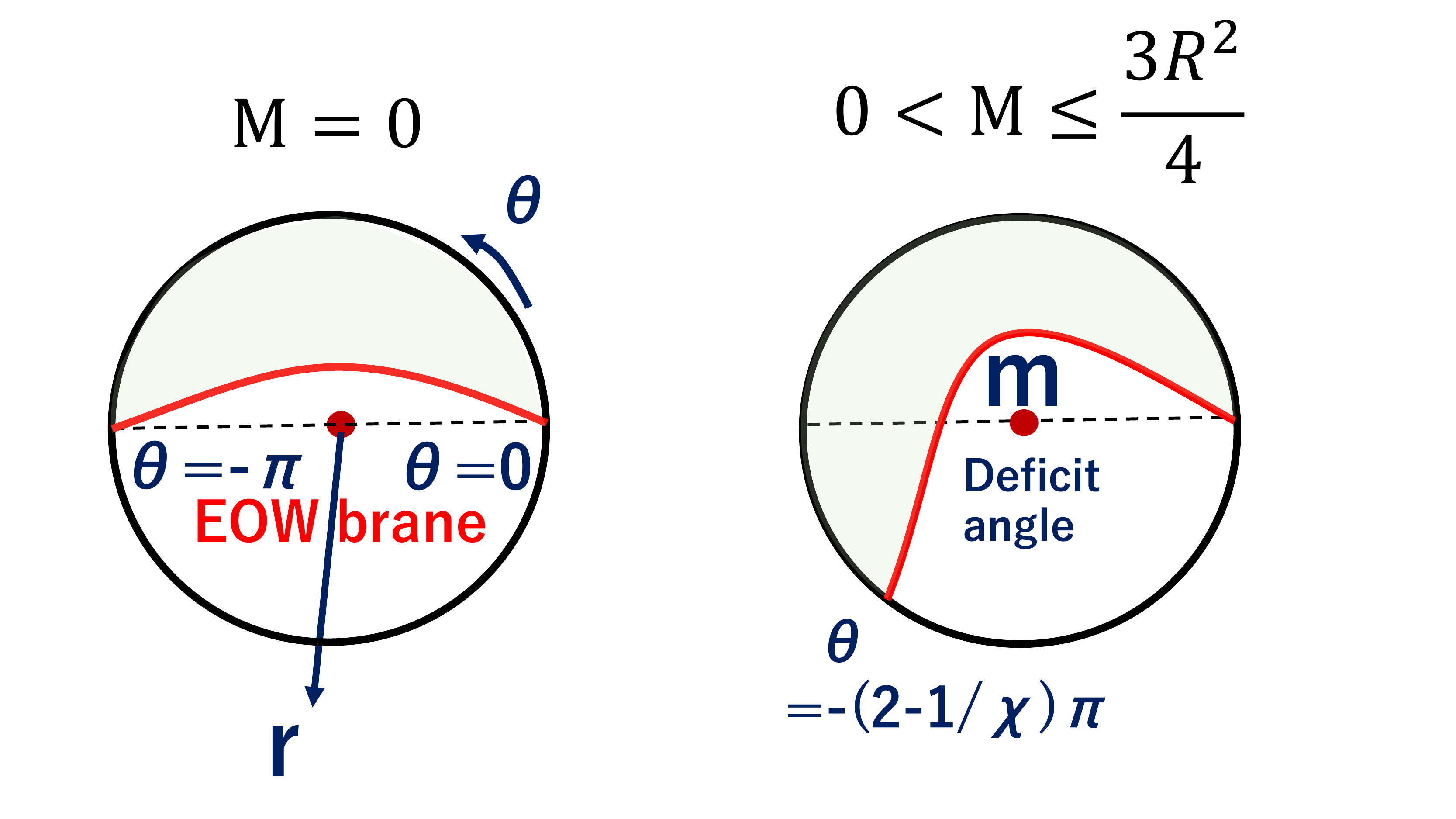}
  \caption{Cross sections at constant $\tau$ for the backreacted geometry with a mass and a negative tension $\lambda<0$. We depicted the EOW branes as the red curves. The light green regions are the gravity duals of the BCFT.}
\label{deformationfig2}
\end{figure}

 For $\lambda<0$, the EOW brane extends from $\theta=0$ to $\theta=\frac{\pi}{\chi}\simeq \frac{\pi}{\chi}-2\pi$ as shown in Fig.\ref{deformationfig2}. In contrast to the $\lambda>0$ case, the dual gravity region does not include the falling particle but is affected by its backreaction.
 
 We would like to note that if $M$ is large enough such that $M>\frac{3}{4}R^2$, then the EOW brane gets folded as in the right panel of Fig.\ref{deformationfig} and the gravity dual does not seem to make sense. As discussed in \cite{Cooper:2018cmb} (see also \cite{Miyaji:2021ktr,Geng:2021iyq}), this kind of folded solution implies that the hard wall approximation is no longer valid in the presence of strong backreaction. Nevertheless, as we will see later, a suitable treatment of the AdS/BCFT yields $M$ is always less than or equal to $\frac{3}{4}R^2$, given a conformal dimension less than the black hole threshold.

\begin{figure}
\centering
    \includegraphics[width=4cm]{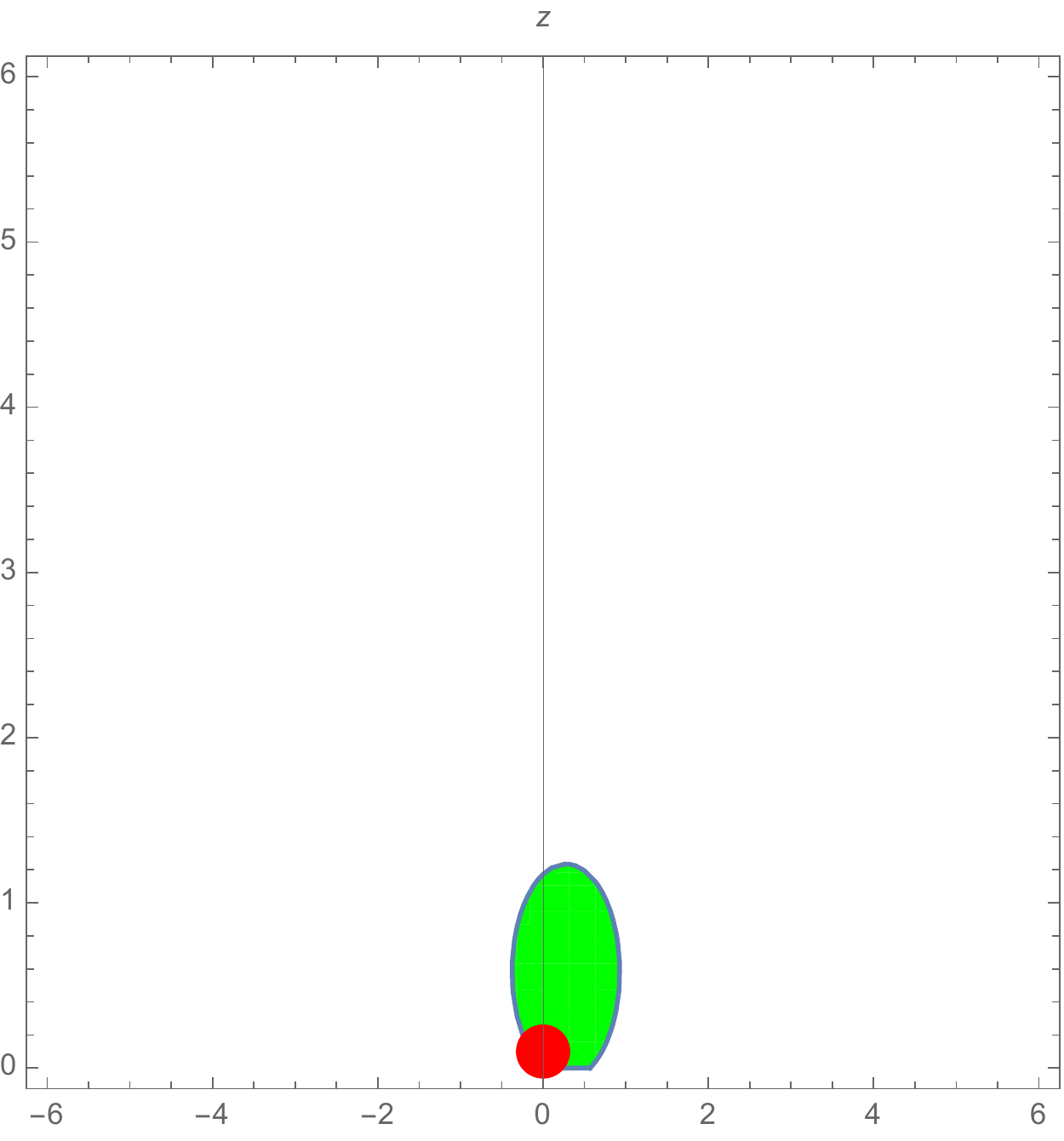}
    \vspace{1cm}
    \includegraphics[width=4cm]{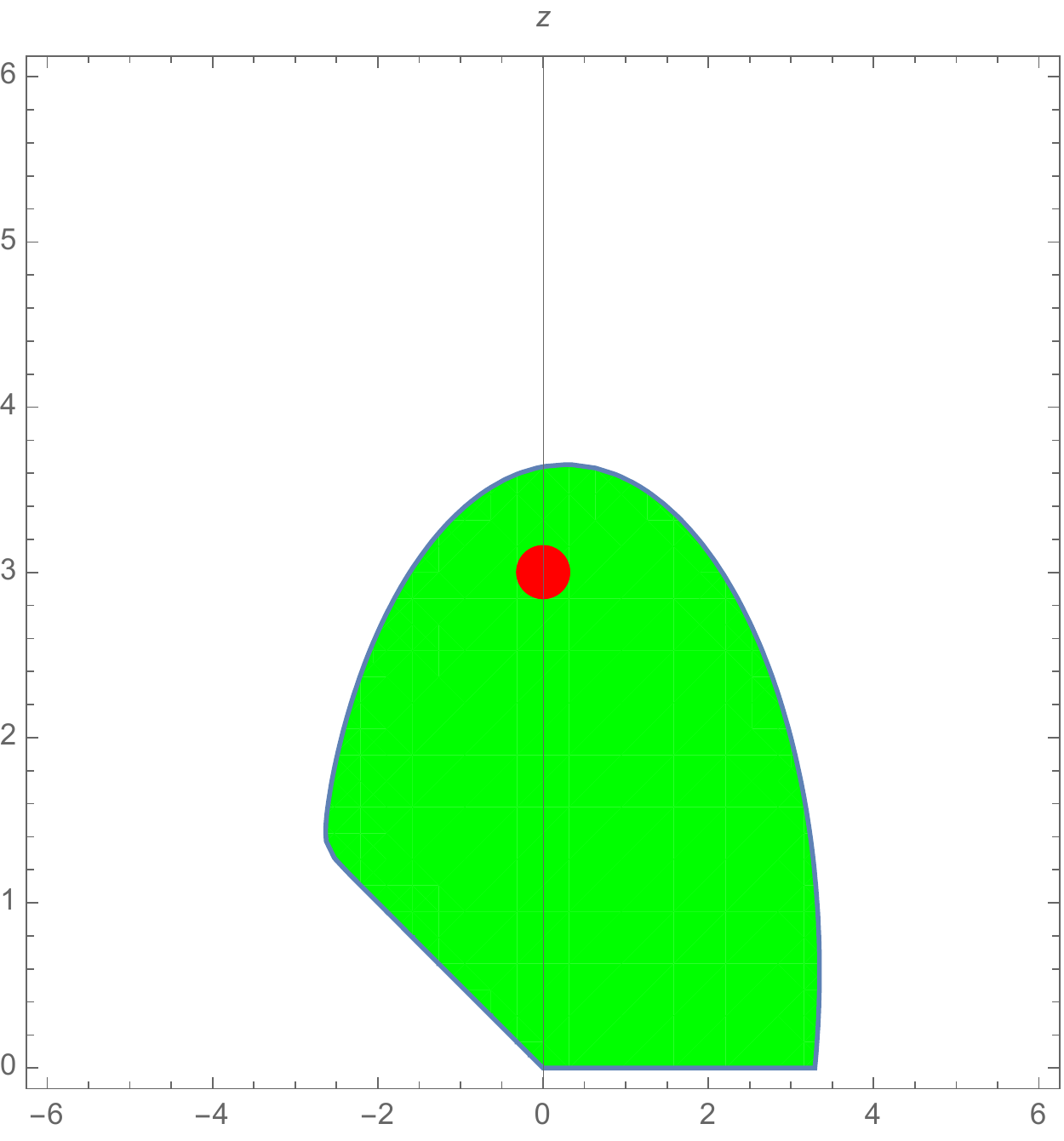}
    \vspace{1cm}
     \includegraphics[width=4cm]{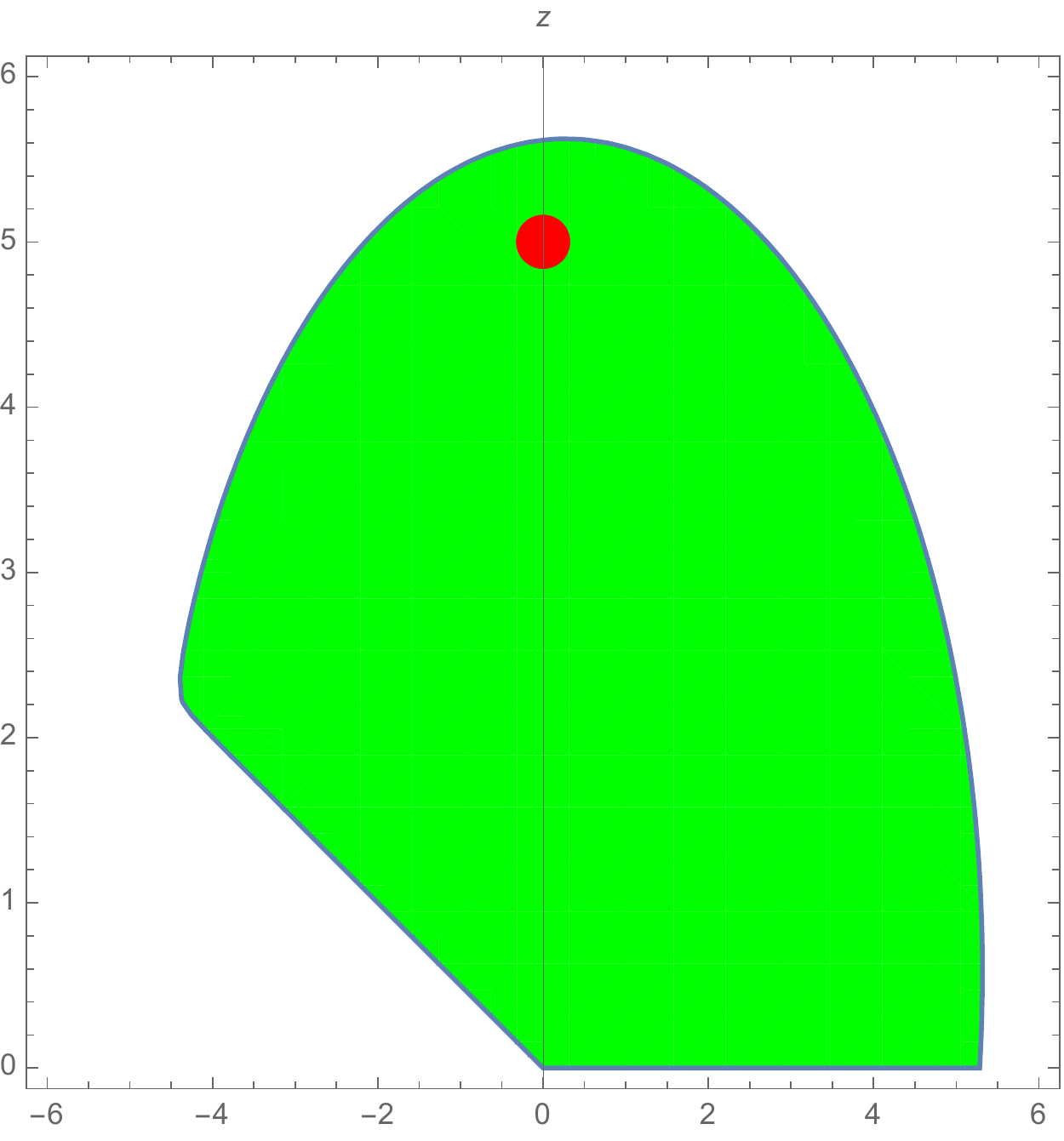}
  \caption{Time evolution of an EOW brane and a massive particle in the Poincare coordinates  with  $\alpha=0.1$, $\chi=0.9$ and $\lambda=2$. The vertical line is $z$ axis (the bulk direction) and the horizontal line is $x$ axis (the boundary spatial direction). The green region represents the gravity dual region of the BCFT  and the red dot represents the position of the massive particle. The EOW branes attaches at the boundary at $x=0$ and $x=Z(t)$. We chose the time to be $t=0,3$ and $5$ in the left, middle and right panel. As the time evolves the EOW brane probes deeper in the bulk.}
  \label{timeetw135}
\end{figure}

By applying the map (\ref{mappo}), assuming $0<M<\frac{3}{4}R^2$, 
we find that the BCFT dual to the Poincare patch has two boundaries which are the two intersections of  the AdS boundary $z=0$ and the EOW brane: $x=0$ and 
\ba
x=\pm\left( \frac{\ap}{\gamma}+\s{\ap^2\left(1+\frac{1}{\gamma^2}\right)+t^2} \right)\equiv \pm Z(t), \quad Z(t) >0,\label{disco}
\ea
where 
\ba
\gamma=\tan\left(\frac{\pi}{\chi}\right). 
\ea
The $+$ sign corresponds to the $\lambda>0$ case while the $-$ sign corresponds to the $\lambda<0$ case. The BCFT lives on the spacetime defined by $0\le x \le Z(t)$ for $\lambda>0$ and $\{ 0\le x, x \le Z(t) \}$ for $\lambda<0$. In the following discussions, we focus on the $\lambda>0$ case, though, the $\lambda<0$ case is treated in the same manner.

We plot the time evolution of a massive particle and the EOW brane, see Fig  \ref{timeetw135}. We can see that the EOW brane bends toward the conformal boundary due to the massive particle.

In summary, the BCFT lives on the spacetime defined by $0\le x \le Z(t)$ (for $\lambda>0$).
This geometry at the AdS boundary is depicted in the left panel of Fig.\ref{deformationfig}. Note that  the location of the second boundary is time dependent and at late times it almost expands at the speed of light. As we will see later, this geometry naturally arises from a conformal transformation in a two-dimensional BCFT.

\begin{figure}
  \centering
  \includegraphics[width=8cm]{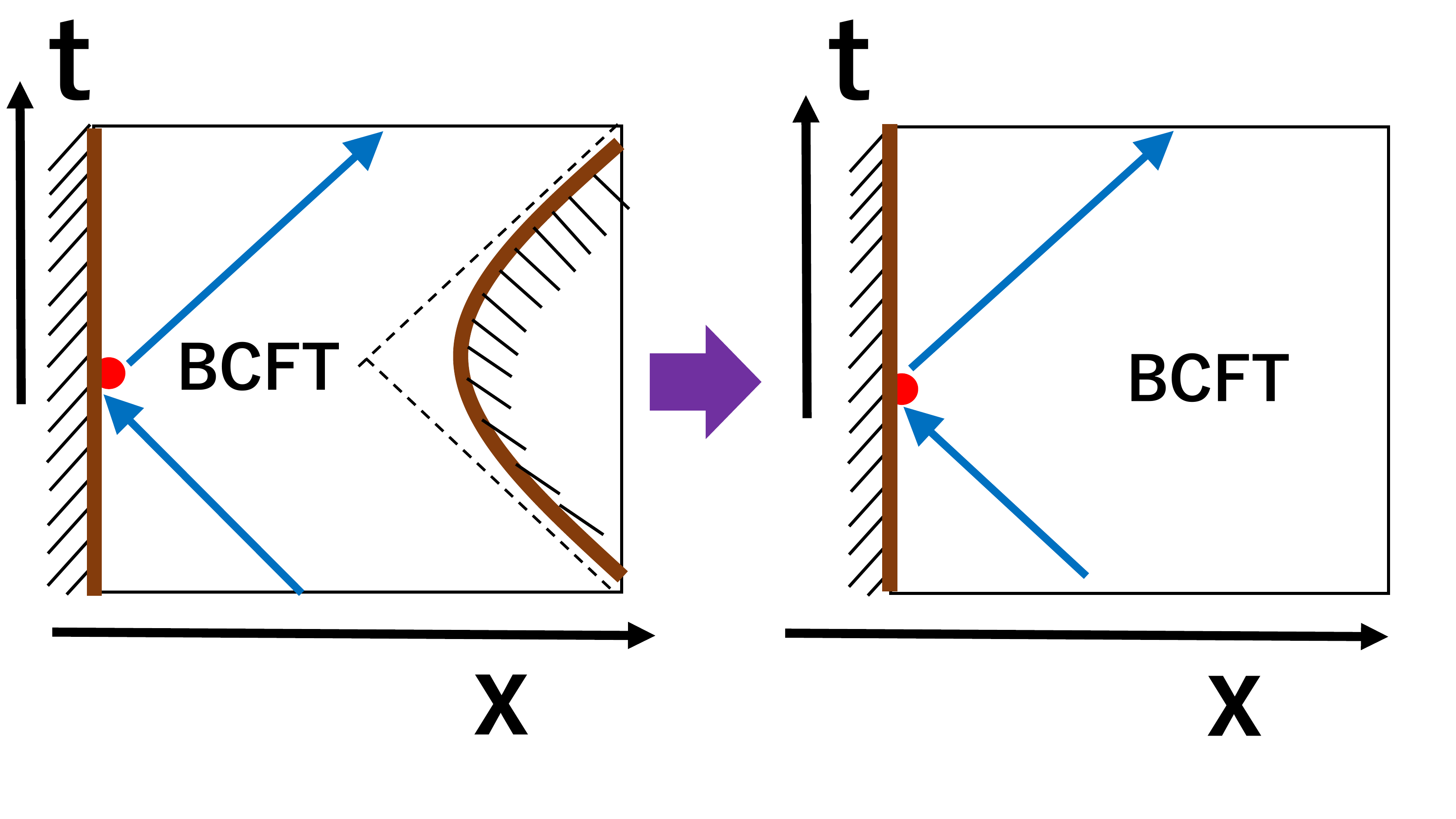}
  \caption{The left picture sketches the setup of BCFT in the presence of the two boundaries: $x=0$ and (\ref{disco}). We may try to remove the right boundary by a coordinate transformation (right picture).}
\label{setupsbdyfig}
\end{figure}

In principle, it is also possible shift the location of the EOW brane for  $a<0$ in (\ref{eomb}).
It is given by the surface
\ba
\frac{\ti{r}}{R}\sin\theta+\frac{a}{R\ap} (\s{R^2+\ti{r}^2}\cos\ti{\tau}+\ti{r}\cos\ti{\theta})+\lambda=0.
\ea
This is dual to a local operator quench (\ref{LOS}) at $x_a=-a$, by shifting the location of boundary from $x=a$ to $x=0$. Since this gives a complicated time-dependent spacetime, we will not discuss this further. 

On the other hand, for $M>R^2$, the spacetime (\ref{GBmet}) describes a BTZ black hole. We can still analytically continue the expression (\ref{teha}) and (\ref{profti}), we obtain
\ba
r\sinh\left(\frac{\s{M-R^2}}{R}\theta\right)=-\lambda \s{M-R^2}.
\ea
This EOW brane extends from the AdS boundary to the black hole horizon. We will not get into this black hole setup in more detail as our comparison with the CFT result can be done with the deficit angle geometry.

\subsection{Coordinate Transformation}
As we have found in the previous section, 
for $0<M<\frac{3}{4}R^2$, the AdS/BCFT setup is given by the boundary surface (\ref{beoms})
in the deficit angle geometry (\ref{GBmet}). Via the coordinate transformation 
(\ref{mappo}), it is mapped into the asymptotically Poincare AdS geometry whose boundaries consist of two segments $x=0$ and $x=Z(t)$ (\ref{disco}) as depicted in the right of Fig.\ref{setupsbdyfig}. The appearance of the second boundary may not be surprising because the massive particle in the center of the global AdS attracts the EOW brane toward the center and this backreaction bends the brane such that its intersection with the AdS boundary gets shifted towards the first boundary $x=0$.  
Originally, however, we have intended a local operator quench in a BCFT on a half place (the right panel of Fig.\ref{setupsbdyfig}), 
instead of the region surrounded by two boundaries 
(the left one of  Fig.\ref{setupsbdyfig}).

To resolve this issue, we would like to perform the following rescaling of the global AdS coordinates:
\ba
\theta'=\eta\theta,\ \ \tau'=\eta\tau,\ \ r'=r/\eta.  \label{rescaleex}
\ea
Applying this to the metric \eqref{GBmet} gives
\ba
ds^2=-(r'^2+R^2-M')d\tau'^2+\frac{R^2}{r'^2+R^2-M'}dr'^2+r'^2d\theta'^2,
\ea
where
\ba
M'=\frac{M}{\eta^2}+R^2\left(1-\eta^{-2}\right).  \label{newmass}
\ea
Note that this is equivalent to 
\begin{equation}
    \chi=\sqrt{\frac{R^2-M}{R^2}}\rightarrow \frac{\chi}{\eta}=\sqrt{\frac{R^2-M^\prime}{R^2}}.
    \label{eq:chi-transform}
\end{equation}
Since the asymptotically AdS region surrounded by the surface $Q$ (\ref{beoms}) is 
given by $0<\theta<\left(2-\frac{1}{\chi}\right)\pi$, if we choose
\ba
\eta\geq \eta_0\equiv \frac{1}{2-{1}/{\chi}},  \label{mineta}
\ea
then the range of the new angular coordinate $\theta'$ takes 
$0<\theta'<\theta'_{max}$ with $\theta'_{max}\geq \pi$  on the asymptotically AdS boundary.\footnote{Notice that even when $\lambda<0$, this rescaling by $\eta\geq\eta_0$ takes the second boundary $\theta=-\left(2-\frac{1}{\chi}\right)\pi$ to $\theta^\prime\leq-\pi$. Thus, by the same coordinate transformation, the BCFT with a negative tension becomes a half space.} When $\eta=\eta_0$, we have $\theta'_{max}=\pi$.

Therefore, if we apply the coordinate transformation\footnote{Here we mean $r'\sin\theta'=\frac{Rt'}{z'}$ for example.} 
(\ref{mappo}) with $(r,\theta,\tau)$ and $(z,x,t)$ replaced with 
$(r',\theta',\tau')$ and $(z',x',t')$ with $\eta\geq\eta_0$, then the resulting asymptotically Poincare AdS, given by the coordinate $(z',x',t')$,
includes only a single boundary $x'=0$ since the Poincare patch only covers the regime $-\pi\le\theta^\prime< \pi$.  While the falling particle trajectory remains the same $z'=\s{t'^2+\ap^2}$, the shape of EOW brane gets modified. In this way,  we can realize the gravity setup dual to the local operator quench on a half plane by choosing $\eta\geq \eta_0$. Especially when $\eta=\eta_0$, the entire BCFT and its gravity dual region is covered even after the Poincare patch. We will justify this correspondence after the rescaling by comparing the bulk calculation with that in the holographic CFT in the subsequent sections.

\subsection{Holographic Energy-momentum Tensor}\label{sec:m-less-r}

One way to check the validity of the dual CFT interpretation is to compute the holographic energy-momentum tensor \cite{Balasubramanian:1999re}.
 The value of the holographic energy-momentum tensor in our setup before we perform the coordinate transformation (\ref{rescaleex})
 is exactly the same as that without the boundaries, which was computed in 
\cite{Nozaki2013}:
\ba
T_{--} (M)=\frac{M\ap^2}{8\pi G_N R\left((t-x)^2+\ap^2\right)^2},\ \ 
T_{++} (M)=\frac{M\ap^2}{8\pi G_N R\left((t+x)^2+\ap^2\right)^2}. \label{eflux}
\ea
The mass $M$ is related to the mass $m$ of the particle via (\ref{relam}).
One might expect that the conformal dimension of the dual operator $O(x)$ for the local quench is given by 
$mR$ via the familiar correspondence rule. However, this is not completely correct due to the reason we will explain soon later. Instead, we introduce $\Delta_{AdS}$ to distinguish this from the correct conformal dimension $\Delta_O$ of the dual primary operator in the BCFT and write as follows:
\ba
mR\simeq \Delta_{AdS}.  \label{dimfo}
\ea

\begin{figure}
  \centering
  \includegraphics[width=8cm]{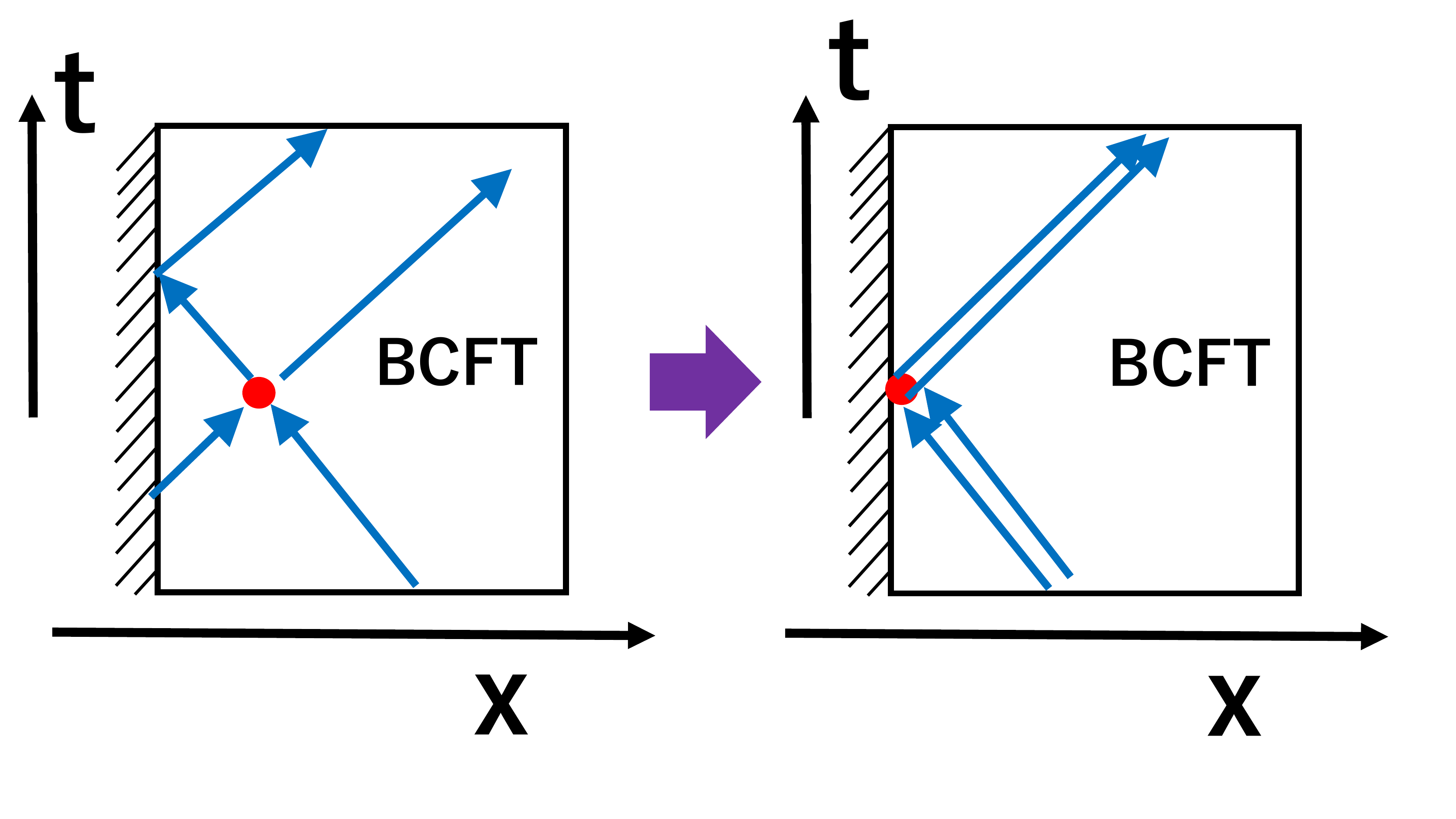}
  \caption{The energy fluxes from an excitation (red point) in the presence of 
  a boundary. When the excitation coincides with the boundary, the energy flux is doubled as in the right picture.}
\label{doublefluxfig}
\end{figure}

Indeed, the result of energy fluxes may look confusing at first because the holographic expression  (\ref{eflux}) looks identical to that without any EOW brane inserted. For a local operator quench in the presence of a boundary, we actually expect that the energy fluxes will be doubled due to the mirror charge effect as explained in Fig.\ref{doublefluxfig}. As we will confirm from the CFT calculation later, even though the boundary $x=0$ produces the doubled flux, the presence of the other boundary (\ref{disco}) reduces the energy fluxes, which is analogous to the Casimir effect.

This also gives another motivation for performing the previous coordinate transformation (\ref{rescaleex}) as this removes the extra boundary from the dual BCFT.  The energy flux after the transformation is simply given by (\ref{eflux}) with $M$ replaced with $M'$ in (\ref{newmass}). This leads to 
a class of asymptotically Poincare AdS solutions with an EOW brane which is specified by the two parameters 
$M$ and $\eta$. The energy flux is obtained by replacing $M$ in (\ref{eflux}) with $M'$ given by (\ref{newmass}), which is a monotonically increasing function of $\eta$.  Note that 
$M$ is related to $\Delta_{AdS}$ via (\ref{dimfo}) and (\ref{relam}):\footnote{It is useful to note that $\chi=\sqrt{\frac{R^2-M}{R^2}}=\sqrt{1-12\frac{\Delta_{AdS}}{c}}$ from \eqref{relationm}.}
\ba
\frac{M}{R^2}=12\frac{\Delta_{AdS}}{c}, \label{relationm}
\ea
which leads to the following holographic energy-momentum tensor in the BCFT language:
\ba
&& T_{\pm\pm} (M^\prime)=s_{AdS}(\Delta_{AdS},\eta)\cdot\frac{\ap^2 }{\pi \left((t\pm x)^2+\ap^2\right)^2},\no
&& s_{AdS}(\Delta_{AdS},\eta)\equiv\frac{\Delta_{AdS}}{\eta^2}+\frac{c}{12}
\left(1-\frac{1}{\eta^2}\right).
\label{efluxx}
\ea
It is also useful to note that the energy density $T_{tt}$ is given by 
\ba
&& T_{tt}=T_{++}+T_{--}=2s_{AdS}(\Delta_{AdS},\eta)\cdot H(t,x),\no
&& H(t,x)=\frac{\ap^2\left((t^2+x^2+\ap^2)^2+4t^2x^2\right)}{\pi\left((x^2-t^2-\ap^2)^2+4\ap^2x^2\right)^2}.\label{energyd}
\ea

On the other hand, the other parameter $\eta$ describes the degrees of freedom of gravitational excitations dual to those of descendants in the BCFT and this affects the shape of EOW brane. In other words, $\eta$ corresponds to a conformal transformation.
Since the cylinder coordinates and the plane coordinates are related by the conformal transformation 
(\ref{confmap}), changing $\eta$ induces further conformal transformation:
\ba
t'\pm x'=\ap\tan\left(\frac{\tau'\pm \theta'}{2}\right)=\ap\tan\left(\frac{\eta(\tau\pm \theta)}{2}\right).
\label{conf-map-eta}
\ea
The energy-momentum tensor is generated from this via the Schwarzian derivative term.
This is dual to the descendant excitations of the two-dimensional CFT. Note that if the two solutions $(M_1,\eta_1)$ and $(M_2,\eta_2)$ satisfies
\ba
\frac{R^2-M_1}{\eta_1^2}=\frac{R^2-M_2}{\eta_2^2},\label{relpa}
\ea
or equally $\chi_1/\eta_1=\chi_2/\eta_2$, then 
the energy-momentum tensors become identical. This means that the asymptotic metric near  the AdS boundary is the same.  However they are different globally,  because of the different deficit angle due to the massive particle.  

In particular, when $\eta=\eta_0$ (\ref{mineta}), 
the energy flux gets minimized among those dual to a geometry with a single boundary $x=0$ and thus we expect $\eta=\eta_0$ corresponds to the BCFT with a local operator excitation without any other excitations. In this case, the energy-momentum tensor is given by (\ref{efluxx}) with
$s_{AdS}(\Delta_{AdS},\eta_0)$ takes 
\ba
s_{AdS}(\Delta_{AdS},\eta_0)=\frac{c}{3}\s{1-\frac{12\Delta_{AdS}}{c}}
\left(1-\s{1-\frac{12\Delta_{AdS}}{c}}\right).
\label{efluxy}
\ea
We argue that in this case $\eta=\eta_0$, $s_{AdS}$ is directly related to the conformal dimension of 
the primary $O(x)$ in the dual BCFT by
\ba
s_{AdS}(\Delta_{AdS},\eta_0)=2\Delta_O.
\label{eq:AdS EM tensor}
\ea
This is because this setup is the BCFT defined on the right half plane $x>0$ and the energy flux should be simply twice of that in the same CFT without any boundary. We will show that this is true from the explicit CFT calculation in \S\ref{sec:EMtensor-holCFT}. It is also useful to note that 
this relation between $\Delta_O$ and $\Delta_{AdS}$ can be expressed as follows: 
\ba
2\s{1-\frac{12\Delta_{AdS}}{c}}-1=\s{1-\frac{24\Delta_O}{c}},\label{relasd}
\ea
and when $\Delta_O,\Delta_{AdS}\ll c$, we find 
\ba
\frac{\Delta_{AdS}}{c}=\frac{\Delta_O}{c}+3\left(\frac{\Delta_O}{c}\right)^2+\ddd.
\ea
It is also helpful to note that the standard range of primary operator below the black hole threshold given by $0<\Delta_O<\frac{c}{24}$ corresponds\footnote{Notice that $\Delta_O$ is the total conformal dimension i.e. the sum of chiral and anti-chiral conformal dimension.} to the range $\frac{1}{2}<\s{1-\frac{12\Delta_{AdS}}{c}}=\chi<1$. This is consistent with the previous observation\footnote{This bound is equivalent to $\Delta_{AdS}<c/16$ given in \cite{Geng:2021iyq} by replacing $\Delta_{\mathrm{bcc}}$ in their paper with $\Delta_{AdS}$.} that the EOW brane configuration makes sense only when $0<M<\frac{3}{4}R^2$  in order to avoid the self-intersection of the brane.\footnote{The self-intersection problem also appears in higher dimensions \cite{Fallows:2022ioc}. Although our analysis focus on $d=2$, a similar analysis may circumvent the problem likewise.}
One important difference, which has been confused previously, is that this bound $\Delta_O<\frac{c}{24}$ is exactly equivalent to the black hole threshold for $2\Delta_O$ through the correct relation \eqref{relasd}. Indeed, the dimension gets actually doubled in the presence of the boundary due to the mirror effect as we will explain from the BCFT viewpoint in section 4. Notice also that we can in principle extend to the heavier excitation with $\Delta_O\geq \frac{c}{24}$ by analytically continuing the formula (\ref{relasd}), where $\Delta_{AdS}$ and therefore the mass $M$ gets complex valued.

\section{Time Evolution of Holographic Entanglement Entropy}\label{sec:holoEE}

In this section, we calculate the time evolution of the holographic entanglement entropy in our gravity dual of the local operator quench in the BCFT. We can analytically obtain results since the holographic setup is related to a global AdS$_3$ with an EOW brane and a deficit angle through the chain of coordinate transformations, as we saw in the previous section. 

Our goal is to calculate the holographic entanglement entropy in the asymptotically Poincare AdS$_3$ background. On this boundary, we define a subsystem $I$ to be an interval $I$ between two points $A$ and $B$ at each time $t$. We write the spatial coordinate of $A$ and $B$ as $x_A$ and $x_B$, respectively, assuming $x_{A}< x_{B}$. Then we follow the time evolution of the entanglement entropy $S_{AB}$ in our BCFT as a function of the boundary time $t$. 

The entanglement entropy $S_{AB}$ is defined as usual by first introducing the (time-dependent) reduced density matrix $\rho_{AB}(t)$ by tracing out the complement $I^c$ of the interval $I=[x_A,x_B]$ from the density matrix for the operator local quench state (\ref{LOS}):
\ba
\rho_{AB}(t)=\mbox{Tr}_{I^c}\left[|\Psi(t)\lb\la\Psi(t)|\right].
\label{eq:reduced-rho}
\ea
The entanglement entropy is defined by the von-Neumann entropy as a function of time $t$:
\ba
S_{AB}(t)=-\mbox{Tr}[\rho_{AB}(t)\log\rho_{AB}(t)].
\ea

\subsection{Holographic Entanglement Entropy in AdS/BCFT}
In AdS$_3/$CFT$_2$, the holographic entanglement entropy is computed by the length of geodesics which anchor the boundary points $A:(z=\epsilon,x=x_{A},t)$ and $B:(z=\epsilon,x=x_{B},t)$, where $\epsilon$ is the UV cutoff surface \cite{Ryu:2006bv,Ryu:2006ef,Hubeny:2007xt}.

\begin{figure} 
  \centering
  \includegraphics[width=6cm]{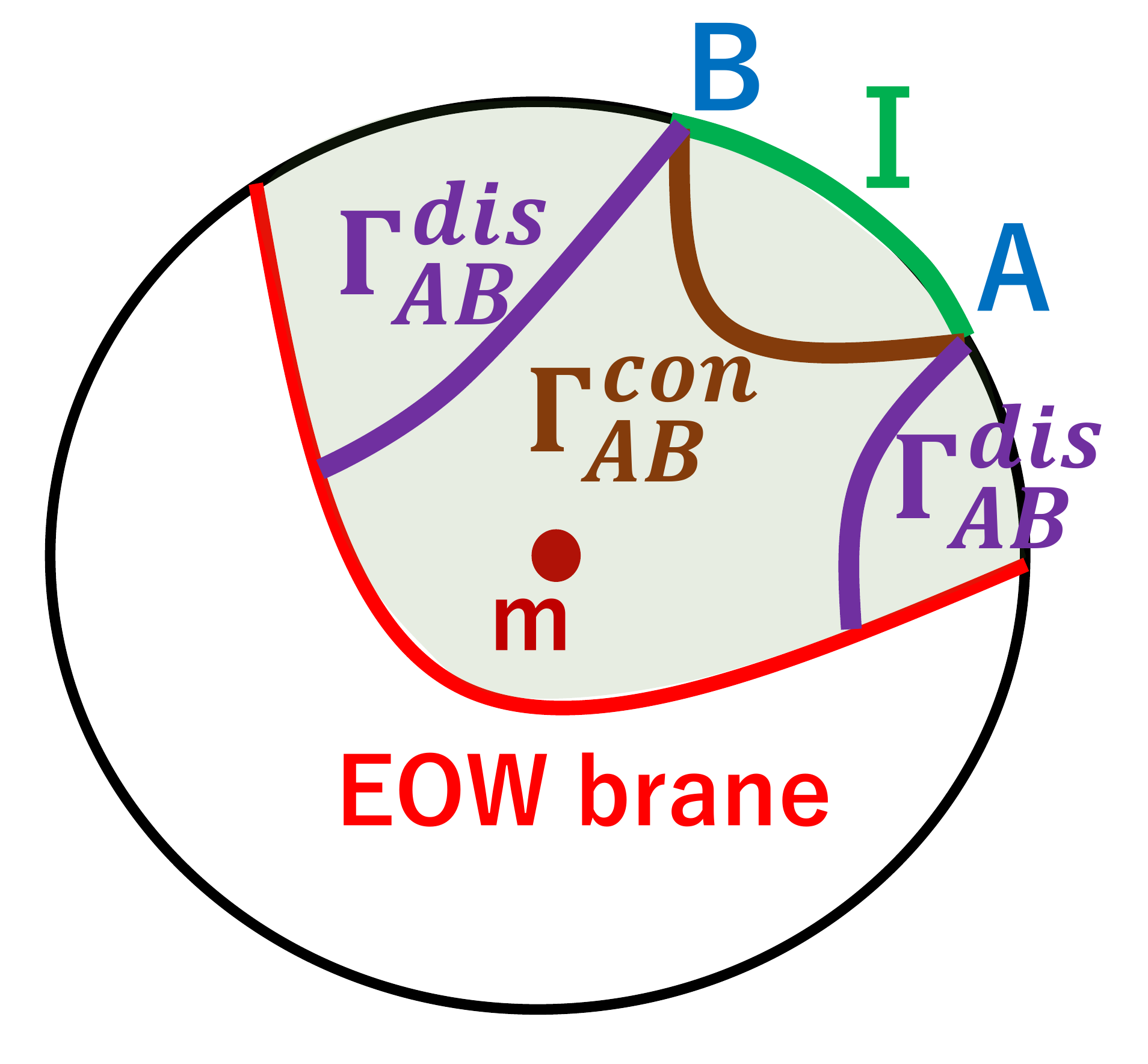}
  \caption{A sketch of calculation of holographic entanglement entropy in AdS/BCFT. We showed that the connected geodesic $\Gamma^{con}_{AB}$ (brown curve) and the disconnected geodesics $\Gamma^{dis}_{AB}$ (purple curves) for a boundary subsystem given by an interval $I$ between $A$ amd $B$.}
\label{HEEfig}
\end{figure}

In the presence of an EOW brane, we need to take into account geodesics which end on the EOW brane \cite{Takayanagi2011,Fujita:2011fp}, which leads to multiple candidates of geodesics. One is the ``connected geodesic" $\Gamma^{con}_{AB}$, which literary connects these two boundary points $A$ and $B$. The other is the ``disconnected geodesics" $\Gamma^{dis}_{AB}$ which consist of two disjoint pieces, one connects the boundary point $A$ and a point on the EOW brane, and the other connects $B$ and ends on the EOW brane. The net result for the holographic entanglement entropy is given by taking the minimum of these two contributions,
\be
S_{AB} ={\rm Min} \left\{ S^{con}_{AB},  S^{dis}_{AB}\right\},
\ee
where
\be
S^{con}_{AB}=\frac{L(\Gamma^{con}_{AB})}{4G_N}, \ \ \ \ \ \ 
S^{dis}_{AB}=\frac{L(\Gamma^{dis}_{AB})}{4G_N}.
\ee
$L(\Gamma)$ denotes the length of the geodesic $\Gamma$ and $G_N$ is the Newton constant 
as sketched in Fig.\ref{HEEfig}.

Below we separately compute these two contributions, then find the actual value of the entropy.  We initially  do not perform the rescaling (\ref{rescaleex}), i.e. we set $\eta=1$, where the gravity dual is given by the BCFT on $0<x<Z(t)$, where 
$Z(t)$ is given in (\ref{disco}). Later, we will extend our analysis to $\eta>1$ case.

\subsection{Connected Entropy}

First, we would like to calculate the connected entropy $S^{con}_{AB}$.
As we review in  \ref{sec:Embedd}, an efficient  way to compute the length of a geodesic in $AdS_{3}$ is using its embedding to a flat space $\mathbb{R}^{2,2}$, where we derive the detailed formula. Below we show only the  results. The geodesic length connecting two boundary points $ A: (\theta_{A}, \tau_{A}, r_{A})$  and $B: (\theta_{B}, \tau_{B}, r_{B})$ in the metric \eqref{GBmet} is obtained by rescaling  (\ref{AppendixA EE}) by \eqref{teha}. The corresponding holographic entanglement entropy reads
\begin{equation}
    S^{con}_{AB} = \frac{c}{6}\log{\left[ \frac{2r_A r_B}{R^2 \chi^2}(\cos(\chi(\tau_A-\tau_B))-\cos(\chi(\theta_A-\theta_B))\right]},
    \label{connected HEE}
\end{equation}
where
\ba
\chi = \sqrt{\frac{R^2-M^2}{R^2}}= \s{1-\frac{12\Delta_{AdS}}{c}},
\ea
as we have introduced 
and $c$ is the central charge $c= \frac{3R}{2G}$ of the dual CFT \cite{Brown:1986nw}.

\begin{figure}[t]
    \centering
    \includegraphics[width = 60mm]{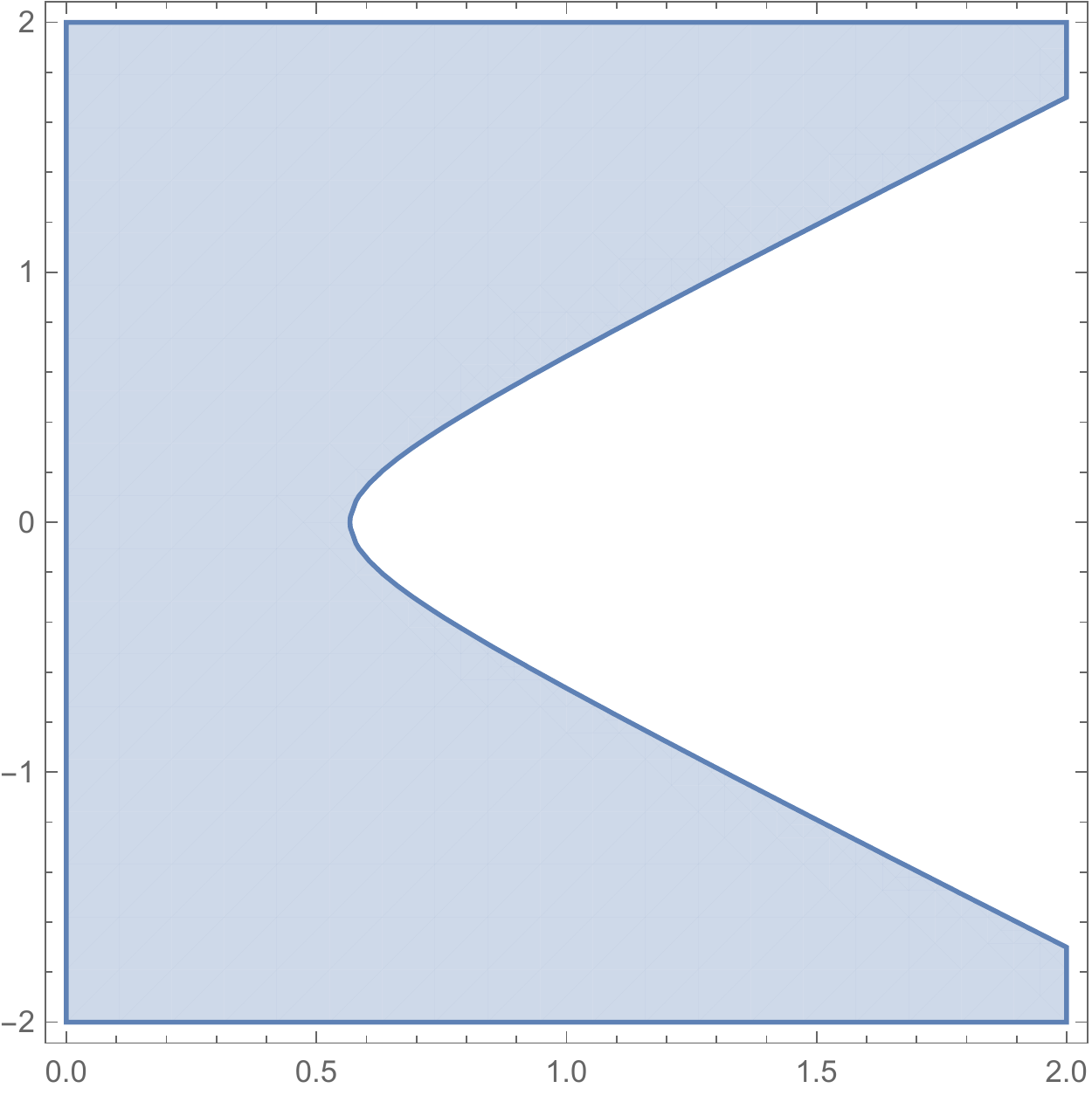}
    \hspace{1cm}
    \includegraphics[width = 60mm]{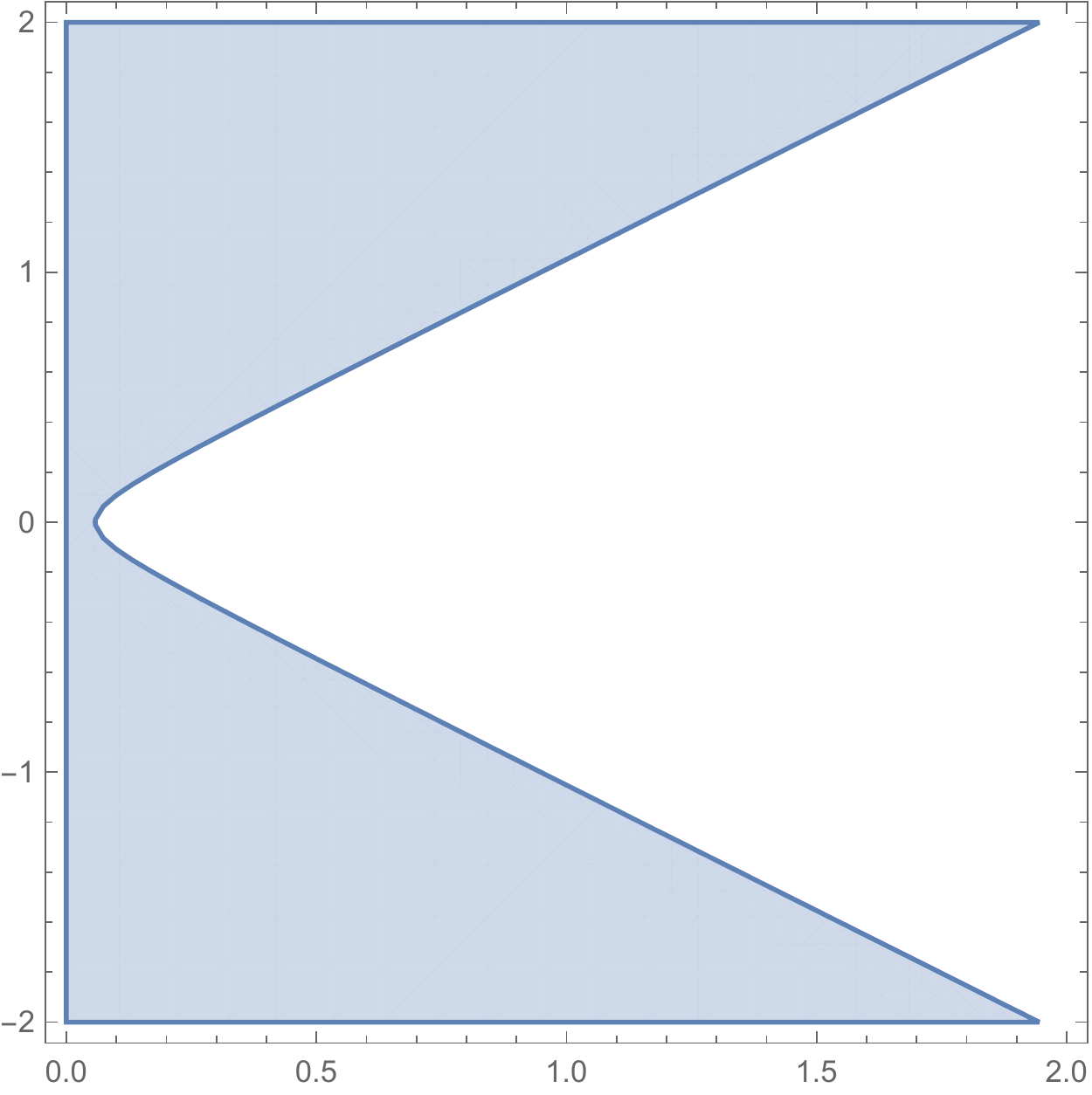}
    \caption{Plot of the physical region $0<x<Z(t)$ of our BCFT in $(x,t)$ plane. We chose $\chi=0.9$ in the left plot and $\chi=0.6$ in the right plot.
    Again we chose $R=1$ and $\ap=0.1$.}
    \label{fig:regionplot}
\end{figure}

Although we  presented the formula for the holographic entanglement entropy in global coordinates, ultimately we are interested in its expression in the Poincare coordinates where the setup of the local quench is  introduced.  By restricting the coordinate transformation (\ref{mappo}) at the AdS boundary $z=\ep$, we obtain the boundary map:
\ba
&& e^{i\tau}=\frac{\ap^2+x^2-t^2+2i\ap t}{\s{(x^2-t^2)^2+2\ap^2(t^2+x^2)+\ap^4}},\no
&& e^{i\theta}=\frac{\ap^2-x^2+t^2+2i\ap x}{\s{(x^2-t^2)^2+2\ap^2(t^2+x^2)+\ap^4}},\no
&& r=\frac{R}{2\ap \ep}\s{(x^2-t^2)^2+2\ap^2(t^2+x^2)+\ap^4}. \label{thrmap}
\ea
Mapping the end points $A=(x_{A},t_{A},\epsilon)$ and $B$ in the Poincare coordinates by the above transformation to those in the global AdS coordinates, we can calculate the geodesic length from the formula 
(\ref{connected HEE}). Note that we need to choose $A$ and $B$ within the region of BCFT i.e. $0<x_A<x_B<Z(t)$. Refer to Fig.\ref{fig:regionplot} for an explicit plot.

\begin{figure}[h]
    \centering
    \includegraphics[width = 60mm]{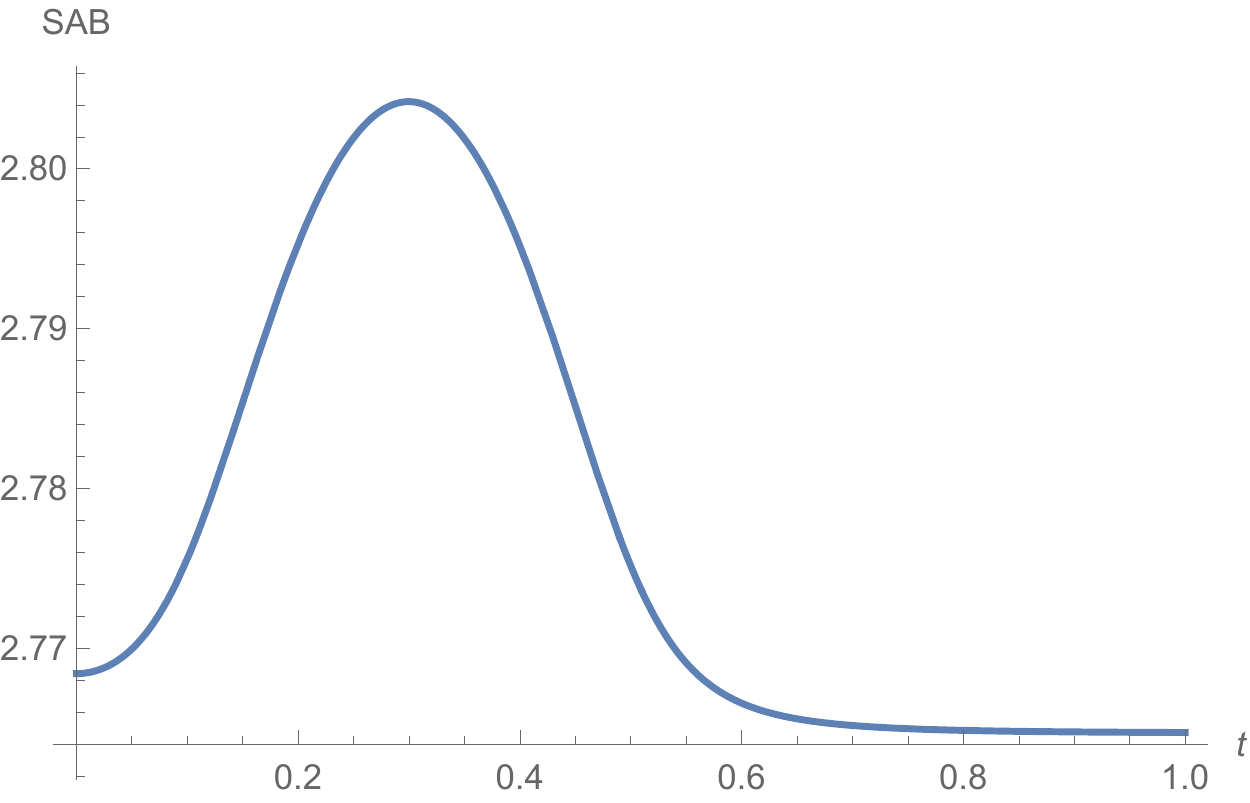}
    \hspace{1cm}
    \includegraphics[width = 60mm]{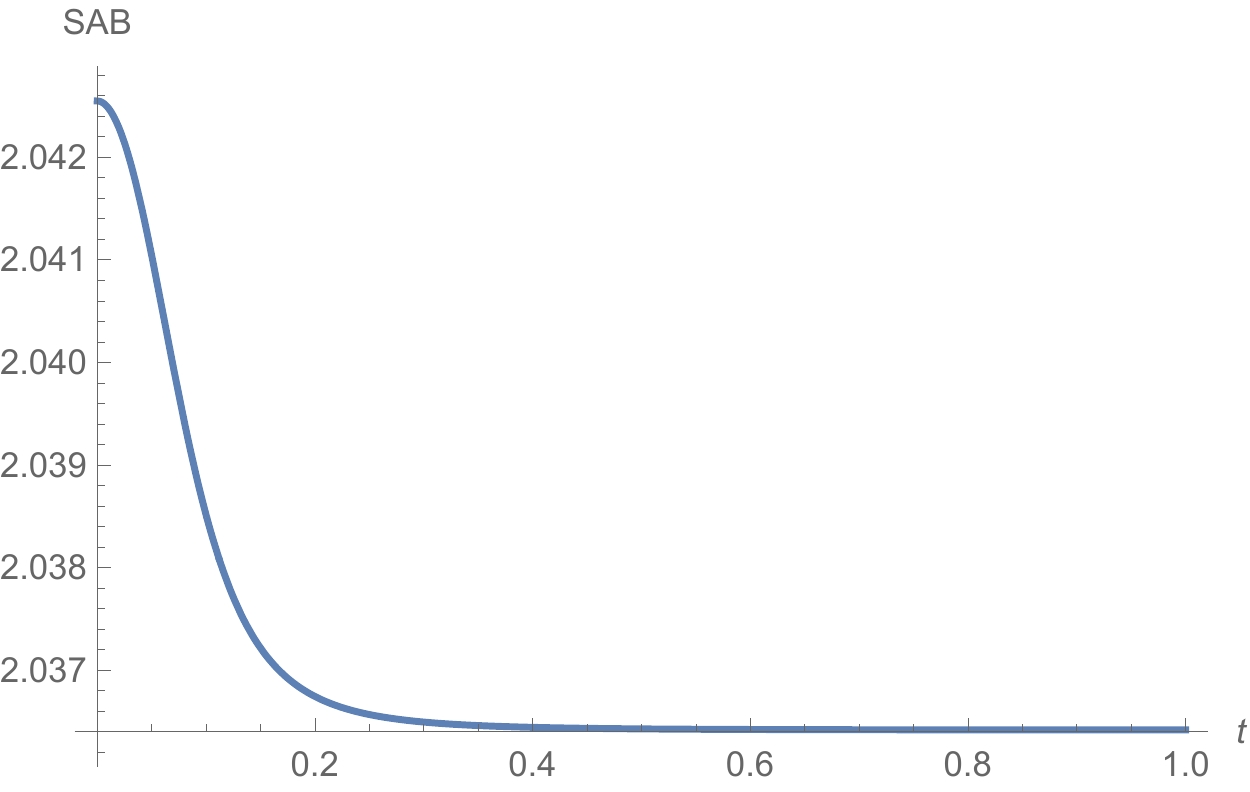}
    \caption{Time evolution of the connected entanglement entropy $S^{con}_{AB}$. In the left plot, we chose $\chi=0.9$ and $(x_A,x_B)=(0.1,0.5)$. In the right one, we chose $\chi=0.6$ and $(x_A,x_B)=(0.005,0.05)$. 
    In both, we took $R=c=1$, $\ep=0.0001$ and $\ap=0.1$.
    }
    \label{fig:connected entropy}
\end{figure}

We plot the time evolution of the connected entanglement entropy with a given interval for $\chi=0.9$ and $\chi=0.6$ in Fig.\ref{fig:connected entropy}. 
We can see that there is a peak when the shock wave from the falling particle hits the center of the interval $I$ i.e. $\s{\ap^2+t^2}\simeq \frac{x_A+x_B}{2}$ \cite{Nozaki2013}. Moreover the final value is given by the vacuum entanglement entropy \cite{Holzhey:1994we}
\begin{equation}
    S_{AB}= \frac{c}{3}\log\left[ \frac{x_B-x_A}{\epsilon}\right].
\end{equation}

It is also useful to examine the first law of entanglement entropy \cite{Bhattacharya:2012mi,Blanco:2013joa}, which states that the growth of the entanglement entropy $\Delta S_{AB}$, defined by the difference of the entanglement entropy of an excited state and that of the CFT vacuum, is directly related to the energy density $T_{tt}$ in the small subsystem limit $|x_A-x_B|\to 0$ via
\ba
T_{tt}(x_A,t)=\lim_{|x_A-x_B|\to 0}\frac{3 }{\pi|x_A-x_B|^2}\cdot\Delta S_{AB}(x_A,x_B,t).  \label{firstL}
\ea
In this short subsystem limit, we find after some algebra that our holographic entanglement entropy (\ref{connected HEE}) behaves as 
\ba
&& \Delta S^{con}_{AB}=\frac{c}{6}\log{\left[ \frac{\cos(\chi(\tau_A-\tau_B))-\cos(\chi(\theta_A-\theta_B))}{\chi^2\left(\cos(\tau_A-\tau_B)-\cos(\theta_A-\theta_B)\right)}\right]}\no
&&\simeq \frac{c}{18}(1-\chi^2)H(t,x_A)(x_A-x_B)^2,
\ea
where $H(t,x)$ is defined in (\ref{energyd}). Thus, we can confirm that the first law (\ref{firstL}) perfectly reproduces the energy-momentum tensor (\ref{efluxx}) and (\ref{energyd}) at $\eta=1$.

\subsection{Disconnected Entropy}

Next, we will consider the disconnected contribution $S^{dis}_{AB}$. We will work in the tilde coordinates again. This time, we consider the geodesics that stick to the EOW brane perpendicularly. The formula for the disconnected entropy is obtained in the tilde Poincare coordinates, where the EOW brane is simply given by 
the plane $\tilde{x}=-\lambda \tilde{z}$. In 
this tilde coordinates, we can employ the known result \cite{Takayanagi2011,Fujita:2011fp}
\ba
S^{dis}_{AB}=\frac{c}{6}\log\frac{2l}{\ep}+S_{bdy},
\ea
for each of the disconnected geodesics, where $S_{bdy}$ is the boundary entropy \cite{Affleck:1991tk} and is given by 
$ \frac{c}{6}\sinh^{-1}\lambda$. 
However, we have to keep in mind that in this tilde Poincare coordinates we need to care about the periodicity of $\theta$ due to the conical defect. A careful consideration results in the following expression,
\begin{equation}\label{disconnected HEE}
  S^{dis}_{AB}= \frac{c}{6} \log{\left(\frac{2r_A}{R\chi} \sin(\chi\theta_A^{\text{min}})\right)}+\frac{c}{6} \log{\left(\frac{2r_B  }{R\chi} \sin(\chi\theta_B^{\text{min}})\right)}+\frac{c}{3}\sinh^{-1}\lambda,
\end{equation}
where $\theta^{\text{min}}$ is defined as the smaller angle measured from $x=0$ or $x=Z(t)$:
\begin{equation}
    \theta^{\text{min}}  = \min{\left[\theta,\left(2-\frac{1}{\chi}\right)\pi - \theta\right] }.
\end{equation}
We note that we should take the minimum of $\theta$ because in the disconnected case we have two extremal values of the geodesics due to the deficit angle at the center.

\begin{figure}[h]
    \centering
    \includegraphics[width = 60mm]{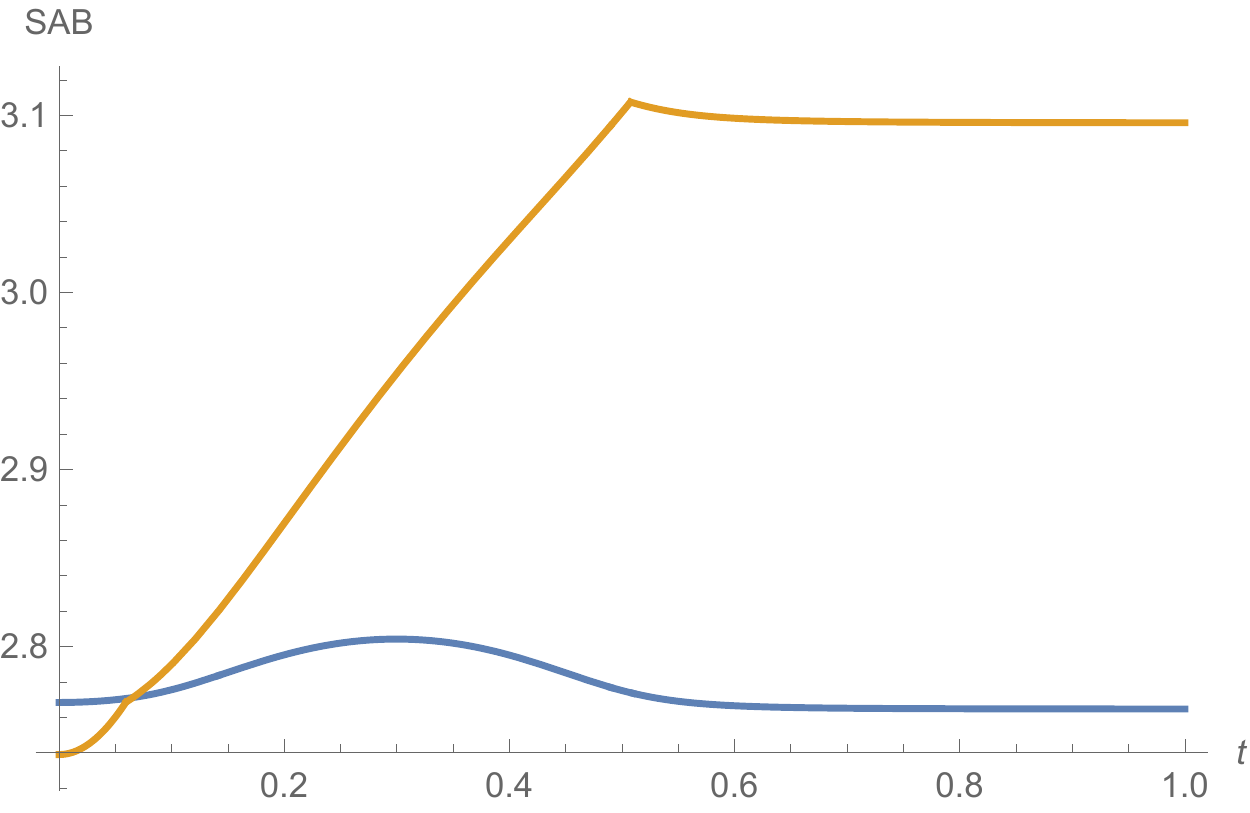}
     \hspace{0.5cm}
    \includegraphics[width = 60mm]{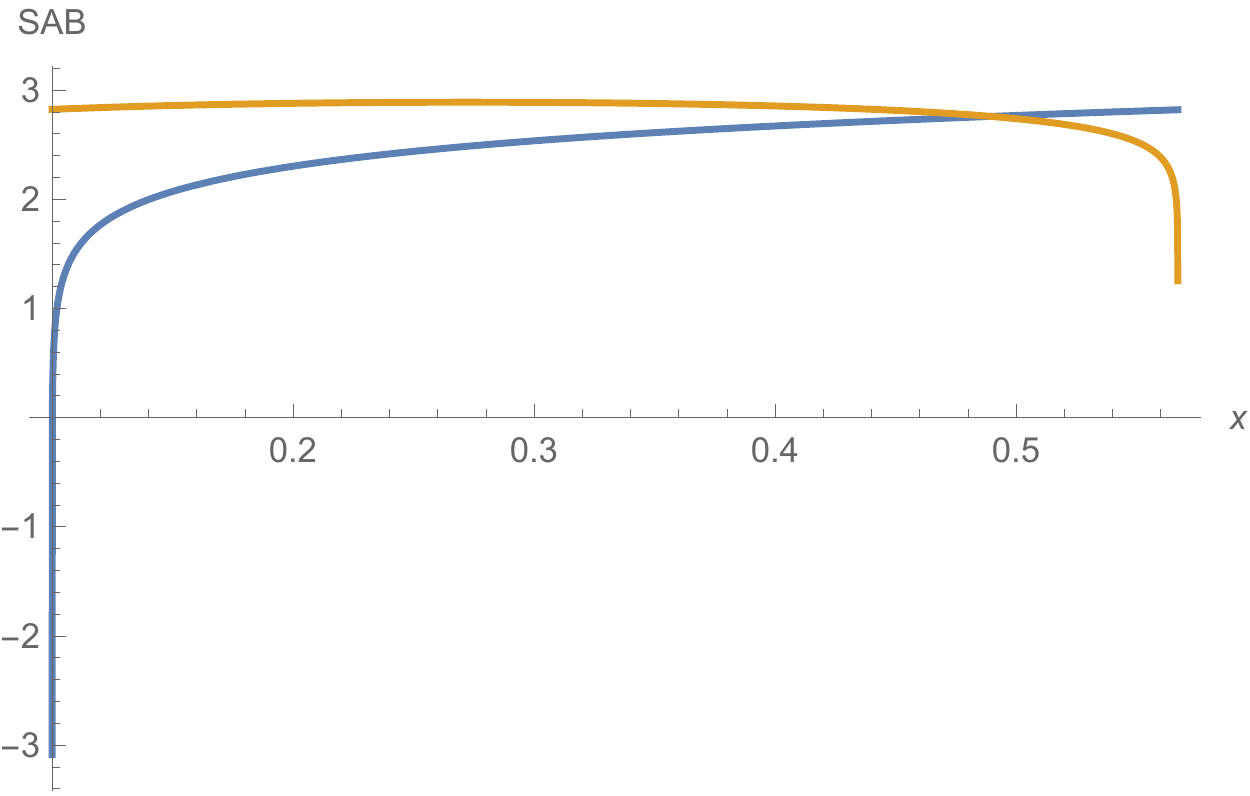}
    \caption{Plots of connected entropy $S^{con}_{AB}$ (blue curves) and disconnected entropy $S^{dis}_{AB}$ (orange curves) for $\chi=0.9$. The left panel shows the time evolution of them for the interval $(x_A,x_B)=(0.1,0.5)$.
    The right one describes their behaviors as functions of 
    $x$ when we chose the subsystem to be $(x_A,x_B)=(0.1,x)$ at $t=0$. In both, we took $R=c=1$, $\ep=0.0001$, $\lambda=1$ and $\ap=0.1$.
    }
    \label{fig:Sdis1}
\end{figure}

\begin{figure}[h]
    \centering
    \includegraphics[width = 60mm]{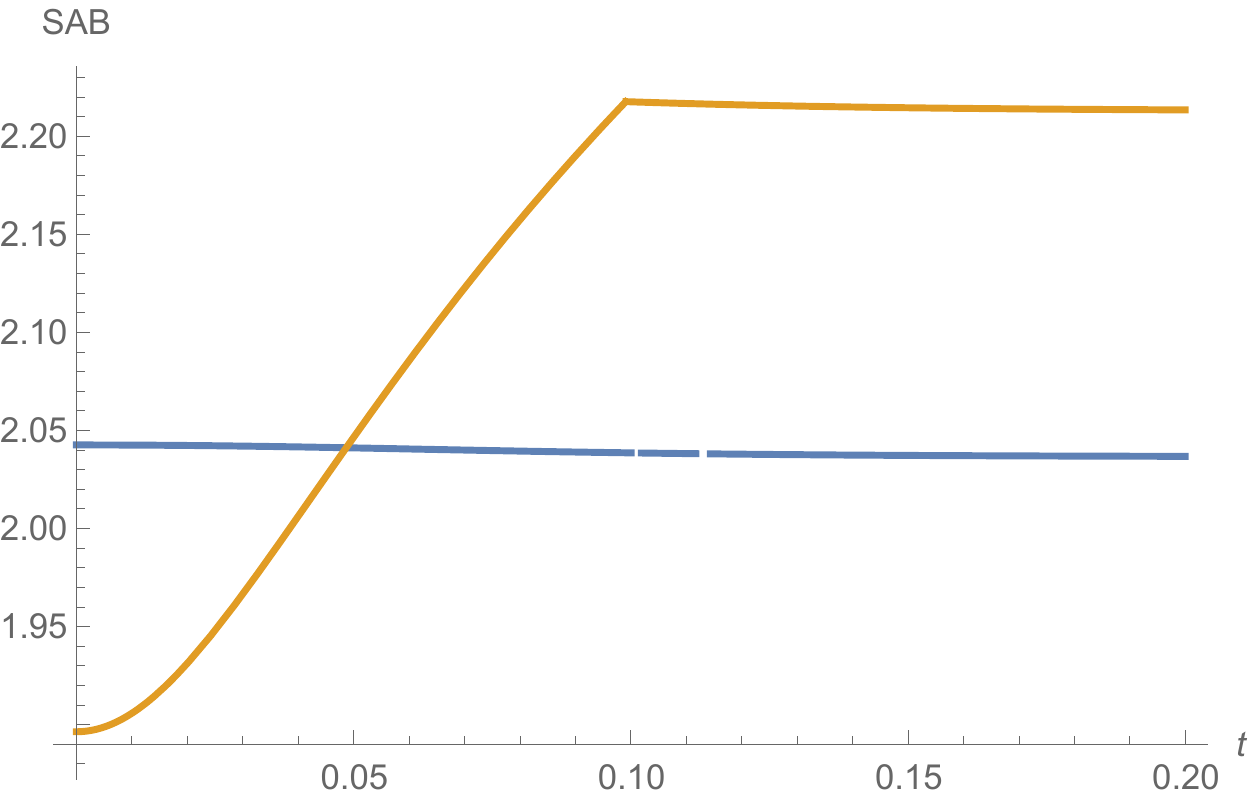}
    \hspace{0.5cm}
    \includegraphics[width = 60mm]{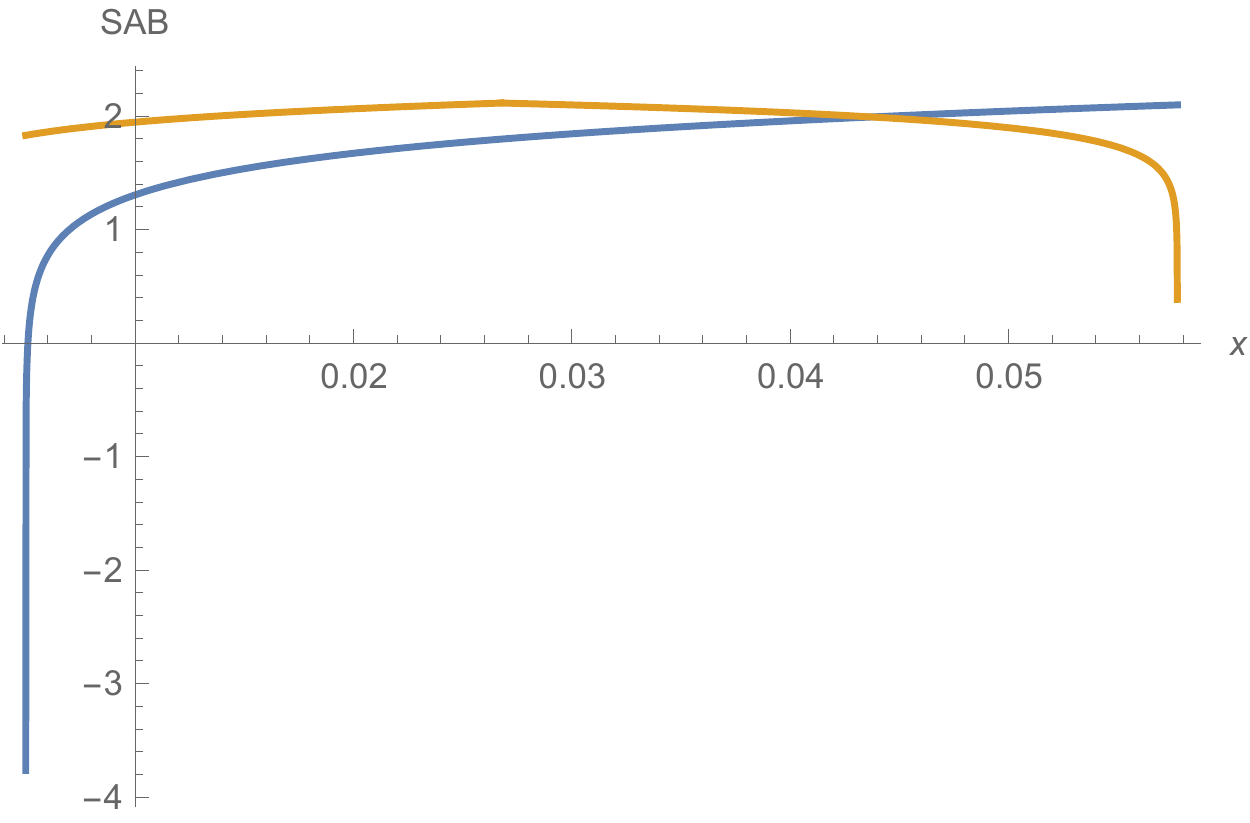}
    \caption{Plots of connected entropy $S^{con}_{AB}$ (blue curves) and disconnected entropy $S^{dis}_{AB}$ (orange curves) for $\chi=0.6$. The left panel shows the time evolution of them for the interval $(x_A,x_B)=(0.005,0.05)$.
    The right one describes their behaviors as functions of 
    $x$ when we chose the subsystem to be $(x_A,x_B)=(0.005,x)$ at $t=0$. In both, we take $R=c=1$, $\ep=0.0001$, $\lambda=1$ and $\ap=0.1$.}
    \label{fig:Sdis2}
\end{figure}

The resulting holographic entanglement entropy is plotted in Fig.\ref{fig:Sdis1} ($\chi=0.9$) and Fig.\ref{fig:Sdis2} ($\chi=0.6$). Under the time evolution, $S^{dis}_{AB}$ gets initially increasing since the EOW brane extends toward the inner region and 
the disconnected geodesics get longer. We can also note a peak around the time $t\simeq \s{x_B^2-\ap^2}$ since the falling particle crosses the disconnected geodesics which extend from $B$.
However, the actual holographic entanglement entropy is dominated by $S^{con}_{AB}$ after some critical time. In early time regime, $S^{dis}_{AB}$ is smaller and this is analogous to the setup of global quantum quenches
\cite{Calabrese:2005in}. We can also understand the plots of holographic entanglement entropy with respect to the subsystem size. It grows initially and reaches a maximum in a middle point, during which $S^{con}_{AB}$ dominates. After that it starts decreasing, and is eventually dominated by $S^{dis}_{AB}$. If we choose $x_A=0$, then we will end up with $S^{dis}_{AB}=0$ and this may be a sort of the Page curve behavior, because the total system is defined as an interval $0<x<Z(t)$.

\subsection{Consideration of the Parameter $\eta$}

As we have seen, We can introduce the parameter $\eta$ in addition to the mass parameter $M$, via the coordinate transformation
 (\ref{rescaleex}). We can calculate the holographic entanglement entropy with $\eta\neq 1$ by shifting $\chi$ into $\chi\vert_{M\rightarrow M^\prime} = \chi/\eta$ in (\ref{connected HEE}) and (\ref{disconnected HEE}). This allows us to realize a gravity dual of local operator quench on a half plane $x>0$, by pushing the second boundary to $x=\infty$. For example, it is straightforward to confirm that the first law relation (\ref{firstL}) perfectly reproduces the energy-momentum tensor (\ref{efluxx}) for any $\eta$.
 
\begin{figure}[ttt]
    \centering
    \includegraphics[width = 60mm]{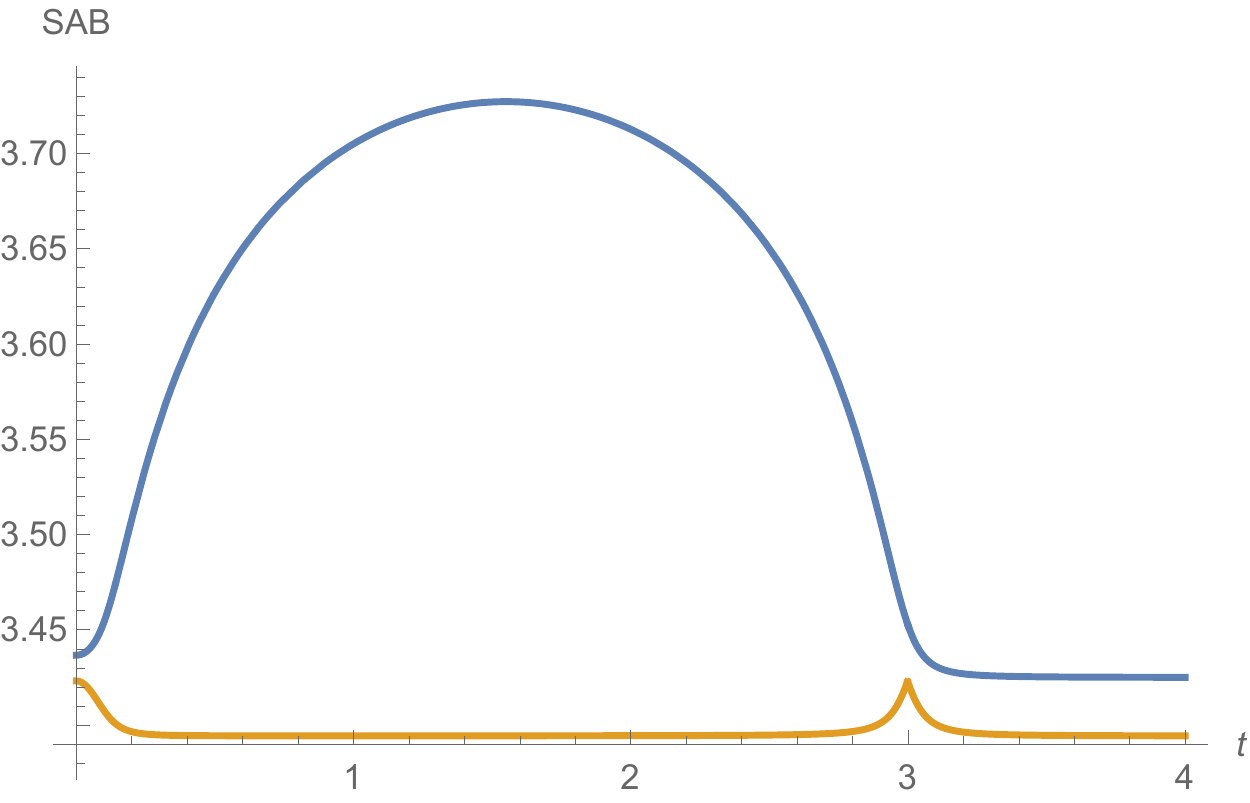}
    \hspace{0.5cm}
    \includegraphics[width = 60mm]{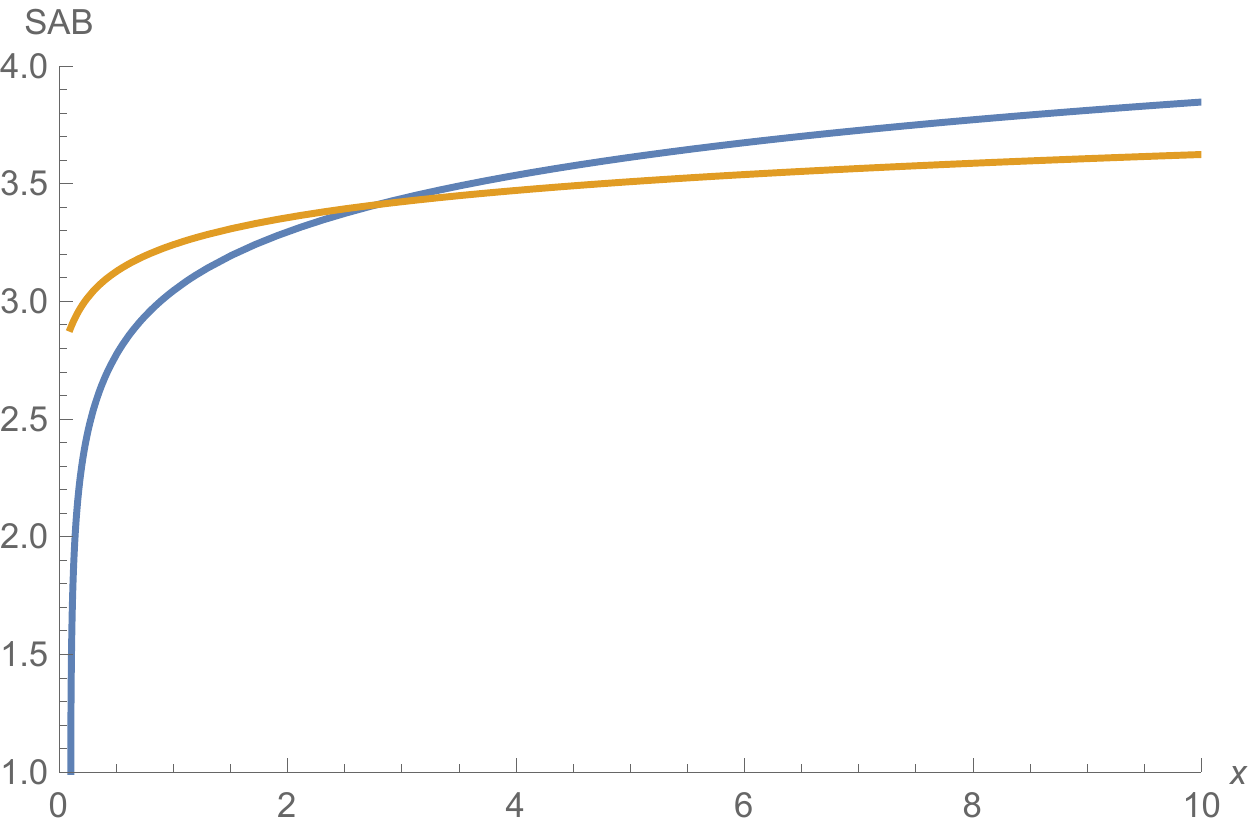}
    \caption{Plots of $S^{con}_{AB}$ (blue) and $S^{dis}_{AB}$ (orange)   for $\eta=\eta_0(=1.125)$ and $\chi=0.9$.    The left panel shows the time evolution of them for the interval $(x_A,x_B)=(0.1,3)$.
    The right one describes their behaviors as functions of 
    $x$ when we chose the subsystem to be $(x_A,x_B)=(0.1,x)$ at $t=0$. In both, we took $R=c=1$, $\ep=0.0001$, $\lambda=1$ and $\ap=0.1$.}
    \label{fig:etaonee}
\end{figure}

 When  $\eta=\eta_0$, we plotted the behavior of the holographic entanglement entropy in Fig.\ref{fig:etaonee}. This setup is expected to be dual to the BCFT only with the excitation by a local operator at $x=0$ and $t=0$.  The profile of $S^{con}_{AB}$ is qualitatively similar to that at $\eta=1$. On the other hand, $S^{dis}_{AB}$ at $\eta=\eta_0$ has two peaks at $t=\s{x_A^2-\ap^2}$
 and  $t=\s{x_B^2-\ap^2}$, which is because the falling massive particle crosses each of the disconnected geodesic. By taking the minimum,  $S^{dis}_{AB}$ is favored. This matches with the BCFT dual because one part of the entangled pair created by the local excitation is reflected at the boundary $x=0$ and merges with the other of the pair. Since both parts come together to the subsystem, the entanglement entropy for the subsystem does not increase except that there is a width $\ap$ of flux of excitations. This means that the entanglement entropy can increase only at the end points $A$ and $B$ for a short time of order $\ap$. This fits nicely with the behavior of $S^{dis}_{AB}$.

\begin{figure}[ttt]
    \centering
    \includegraphics[width = 60mm]{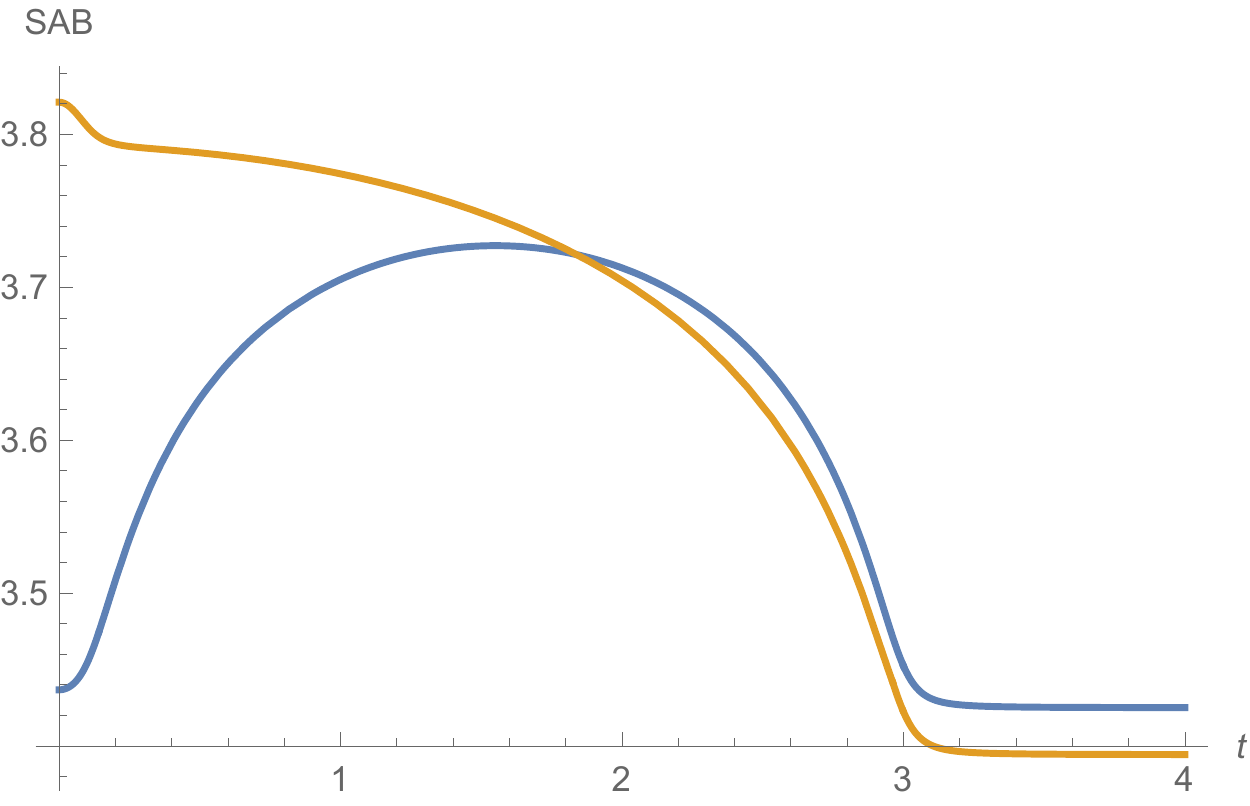}
    \hspace{0.5cm}
    \includegraphics[width = 60mm]{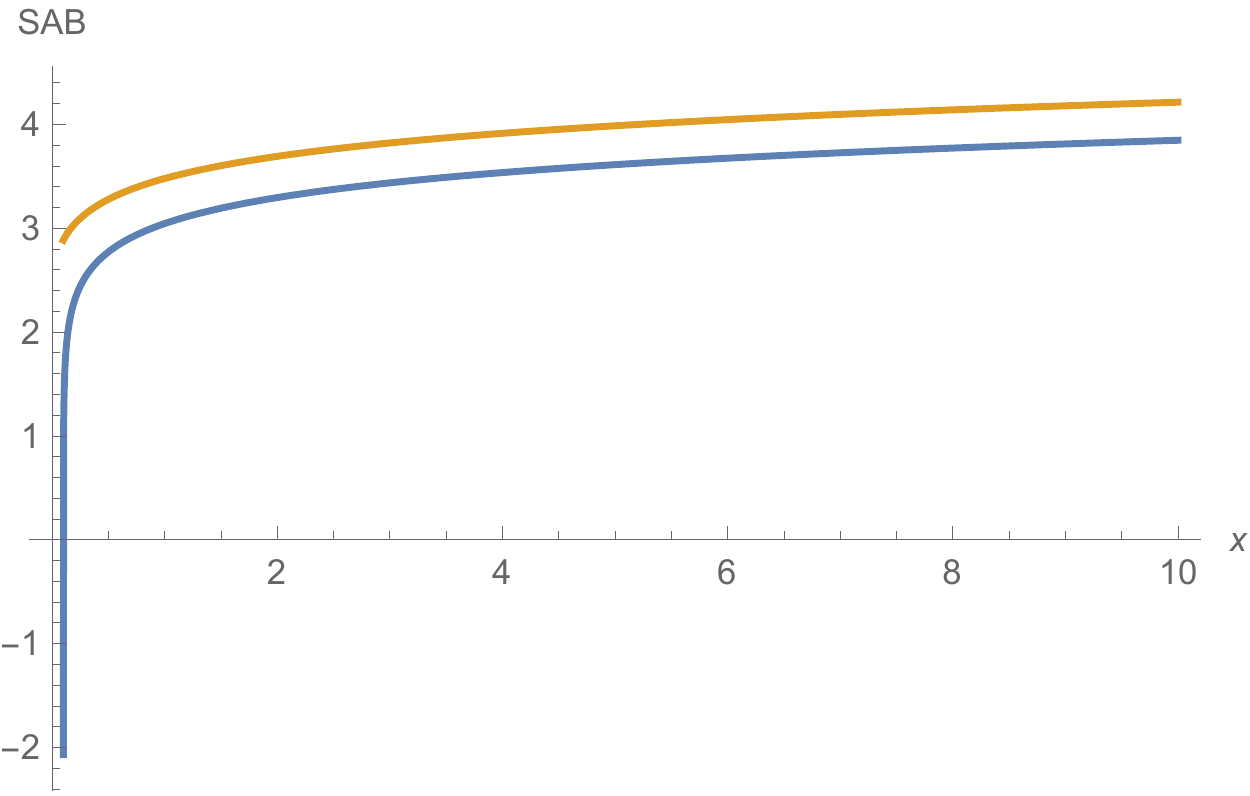}
    \caption{Plots of $S^{con}_{AB}$ (blue) and $S^{dis}_{AB}$ (orange)   for $\eta=1.1\eta_0(\simeq 1.12375)$ and $\chi=0.99$.  The graph of 
    $S^{con}_{AB}$ is identical to that in Fig.\ref{fig:etaonee}, while $S^{dis}_{AB}$ is not.  The left panel shows the time evolution of them for the interval $(x_A,x_B)=(0.1,3)$. 
    The right one describes their behaviors as functions of 
    $x$ when we chose the subsystem to be $(x_A,x_B)=(0.1,x)$ at $t=0$. In both, we took $R=c=1$, $\ep=0.0001$, $\lambda=1$ and $\ap=0.1$.}
    \label{fig:etaoneea}
\end{figure}

As we noted in (\ref{relpa}), there is one parameter family of $(M,\eta)$ which gives the same energy-momentum tensor.  As such an example, we can consider the case $\chi=0.99$ and $\eta=1.1\eta_0$, which should have the same energy flux in the case $\chi=0.9$ and $\eta=\eta_0$. Indeed, we plotted in Fig.\ref{fig:etaoneea}, the connected entropy $S^{con}_{AB}$ is precisely identical to that in Fig.\ref{fig:etaonee}.
This confirms that the three-dimensional metric in a neighborhood of the AdS boundary coincides. However, this is not actually physically equivalent because the global structure, especially the monodromy around the massive particle is different. This is simply because the mass of the particle is determined by $M$.  Therefore, if we consider a geodesic which goes around the particle, its geodesic length depends on the value of $M$. On the other hand if we consider a geodesic which does not go around the particle, its length does not depend on $M$. This monodromy clearly affects only the disconnected geodesics. This explains the reason why $S^{dis}_{AB}$ in Fig.\ref{fig:etaoneea} is more enhanced than that in  Fig.\ref{fig:etaonee}. In the latter, the disconnected geodesic length is reduced due to the larger deficit angle.
By taking a minimum between them, the resulting holographic entanglement entropy gets largely increased in the early time regime. In the BCFT side, this enhancement is due to the descendant excitations.

 \section{The BCFT Calculation -- Energy-momentum Tensor}
 
 In the last section we studied  the local quench process in the presence of boundaries through the dual gravitational setup. 
 In this section, we study the same quench process by purely boundary  CFT means. In particular, we perform the BCFT calculations of the energy-momentum tensor.
 In the calculations of the previous section, we made use of the fact that the bulk geometry with an infalling particle is related to the pure $AdS_{3}$  via a simple diffeomorphism. In this section and the next we study the quantities of our interest directly in the BCFT side by relating the  Euclidean setup with the time-dependent boundary to the upper half plane via a conformal mapping.  This makes explicit the one-to-one correspondence between  each step of calculations on the gravity side and the CFT side.

 Since the BCFT setup involves an additional boundary $x=Z(t)$ \eqref{disco}, we consider the BCFT between these two boundaries. This setup is  related to the upper half plane (UHP), as we summarize in Fig.\ref{setupsbdyfig}). To see this, we conformally map the original region 
 to a wedge-like region (the third figure in Fig.\ref{setupsbdyfig}), which is further mapped to the UHP. 
 
 \begin{figure}
    \centering
    \hspace*{-1.8cm}
    \includegraphics[width = 20.5cm]{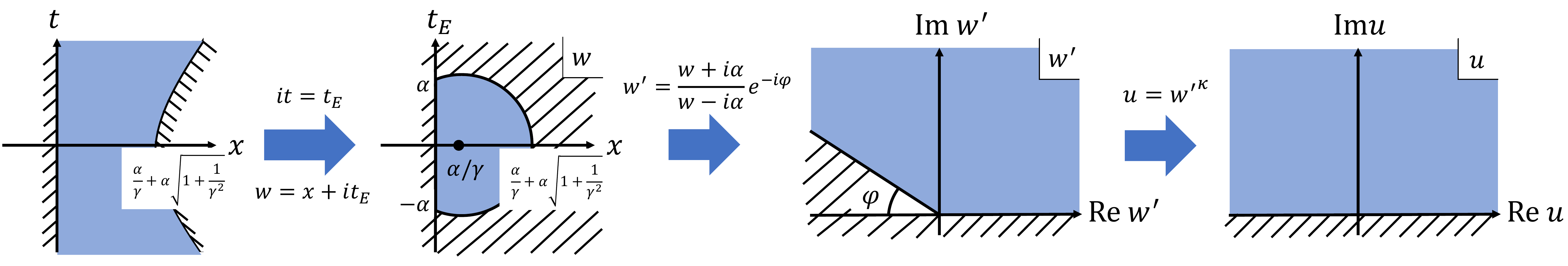}
    \caption{A sequence of conformal maps from the original semi-disk region to the upper half plane: The leftmost figure shows the original Lorentzian region of the BCFT. The second leftmost figure shows the corresponding region on the $w$-plane after Wick rotation. The third figure from the left shows the wedge-like region on the $w^\prime$-plane obtained by a global conformal transformation. The rightmost figure shows the upper half plane (UHP) on the $u$-plane obtained by a further local conformal transformation. A similar sequence of conformal transformations is discussed in \cite{Geng:2021iyq}.}
    \label{fig:conf-map}
\end{figure}
 
\subsection{Conformal Map to the Upper Half Plane}

The region on which we define the BCFT is surrounded by two boundaries. The left boundary is static
\ba
x = 0,
\ea
whereas the right boundary $x=Z(t)$ is time dependent, given by
\ba
\left(x-\frac{\alpha}{\gamma}\right)^2 = \alpha^2\left(1+\frac{1}{\gamma^2}\right)+t^2.
\label{bcft-region1}
\ea
This corresponds to the leftmost figure in Fig.\ref{fig:conf-map}.

A BCFT  on this region can be efficiently studied by mapping the region to the UHP. To do this, we perform the Wick rotation $it=t_E$ to make the hyperbolic boundary into a semi-circle. \eqref{bcft-region1} is now given by
\ba
\left(x-\frac{\alpha}{\gamma}\right)^2+t_E^2 =
\alpha^2\left(1+\frac{1}{\gamma^2}\right)
\ea
and is depicted in the second leftmost figure in Fig.\ref{fig:conf-map}. We denote this region by SD (semi-disk). The semicircle intersects with $x=0$ at $t_{E} =\pm \alpha$. 

We then  perform two conformal transformations to map the SD to the UHP. First, we consider a global conformal transformation such that in the complex coordinates $w=x+ i t_{E}, \bar{w}=x- i t_{E}$,  one of the intersections of two boundaries  $w=i\alpha$ is mapped to the infinity, the other $w=-i\alpha$ is mapped to the origin. We also require  the circle boundary comes  to the real axis. Such a map is given by
\ba
w^\prime=\frac{w+i\alpha}{w-i\alpha}e^{-i\varphi},
\label{conf-map-w}
\ea
where
\ba
\tan \varphi=\gamma.
\ea
The resulting region is shown in the second rightmost figure of Fig.\ref{fig:conf-map}. 
Although there is an ambiguity of choosing $\varphi$, we choose $0\le \varphi <\pi$ so that the $\gamma\rightarrow 0 \Leftrightarrow M\rightarrow 0$ limit covers the whole region as expected. As we discussed in Sec.\ref{sec:m-less-r} (and \eqref{restrictionM}), we have $1/2< \chi=\sqrt{(R^2-M^2)/R^2} \le 1$.\footnote{Note that due to this condition, we always have a finite region after this conformal transformation.} This determines the value of  $\varphi$ uniquely as
\ba
\varphi=\pi\left(\sqrt{\frac{R^2}{R^2-M}}-1\right)=\pi\left(\frac{1}{\chi}-1\right).
\label{eq:vp}
\ea

We then map the wedge region to UHP. This is achieved by
\ba
u=w^{\prime\, \kappa}
\label{conf-map-u}
\ea
by suitably choosing $\kappa$.
In order to obtain a unique $\kappa$, we need to specify the branch. Let us take $0\le \arg u \le \pi$. Then, since $0\le \varphi<\pi$, $\kappa$ is determined as
\ba
\kappa (\pi-\varphi)=\pi\quad \Leftrightarrow\quad \kappa=\frac{1}{1-\frac{\varphi}{\pi}}.
\label{eq:kappa}
\ea
Note that we simply find $\kappa=\eta_0$, where $\eta_0$ was introduced as a critical rescaling parameter in (\ref{mineta}). The conformal map (\ref{conf-map-u}) creates a deficit angle at $w=0$ and $w=\infty$, which indeed corresponds to the intersection of the deficit angle in the AdS$_3$ with the AdS boundary in our gravity dual.

Finally, we comment on the case with $\lambda<0$. Even in this case, the same conformal transformation can be used. Since the original BCFT region with $\lambda<0$ is given by the complement region with the parity in the $x$ direction reversed in Fig\ref{fig:conf-map}, the resulting region becomes the lower half plane instead of the UHP. Therefore, the calculation of the energy-momentum tensor and entanglement entropy in the subsequent sections is exactly the same for $\lambda<0$.




\subsection{Energy-momentum Tensor without a Local Operator}
In this section, we  compute  the one-point function of the energy-momentum tensor at $(w,\bar{w})=(x-t,x+t)$ in the SD and intend to compare it with the  holographic result \eqref{eflux}. We only take the boundary effect into account in this section and no local operator is being inserted. The conformal transformations (Fig.\ref{fig:conf-map}) yield
\ba
\langle T(w) \rangle_\mathrm{SD}=
\left(\frac{dw^\prime}{dw}\right)^2
\langle T(w^\prime) \rangle_\mathrm{wedge}
\ea
and
\ba
0=\langle T(u) \rangle_\mathrm{UHP}=\left(\frac{dw^\prime}{du}\right)^2 \left( \langle T(w^\prime) \rangle_\mathrm{wedge} -\frac{c}{12}\{u;w^\prime\} \right),
\ea
where
\ba
\{u;w^\prime\}=\frac{\de_{w^\prime}^3 u}{\de_{w^\prime} u} - \frac{3}{2} \left(\frac{\de_{w^\prime}^2 u}{\de_{w^\prime} u}\right)^2
\ea
is the Schwarzian derivative.
From this we can read off the holomorphic part of the energy-momentum tensor as
\ba
\langle T(w) \rangle_\mathrm{SD}=\frac{c}{6}\ap^2 (\kappa^2-1) \frac{1}{((x-t)^2+\ap^2)^2},
\ea

Next, we would like to compare it with the holographic result \eqref{eflux} for $ T_{--}$. The holomorphic part of the energy-momentum tensor  computed above is related to the holographic convention by  $\langle T(w) \rangle = -2\pi T_{--}$.\footnote{The energy-momentum tensor is here defined as $T_{\mu\nu}=\frac{2}{\sqrt{|g|}}\frac{\delta W}{\delta g^{\mu\nu}}$, where $W$ is the free energy defined via the partition function $Z=e^{-W}$. The convention here follows \cite{DiFrancesco:1997nk} and same as \cite{Nozaki2013}.} 
Therefore, we find the energy flux reads
\ba
  T^{{\rm BCFT}}_{\pm\pm}=-\frac{c}{12}(\kappa^2-1)\cdot \frac{\ap^2}{\pi((x\pm t)^2+\ap^2)^2}.  \label{eq:EMtensor-bdy}
\ea
Note that the coefficient is negative and this is because the existence of two boundaries leads to the Casimir energy. The functional form of (\ref{eq:EMtensor-bdy}) agrees with the holographic result (\ref{eflux}). We will see in the next subsection, the coefficient also perfectly agrees with the holographic result.


\subsection{Energy-momentum Tensor with a Local Excitation}\label{sec:op-exc}
In the previous subsection, we computed the one-point function of the energy-momentum tensor without any insertions of local operators. In this subsection, we evaluate the one-point function of the energy-momentum tensor of an excited state by a primary operator $O(x)$, in the BCFT setup.  We denote  its conformal weight  $(h,\bar{h})$. For simplicity, we focus on a scalar operator, i.e. $h=\bar{h}=\Delta_O/2$. 
Such a state has the following form,
\begin{align}
    |\Psi (t)\rangle &= \mathcal{N} e^{-itH} e^{-\ap H} O(t=0,x=\ti{\epsilon}) e^{\ap H} |0\rangle \nonumber\\
    &= \mathcal{N} e^{-itH} O(t_E=-\ap,x=\ti{\epsilon}) |0\rangle, 
\end{align}
where  $\mathcal{N}$ is a  normalization coefficient to ensure $\langle \Psi (t) | \Psi(t) \rangle=1$, and $\alpha$ plays a role of a UV regulator. Note that as mentioned in the footnote 1, the UV regulator $\ap$ for the quench is written as the imaginary time position of the operator while the (Lorentzian) time evolution cannot as $H|0\rangle\neq 0$ due to the existence of the time-dependent boundary.\footnote{This is apparent if we consider the Euclidean path integral representation of the state.} The spatial location of the local operator is arbitrary, however  for a comparison with the holographic setup,  we place it at $x=\ti{\epsilon} \ll 1$, as we are interested in the limit where the local operator is inserted at the boundary $x=0$. 
Thus the location of the operator insertion on the Euclidean SD is given by $w_2=\ti{\epsilon}+it_{E\, 2}=\ti{\epsilon}-i\ap$. Similarly, the bra state is given by
\begin{align}
    \langle \Psi (t)| &= \mathcal{N} \langle 0 | e^{\ap H} O (t=0,x=\ti{\epsilon}) e^{-\ap H} e^{itH}  \nonumber\\
    &=\mathcal{N}  \langle 0 | O(t_E=\ap,x=\ti{\epsilon}) e^{itH}.
\end{align}
We denote the insertion point corresponding to this as $w_1=\ti{\epsilon}+ i\ap$. Let us consider measuring the holomorphic energy momentum tensor at $w=x-t$, that is, discussing its time evolution in the Heisenberg picture.
By employing  the series of conformal transformations, we have
\begin{align}
    &\langle \Psi (t=0) |T(w=x-t)|\Psi(t=0)\rangle \nonumber\\
    =&\frac{\langle O(w_1,\bar{w}_1) T(w) O(w_2,\bar{w}_2)\rangle_{\mathrm{SD}}}{\langle O(w_1,\bar{w}_1) O(w_2,\bar{w}_2)\rangle_{\mathrm{SD}}} \nonumber\\
    =&
    \left( \frac{dw^\prime}{dw} \right)^2 
    \left[
    \left( \frac{du}{dw^\prime} \right)^2 \frac{\langle O(u_1,\bar{u}_1) T(u) O(u_2,\bar{u}_2) \rangle_{\mathrm{UHP}}}{\langle O(u_1,\bar{u}_1) O(u_2,\bar{u}_2) \rangle_{\mathrm{UHP}}}
    +\frac{c}{12} \{u;w^\prime\}
    \right],
\end{align}
where $u_i=u(w_i)$ and $\bar{u}_i=\bar{u}(\bar{w}_i)$.
To compute the first term in the brackets, we employ the conformal Ward identity in the UHP~\cite{Cardy:2004hm}
\begin{align}
    &\left\langle T(u) \prod_j O(u_j,\bar{u}_j) \right\rangle_{\mathrm{UHP}} \nonumber\\
    &= \sum_j \left(
    \frac{h}{(u-u_j)^2} + \frac{1}{u-u_j} \de_{u_j} + \frac{\bar{h}}{(u-\bar{u}_j)^2} + \frac{1}{u-\bar{u}_j} \de_{\bar{u}_j}
    \right) \left \langle \prod_j O(u_j,\bar{u}_j) \right \rangle_{\mathrm{UHP}} .\label{yyy}
\end{align}

This expression is further simplified by using the doubling trick~\cite{Cardy:1984bb,Recknagel:2013uja}. This  trick relates  a correlator on the UHP  to a chiral correlator on the entire plane $\mathbb{C}$.  For example, we have
\ba
\langle O(u_{1},\bar{u}_{1})  O(u_{2},\bar{u}_{2})\rangle_{\mathrm{UHP}} =\langle O(u_1) O(\bar{u}_1) O(u_2) O(\bar{u}_2) \rangle_{\mathbb{C}}^{c}.
\ea
The superscript $c$ means the correlator is chiral.

Using this relation as well as the detailed expression of the conformal map,  we get
\begin{align}
    & \langle \Psi (t=0) |T(w=x-t)|\Psi(t=0)\rangle \nonumber\\
    =& -\frac{4\ap^2 \kappa^2}{(w^2+\ap^2)^2} \frac{1}{\langle O(u_1) O(\bar{u}_1) O(u_2) O(\bar{u}_2) \rangle_{\mathbb{C}}^{c}} \nonumber\\
    & \left[
    u^2 \sum_{i=1}^2 
    \left\{ h \left( \frac{1}{(u-u_i)^2} + \frac{1}{(u-\bar{u}_i)^2} \right) + \frac{1}{u-u_i}\de_{u_i} + \frac{1}{u-\bar{u}_i}\de_{\bar{u}_i} 
    \right\}
    -\frac{c}{24} \left(1-\frac{1}{\kappa^2}\right)
    \right] \nonumber\\
    & \langle O(u_1) O(\bar{u}_1) O(u_2) O(\bar{u}_2) \rangle_{\mathbb{C}}^{c}.
    \label{em-tensor-op}
\end{align}

\subsubsection{Energy-momentum Tensor in Holographic CFT} \label{sec:EMtensor-holCFT}

\begin{figure}
    \centering
    \includegraphics[width=15cm]{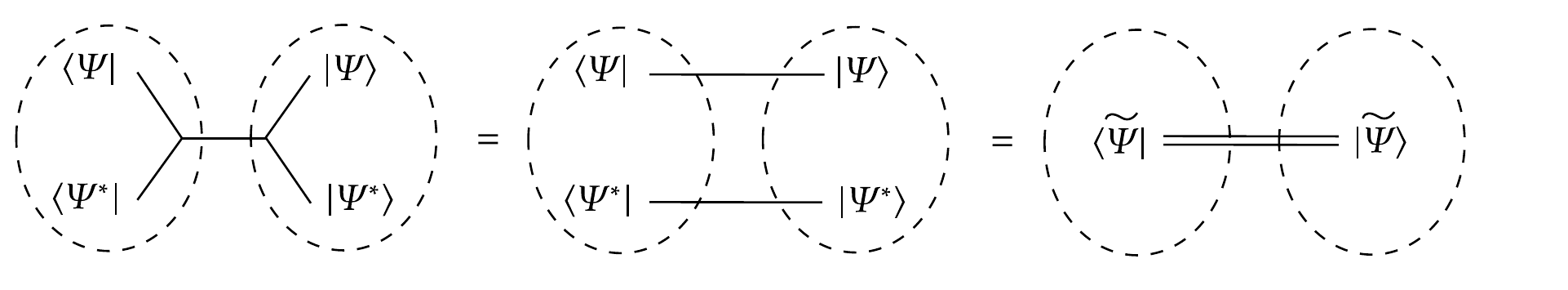}
    \caption{Left: The two-point function in the BCFT equals to the chiral four-point function. Middle: Since the holographic dual in concern has a vanishing one-point function, the contractions of operators in dashed circles cannot contribute to the correlator on their own. Right: The primary operator $\Psi$ and its mirror operator $\Psi^\ast$ should be regarded as a single operator $\tilde{\Psi}$, whose conformal dimension is $\tilde{\Delta}=2\Delta_O$.
    }
    \label{fig:ope}
\end{figure}

Let us evaluate the energy-momentum tensor in the holographic CFT dual to our AdS/BCFT model. In our analysis of  the dual gravitational setup, we assumed that the EOW brane does not have any source to the  bulk scalar field dual to the local operator $O(u)$ 
(for the analysis in the presence of non trivial scalar field configuration, refer to \cite{Suzuki:2022xwv}).  This means that  its  one-point function in the UHP vanishes: 
$\langle O(u,\bar{u}) \rangle_{\mathrm{UHP}} =0$. Therefore  $\langle O(u_1) O(\bar{u}_1) O(u_2) O(\bar{u}_2) \rangle_{\mathbb{C}}^{c}$  is factorized into the product of two point functions $\la O(u_1)O(u_2)\lb \cdot \la O(\bar{u}_1)O(\bar{u}_2)\lb$ for any parameter region of $u_1$ and $u_2$. This leads to the conclusion
\ba
\langle O(u_1) O(\bar{u}_1) O(u_2) O(\bar{u}_2) \rangle_{\mathbb{C}}^{c}
\propto \frac{1}{|u_1-u_2|^{\Delta_O} |\bar{u}_1-\bar{u}_2|^{\Delta_O}}.
\label{holcorb}
\ea
Notice that this is equivalent to the chiral two-point function with the doubled conformal dimension $\tilde{\Delta}\equiv 2\Delta_O$. (Fig.\ref{fig:ope}) This can be understood as follows.
The primary operators $O$ at the origin and the infinity constituting the ket and bra states are doubled via the doubling trick, up to the normalization. Let us denote these doubled states as $|\Psi\rangle \rightarrow |\Psi \Psi^\ast\rangle$ and $\langle \Psi| \rightarrow \langle\Psi \Psi^\ast|$.
Then, the equivalence between the four-point correlator and the two-point correlator with a doubled conformal dimension implies that we should regard the local operator and its mirror, $|\Psi\Psi^\ast\rangle$ and $\langle \Psi\Psi^\ast |$, as composite operators $|\tilde{\Psi}\rangle$ and $\langle \tilde{\Psi}|$, where tilded operators have a double conformal dimension $\tilde{\Delta}=2\Delta_O$ than the original ones. As we will see, this identification is necessary for the first law of entanglement entropy and the correct correspondence with the holographic calculation.

Since we take the limit $\ti{\ep}\to 0$ which leads to
$w_1\to-i\ap$ and $w_2\to i\ap$ in the $w$-plane, we find $u_1\to 0$ and $u_2\to \infty$. 
Thus, this leads to the following result at any time $t$ by plugging (\ref{holcorb}) into 
(\ref{em-tensor-op}):
\ba
\la T(w)\lb=\frac{c\ap^2 (\kappa^2-1)}{6(w^2+\ap^2)^2}
-\frac{4\Delta_O \ap^2 \kappa^2}{(w^2+\ap^2)^2}.
\ea
This leads to the energy-momentum tensor in Lorentzian signature via
$T_{--}=-\frac{1}{2\pi}T(w)$:
\ba
&& T_{--}=s_{BCFT}(\Delta_O)\cdot \frac{\ap^2}{\pi((x-t)^2+\ap^2)^2},\no
&& s_{BCFT}(\Delta_O)\equiv-\frac{c}{12}(\kappa^2-1)+2\kappa^2\Delta_O
=\kappa^2 \tilde{\Delta} +\frac{c}{12}(1-\kappa^2).  \label{EMgencft}
\ea
Notice that the part $-\frac{c}{12}(\kappa^2-1)$ in $s_{BCFT}$ can be regarded as the Casimir energy part which is negative and the other part $2\kappa^2\Delta_O$ corresponds to the local operator excitation. It is also useful to note that when the local excitation is light $\Delta_O\ll c$, we find
\ba
s_{BCFT}(\Delta_O)\simeq \Delta_O.
\ea

Now we would like to compare the BCFT result (\ref{EMgencft}) with the gravity dual result (\ref{efluxx}) at $\eta=1$. It is straightforward to see that these two are identical $s_{BCFT}(\Delta_O)=s_{AdS}(\Delta_{AdS},1)=\Delta_{AdS}$ via the relation (\ref{relasd}). Furthermore, by solving \eqref{EMgencft} for $\tilde{\Delta}=2\Delta_O$ and using the fact $\kappa=\eta_0$, we obtain\footnote{It is worth to note that $s_{BCFT}$ and $s_{AdS}|_{\eta=\eta_0}$ as a function of $\tilde{\Delta}$ and $\Delta_{AdS}$ respectively are inverse to each other.}
\begin{equation}
    2\Delta_O=\tilde{\Delta}=s_{AdS}(s_{BCFT}(\Delta_O),\eta_0) =s_{AdS}(\Delta_{AdS},\eta_0).
\end{equation}
This is nothing but \eqref{eq:AdS EM tensor}. In this way, we can perfectly reproduce the previous holographic energy-momentum tensor from the present BCFT approach.

\subsubsection{Energy-momentum Tensor in the Free Scalar CFT}

It is also helpful to compare our energy flux in the holographic CFT with that in a free scalar CFT.  Consider the $c=1$ CFT of an uncompactified scalar in two dimensions. The real scalar field $\phi(u,\bar{u})$ has the operator product expansion (OPE) in the presence of the boundary (the real axis in the $u$ coordinates) with the Neumann boundary condition:
\ba
\la \phi(u,\bar{u}) \phi(u',\bar{u}') \lb_N=-\log |u-u'|^2-\log |u-\bar{u}'|^2.
\ea
For the Dirichlet boundary condition, we have 
\ba
\la \phi(u,\bar{u}) \phi(u',\bar{u}') \lb_D=-\log |u-u'|^2+\log |u-\bar{u}'|^2.
\ea

For the local primary operator we choose,
\ba
O(u_1,\bar{u}_1)=e^{ik\phi(u_1,\bar{u}_1) },\ \ \ \ O^\dagger(u_2,\bar{u}_2)=e^{-ik\phi(u_2,\bar{u}_2) },
\ea
both of which have the conformal dimension $\Delta_O=2h=k^2$.

The two-point functions for these  boundary conditions read
\ba
&& \la O(u_1,\bar{u}_1)O^\dagger(u_2,\bar{u}_2)\lb_N
=\frac{|u_1-\bar{u}_1|^{\Delta_O}|u_2-\bar{u}_2|^{\Delta_O}}
{|u_1-u_2|^{2\Delta_O}|u_1-\bar{u}_2|^{2\Delta_O}},\no
&& \la O(u_1,\bar{u}_1)O^\dagger(u_2,\bar{u}_2)\lb_D
=\frac{|u_1-\bar{u}_2|^{2\Delta_O}}{|u_1-u_2|^{2\Delta_O}|u_1-\bar{u}_1|^{\Delta_O}|u_2-\bar{u}_2|^{\Delta_O}}.\label{twoptsc}
\ea

The expectation value of the energy-momentum tensor $T(u)=-\frac{1}{2}\de_u\phi\de_u\phi$ 
is evaluated for each boundary condition as follows
\ba
&& \la T(u)\lb_N=\frac{\Delta_O}{2}\left(\frac{1}{u-u_1}+\frac{1}{u-\bar{u}_1}
-\frac{1}{u-u_2}-\frac{1}{u-\bar{u}_2}\right)^2,\no
&& \la T(u)\lb_D=\frac{\Delta_O}{2}\left(\frac{1}{u-u_1}-\frac{1}{u-\bar{u}_1}
-\frac{1}{u-u_2}+\frac{1}{u-\bar{u}_2}\right)^2.
\ea
They perfectly agree with those obtained from the two point functions (\ref{twoptsc}) by applying the conformal Ward identity (\ref{yyy}). 

By taking our limit $u_1\to 0$ and $u_2\to \infty$, we obtain 
\be
\la T(u)\lb_N\simeq \frac{2\Delta_O}{u^2},\ \ \ \ \ 
\la T(u)\lb_D\simeq 0.
\ee
They take different values when compared with the previous holographic BCFT result (obtained by plugging \eqref{holcorb} in \eqref{em-tensor-op} and take the $u_1\rightarrow 0$ and $u_2\rightarrow \infty$ limits):
\ba
 \la T(u)\lb_{Hol}\simeq \frac{\Delta_O}{u^2}.  \label{holbcft}
\ea

Finally, by using the conformal map of energy-momentum tensor 
(\ref{em-tensor-op}), or explicitly 
\ba
\la T(w)\lb=-\frac{4\ap^2\kappa^2}{(w^2+\ap^2)^2}\left[u^2\la T(u)\lb-\frac{c}{24}(1-\kappa^{-2})\right],
\ea
we obtain the physical energy-momentum tensor $\la T(w)\lb$. In this way we found that the energy flux is sensitive to both the types of CFTs and the boundary conditions.

\subsection{Energy-momentum Tensor after Rescaling by $\eta$}
The energy-momentum tensor in the holographic CFT matches with the holographic computation given in \eqref{efluxx} even after the rescaling by an arbitrary $\eta$ \eqref{rescaleex}.

The rescaling by $\eta$ maps $\chi$ and $\kappa=\eta_0$ to $\chi/\eta$ and $\kappa_{\eta}\equiv\kappa/\eta$ respectively since $M$ in $\chi$ becomes $M^\prime$ \eqref{newmass} and the location of the induced boundary $\theta={\pi}/{\kappa}$ becomes $\theta^\prime=\eta{\pi}/{\kappa}$.\footnote{Note that even though $\kappa_{\eta=1}\equiv\kappa$ is written as a function of $\chi$ (\eqref{eq:vp} and \eqref{eq:kappa}), $\kappa_\eta$ does not equal to $\kappa|_{\chi\rightarrow\chi/\eta}$. This is because $\kappa$ is defined as a location of the boundary, which is directly rescaled by $\eta$, not a mere function of $\chi$.} This is consistent with \eqref{Identifyap}.

After replacing $\kappa$ in \eqref{EMgencft} by $\kappa_\eta={\kappa}/{\eta}$,
we exactly reproduce \eqref{efluxx} through \eqref{eq:AdS EM tensor}.

Alternatively, this can be explicitly confirmed by the conformal mapping \eqref{conf-map-eta}
\begin{equation}
    w_{\eta}=\ap\tan\left(\eta \arctan \left(\frac{w}{\ap}\right)\right),
    \quad
    \bar{w}_{\eta}=\ap\tan\left(\eta \arctan \left(\frac{\bar{w}}{\ap}\right)\right)
\end{equation}
where $w_{\eta}=x^\prime-t^\prime$, the $w$ coordinates after the rescaling by $\eta$ (same for the antiholomorphic one).
Then,
\begin{align}
    \langle T(w_\eta) \rangle &= \left(\frac{dw}{dw_\eta}\right)^2 \langle T(w) \rangle +\frac{c}{12} \{w\; w_\eta \} \nonumber\\
    &=\left[\frac{\kappa^2}{\eta^2}\tilde{\Delta}+\frac{c}{12}\left(1-\frac{\kappa^2}{\eta^2}\right)\right]\frac{-2\ap^2}{(w_\eta^2+\ap^2)^2}.
\end{align}
This is nothing but $\kappa\rightarrow\frac{\kappa}{\eta}$ in \eqref{EMgencft} as mentioned above.

In particular, when $\eta=\eta_0=\kappa$,
\begin{equation}
    s_{BCFT}(\Delta_O)=2\Delta_O
\end{equation}
and \eqref{eq:AdS EM tensor} is readily confirmed.

\section{The BCFT Calculation -- Entanglement Entropy}
In this section, we calculate entanglement entropy in the holographic BCFT of concern. In the first part, we compute the vacuum entanglement entropy with the induced boundary $x=Z(t)$ \eqref{disco} as well as the original boundary $x=0$. Then, in the second part, we discuss entanglement entropy with both the boundaries and a local operator insertion. Finally, we comment on the behavior of the entanglement entropy after the generic rescaling by $\eta$ \eqref{rescaleex}. We will see that all of these calculations perfectly reproduce the holographic results obtained in section 3.

\subsection{Entanglement Entropy without a Local Operator}
In this section, we  compute the vacuum entanglement entropy on the SD, i.e. only take the (induced) boundary effect into  account, and compare it with the holographic result discussed in Sec.\ref{sec:holoEE}. By using the conformal map  
(Fig.\ref{fig:conf-map}), entanglement entropy of an interval $I=[x_A,x_B]$ ($0<x_A<x_B<Z(t)$) at time $t$ is given by
\begin{equation}
    S_{AB\ \mathrm{no\, op.}} =\lim_{n\rightarrow 1} \frac{1}{1-n} \log \mbox{Tr} \rho_{AB\ \mathrm{no\, op.}}^n,
\end{equation}
where
\begin{align}
    \mbox{Tr}\rho_{AB\ \mathrm{no\, op.}}^n &=\langle 0 | \sigma_n (w_A,\bar{w}_A) {\sigma}_n (w_B,\bar{w}_B) |0\rangle_{\mathrm{SD}} \nonumber \\
    &=\left(
    \left.\frac{du}{dw}\right\vert_{u=u_A} \left.\frac{d\bar{u}}{d\bar{w}}\right\vert_{\bar{u}=\bar{u}_A}
    \left.\frac{du}{dw}\right\vert_{u=u_B} \left.\frac{d\bar{u}}{d\bar{w}}\right\vert_{\bar{u}=\bar{u}_B}
    \right)^{\Delta_n /2}
    \langle 0 | \sigma_n(u_A,\bar{u}_A) {\sigma}_n(u_B,\bar{u}_B) |0\rangle_\mathrm{UHP}
\label{twist-correlator}
\end{align}
in terms of the reduced density matrix defined in \eqref{eq:reduced-rho}. $\Delta_n=\frac{c}{12}\left(n-\frac{1}{n}\right)$ is the conformal weight of the twist operators $\sigma_n$. Each twist operator is inserted at the endpoints of the interval $I$: $w_{A,B}=x_{A,B}+it_{E}=x_{A,B}-t$, $\bar{w}_{A,B}=x_{A,B}+t$. The $u$'s are coordinates after the conformal transformation, i.e. $u_{A,B}=u(w_{A,B})$ and $\bar{u}_{A,B}=\bar{u}(\bar{w}_{A,B})$.

To compute a correlation function on the UHP, we use the doubling trick~\cite{Cardy:1984bb,Recknagel:2013uja}:
\ba
\langle \sigma_n(u_A,\bar{u}_A) {\sigma}_n(u_B,\bar{u}_B) \rangle_\mathrm{UHP}=\langle \sigma_n(u_A) \sigma_n(\bar{u}_A) {\sigma}_n(u_B) {\sigma}_n(\bar{u}_B) \rangle_\mathrm{\mathbb{C}}^{c}.
\label{doubling}
\ea
Assuming the CFT is holographic, this four-point function can be evaluated via the vacuum conformal block in the large central charge limit~\cite{Hartman:2013mia,Asplund:2014coa}. The two possible OPE channels correspond to the connected and disconnected geodesics in the holographic calculation.



The entanglement entropy reads $S_{AB\ \mathrm{no\, op.}}=\min \{S^{con}_{AB\ \mathrm{no\, op.}},S^{dis}_{AB\ \mathrm{no\, op.}}\}$, where
\ba
&& S^{con}_{AB\ \mathrm{no\, op.}}=\frac{c}{6}\log\frac{|u_A-u_B|^2}{\ep^2|u'_A||u'_B|},\no
&& S^{dis}_{AB\ \mathrm{no\, op.}}=\frac{c}{6}\log\frac{|u_A-\bar{u}_A||u_B-\bar{u}_B|}{\ep^2|u'_A||u'_B|}+2S_{bdy},
\label{eq:EEwoop}
\ea
where $u_{A,B}=u(w_{A,B})$ and  $u'_{A,B}=\left.\frac{du}{dw}\right|_{w=w_{A,B}}$. $S_{bdy}$ is the boundary entropy and $\ep$ is the UV cutoff. 

\begin{figure}
      \includegraphics[width=5cm]{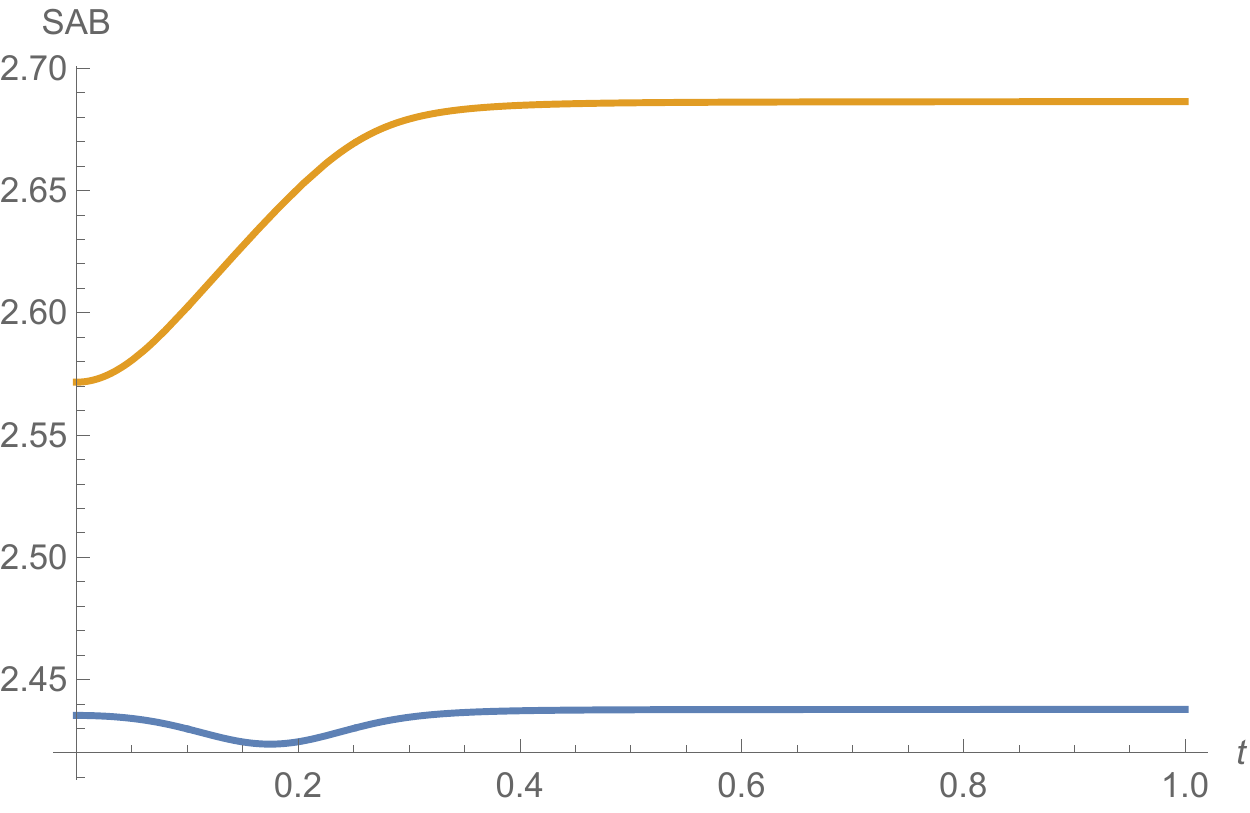}
      \includegraphics[width=5cm]{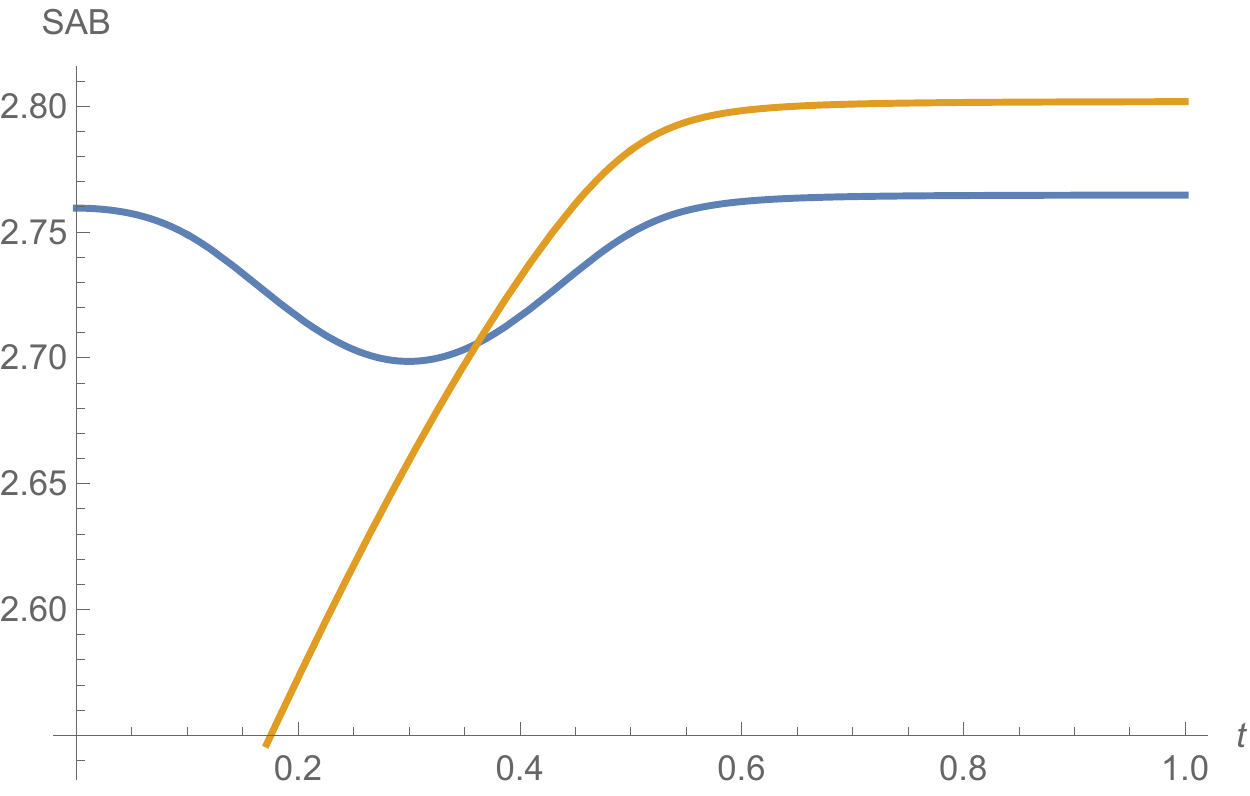}
         \includegraphics[width=5cm]{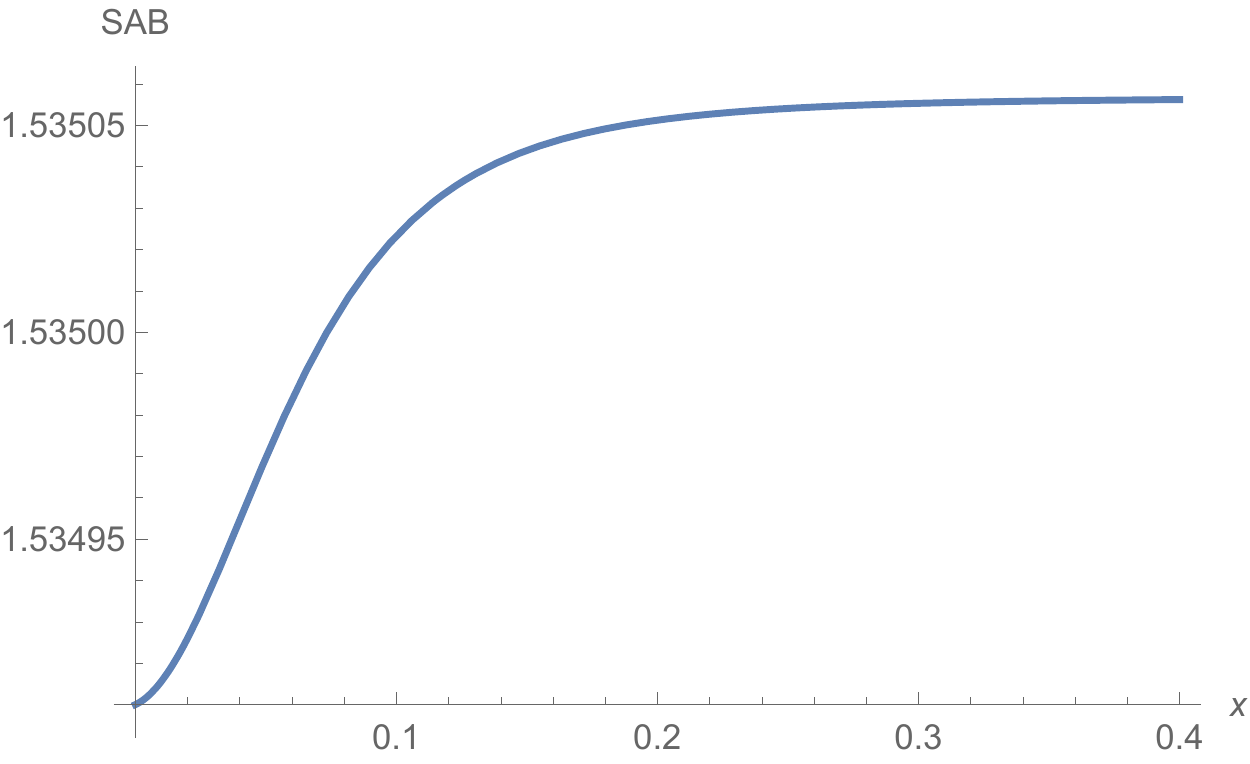}
        \caption{Plots of the entanglement entropy \eqref{eq:EEwoop} in the BCFT.
        We chose $\ap=0.1$,  $\chi=0.9$, and $\ep=0.0001$, where we have $Z(0)=\frac{\ap}{\tan\frac{\vp}{2}}\simeq 0.567$. In the left and middle plot, 
        we chose $I=[0.1,0.25]$ and $I=[0.1,0.5]$, respectively and we showed the entanglement entropy as a function of the time $t$. The blue and orange graph describe $S^{con}_{AB}$ and $S^{dis}_{AB}$ (we set $S_{bdy}=0$), respectively. In the right panel we plotted $S^{con}_{AB}$ at $t=0$ for the subsystem $I=[x,x+0.01]$ as a function of $x$. In these plots, we set $c=1$ and $S_{bdy}=0$.
        }
    \label{fig:heevac}
\end{figure}

We plotted the behavior of the entanglement entropy in Fig.\ref{fig:heevac}. The connected entanglement entropy $S^{con}_{AB}$ for an interval $[x_A,x_B]$ shows a dip around the time $t\sim\frac{x_A+x_B}{2}$. This can be interpreted as the shock wave of negatve energy flux emitted by the falling massive object in AdS, whose trajectory looks like 
$x\sim t$. On the other hand, the disconnected entanglement entropy $S^{dis}_{AB}$ increases until the massive falling particle crosses the second minimal surface when $t\sim x_B$.

The rightmost plot is proportional to the behavior of the negative energy flux at $t=0$ as predicted by the first law of entanglement entropy (\ref{firstL}). Indeed we can confirm that in the small subsystem size  limit $|x_A-x_B|\to 0$, we have 
\ba
 \Delta S^{con}_{AB}\simeq \frac{c}{18}(1-\kappa^2)H(t,x_A)(x_A-x_B)^2,
\ea
which is completely  in accord with the result for the  energy momentum tensor (\ref{eq:EMtensor-bdy}).\footnote{To derive this, it is useful to see $H(t,x)=\frac{\ap^2}{\pi}\left(\frac{1}{(w^2+\ap^2)^2}+\frac{1}{(\bar{w}^2+\ap^2)^2}\right)$. Note that since $\Delta S_{AB}^{con}=-\frac{c}{72}|x_A-x_B|^2\left(\{u_A\;w_A\}+\{\bar{u}_A\;\bar{w}_A\}\right)$, the first law holds for any conformal map $u(w)$. Thus, even after the rescaling by $\eta$, it continues to hold.}

\subsection{Entanglement Entropy with a Local Excitation}

In this section, we discuss entanglement entropy of a single interval $I=[x_A,x_B]$, taking the effect of the local operator insertions as well as the induced boundary into the account. 

\subsubsection{Factorization of Entanglement Entropy}

Entanglement entropy is calculated from
\begin{align}
    \mbox{Tr} \rho_{AB}^n &= \bra{\Psi}\sigma (w_A,\bar{w}_A) \sigma (w_B,\bar{w}_B) \ket{\Psi}_{\mathrm{SD}} \nonumber\\
    &=\left[(2\kappa \ap)^4 \frac{u_A\bar{u}_A u_B\bar{u}_B}{(w_A^2+\ap^2)(\bar{w}_A^2+\ap^2)(w_B^2+\ap^2)(\bar{w}_B^2+\ap^2)} \right]^{\Delta_n /2} \bra{\Psi}\sigma (u_A,\bar{u}_A) \sigma (u_B,\bar{u}_B) \ket{\Psi}_{\mathrm{UHP}} \nonumber\\
    &=\left[(2\kappa \ap)^4 \frac{u_A\bar{u}_A u_B\bar{u}_B}{(w_A^2+\ap^2)(\bar{w}_A^2+\ap^2)(w_B^2+\ap^2)(\bar{w}_B^2+\ap^2)} \right]^{\Delta_n /2} \bra{\Psi\Psi^\ast} \sigma (u_A) \sigma (u_B) \sigma (\bar{u}_A) \sigma (\bar{u}_B) \ket{\Psi\Psi^\ast}_{\mathbb{C}}^c.
    \label{eq:2twists}
\end{align}

\eqref{eq:2twists} is written in terms of the chiral eight-point function and we cannot obtain a simple analytic form even in the large central charge limit in general. 
However, when the subregion size is sufficiently small or large,
there exists two phases corresponding to the connected and disconnected geodesics in the holographic entanglement entropy. 
The holographic entanglement entropy predicts the connected entropy dominates at early time and the disconnected entropy dominates at late time (Fig.\ref{fig:Sdis1}, \ref{fig:Sdis2}). The former corresponds to $x_A\ (x_B)\gg x_B-x_A$ or equivalently, $u_A\sim u_B$ (and $\bar{u}_A\sim\bar{u}_B$), while the latter corresponds to $x_A\ (x_B)\ll x_B-x_A$ or equivalently, $u_{A,B}\sim \bar{u}_{A,B}$. In terms of the OPE channels, they correspond to
\begin{align}
&\tikzset{every picture/.style={line width=0.75pt}}
\begin{tikzpicture}[baseline={([yshift=-3.5ex]current bounding box.center)},x=0.75pt,y=0.75pt,yscale=-1,xscale=1]
\draw    (42.67,44.6) -- (63.67,75.6) ;
\draw    (42.67,104.6) -- (63.67,75.6) ;
\draw    (63.67,75.6) -- (103.67,75.6) ;
\draw    (103.8,51.4) -- (83,25.4) ;
\draw    (103.8,51.4) -- (103.67,75.6) ;
\draw    (103.8,51.4) -- (126.6,25.8) ; 
\draw    (103.67,75.6) -- (205.67,75.6) ;
\draw    (205.8,51.4) -- (185,25.4) ;
\draw    (205.8,51.4) -- (205.67,75.6) ; 
\draw    (205.8,51.4) -- (228.6,25.8) ;
\draw    (265.31,106.13) -- (244.08,75.29) ;
\draw    (264.86,46.13) -- (244.08,75.29) ;
\draw    (244.2,75.4) -- (205.67,75.6) ;
\draw (6,32.6) node [anchor=north west][inner sep=0.75pt]    {$\langle \Psi |$};
\draw (1,92.6) node [anchor=north west][inner sep=0.75pt]    {$\langle \Psi ^{\ast } |$};
\draw (59.6,2.6) node [anchor=north west][inner sep=0.75pt]    {$\sigma ( u_{A})$};
\draw (166.6,2.6) node [anchor=north west][inner sep=0.75pt]    {$\sigma (\bar{u}_{A})$};
\draw (115.6,2.6) node [anchor=north west][inner sep=0.75pt]    {$\sigma ( u_{B})$};
\draw (221.4,2.6) node [anchor=north west][inner sep=0.75pt]    {$\sigma (\bar{u}_{B})$};
\draw (269.4,33.2) node [anchor=north west][inner sep=0.75pt]    {$|\Psi \rangle $};
\draw (270.8,93.2) node [anchor=north west][inner sep=0.75pt]    {$|\Psi ^{\ast } \rangle $};
\end{tikzpicture}
\quad\text{(connected phase)}
\intertext{and}
&
\tikzset{every picture/.style={line width=0.75pt}}
\begin{tikzpicture}[baseline={([yshift=-3.5ex]current bounding box.center)},x=0.75pt,y=0.75pt,yscale=-1,xscale=1]
\draw    (42.67,44.6) -- (63.67,75.6) ;
\draw    (42.67,104.6) -- (63.67,75.6) ;
\draw    (63.67,75.6) -- (103.67,75.6) ;
\draw    (103.8,51.4) -- (83,25.4) ;
\draw    (103.8,51.4) -- (103.67,75.6) ;
\draw    (103.8,51.4) -- (126.6,25.8) ; 
\draw    (103.67,75.6) -- (205.67,75.6) ;
\draw    (205.8,51.4) -- (185,25.4) ;
\draw    (205.8,51.4) -- (205.67,75.6) ; 
\draw    (205.8,51.4) -- (228.6,25.8) ;
\draw    (265.31,106.13) -- (244.08,75.29) ;
\draw    (264.86,46.13) -- (244.08,75.29) ;
\draw    (244.2,75.4) -- (205.67,75.6) ;
\draw (6,32.6) node [anchor=north west][inner sep=0.75pt]    {$\langle \Psi |$};
\draw (1,92.6) node [anchor=north west][inner sep=0.75pt]    {$\langle \Psi ^{\ast } |$};
\draw (59.6,2.6) node [anchor=north west][inner sep=0.75pt]    {$\sigma ( u_{A})$};
\draw (166.6,2.6) node [anchor=north west][inner sep=0.75pt]    {$\sigma ( u_{B})$};
\draw (115.6,2.6) node [anchor=north west][inner sep=0.75pt]    {$\sigma (\bar{u}_{A})$};
\draw (221.4,2.6) node [anchor=north west][inner sep=0.75pt]    {$\sigma (\bar{u}_{B})$};
\draw (269.4,33.2) node [anchor=north west][inner sep=0.75pt]    {$|\Psi \rangle $};
\draw (270.8,93.2) node [anchor=north west][inner sep=0.75pt]    {$|\Psi ^{\ast } \rangle $};
\end{tikzpicture}
\quad\text{(disconnected phase).}
\end{align}

\vspace{0.8cm}

As we discussed in Fig.\ref{fig:ope}, we can regard $|\Psi\Psi^\ast\rangle$ and $\langle \Psi\Psi^\ast |$ as excited states created by a single primary operator $|\tilde{\Psi}\rangle$ and $\langle \tilde{\Psi}|$, whose conformal dimension is $\tilde{\Delta}=2\Delta_O$  as we are considering the BCFT dual of the spacetime without the bulk matter profile and there is no source for the matter field on the EOW brane.
The limits $x_A\ (x_B)\gg x_B-x_A$ and $x_A\ (x_B)\ll x_B-x_A$ imply $\sigma(u_A) \sigma(u_B)\sim \mathbf{1}$ and $\sigma(u_A) \sigma(\bar{u}_A)\sim \mathbf{1}$ respectively. Then, the orthogonality of the two point function in any CFTs leads~\cite{Caputa:2015waa}
\begin{align}
    \bra{\Psi\Psi^\ast} \sigma (u_A) \sigma (u_B) \sigma (\bar{u}_A) \sigma (\bar{u}_B) \ket{\Psi\Psi^\ast}_{\mathbb{C}}^c 
    &= \bra{\tilde{\Psi}} \sigma (u_A) \sigma (u_B) \sigma (\bar{u}_A) \sigma (\bar{u}_B) \ket{\tilde{\Psi}}_{\mathbb{C}}^c \nonumber\\
    &= \sum_\ap \bra{\tilde{\Psi}} \sigma (u_A) \sigma (u_B) \ket{\ap}^c_{\mathbb{C}} \!\bra{\ap} \sigma (\bar{u}_A) \sigma (\bar{u}_B) \ket{\tilde{\Psi}}_{\mathbb{C}}^c \nonumber\\
    &\approx 
    \bra{\tilde{\Psi}} \sigma (u_A) \sigma (u_B) 
    \ket{\tilde{\Psi}}_{\mathbb{C}}^c
    \bra{\tilde{\Psi}}
    \sigma (\bar{u}_A) \sigma (\bar{u}_B) \ket{\tilde{\Psi}}_{\mathbb{C}}^c 
    \label{eq:conn-HHLL}
\end{align}
for the connected entropy and
\begin{align}
    \bra{\Psi\Psi^\ast} \sigma (u_A) \sigma (u_B) \sigma (\bar{u}_A) \sigma (\bar{u}_B) \ket{\Psi\Psi^\ast}_{\mathbb{C}}^c 
    &= \bra{\tilde{\Psi}} \sigma (u_A) \sigma (\bar{u}_A) \sigma (u_B) \sigma (\bar{u}_B) \ket{\tilde{\Psi}}_{\mathbb{C}}^c \nonumber\\
    &=\sum_\ap \bra{\tilde{\Psi}} \sigma (u_A) \sigma (\bar{u}_A) \ket{\ap}^c_{\mathbb{C}} \!\bra{\ap} \sigma (u_B) \sigma (\bar{u}_B) \ket{\tilde{\Psi}}_{\mathbb{C}}^c \nonumber\\
    &\approx 
    \bra{\tilde{\Psi}} \sigma (u_A) \sigma (\bar{u}_A)
    \ket{\tilde{\Psi}}_{\mathbb{C}}^c
    \bra{\tilde{\Psi}}
    \sigma (u_B) \sigma (\bar{u}_B) \ket{\tilde{\Psi}}_{\mathbb{C}}^c 
\end{align}
for the disconnected entropy, where $\sum_\ap \ket{\ap}\!\bra{\ap}=\mathbf{1}$. After all, when we take $n \rightarrow 1$ the chiral eight-point function factorizes into the product of the heavy-heavy-light-light (HHLL) correlator in the connected and disconnected limits.

Note that \eqref{eq:conn-HHLL} is a chiral correlator multiplied by its conjugate. This equals to the nonchiral, ordinary HHLL correlator, which indicates that only the operator insertion affects EE in the connected phase and the boundary effect plays no role other than doubling the conformal dimension due to the mirror operator. This picture is consistent with the holographic result.

\subsubsection{Computation of the Chiral Identity Conformal Block}
In the previous subsection, we have seen the entanglement entropies in the connected/disconnected limits are given in terms of the chiral HHLL correlator $\bra{\tilde{\Psi}}\sigma_n \sigma_n \ket{\tilde{\Psi}}_{\mathbb{C}}^c$. In this subsection, we employ the identity block approximation as we discuss entanglement entropy, where $n\rightarrow 1$
\cite{Asplund:2014coa,Caputa:2015waa,Sully:2020pza}. We denote the chiral identity conformal block as $G_n(z)\equiv \bra{\tilde{\Psi}}\sigma(z) \sigma(1) \ket{\tilde{\Psi}}_{\mathbb{C}}^c \propto \langle {\tilde{\Psi}}(\infty) \sigma(z) \sigma(1) {\tilde{\Psi}}(0)\rangle ^c _\mathbb{C}$.\footnote{$z$ in this subsection is the cross ratio and obviously different from $z$ in the Poincare coordinates.}

We will hereon focus on $\Delta \left.S_{AB}^{(n)}\right|_{\mathrm{SD}}$, the difference of (R\'enyi) entanglement entropy from that of the vacuum on the SD region $S_{AB\ \mathrm{no\, op.}}$. The vacuum entanglement entropy in the SD (i.e. the entanglement entropy without a local operator but with the induced boundary) is already computed in the previous section \eqref{eq:EEwoop}.\footnote{Note that this is different from the vacuum entanglement entropy in the CFT without boundaries; $\Delta S_{AB}\neq \Delta S_{AB}|_{\mathrm{SD}}$. This distinction becomes crucial when we discuss the first law of entanglement.}
The conformal factors cancel out in $\Delta \left.S_{AB}^{(n)}\right|_{\mathrm{SD}}$
and is given by the logarithm of the ratio of the chiral identity conformal blocks: 
\begin{align}
    \Delta \left.S_{AB}^{(n)}\right|_{\mathrm{SD}}&=\frac{1}{1-n}\log \frac{\mbox{Tr}\left.\rho_{AB}^n\right|_{\mathrm{SD}}}{\mbox{Tr}\left.\rho_{AB,\mathrm{vac}}^n\right|_{\mathrm{SD}}} \nonumber\\
    &=\frac{1}{1-n}\log 
    \begin{dcases}
    \left\vert\frac{G_n(z_{{con}})}{G_n^{(0)}(z_{{con}})}\right\vert^2\quad &\text{(connected phase)}\\
    \frac{G_n(z_A)}{G_n^{(0)}(z_A)}\frac{G_n(z_B)}{G_n^{(0)}(z_B)}\quad &\text{(disconnected phase)}
    \end{dcases}
    ,
    \label{eq:EE-op}
\end{align}
where $G_n^{(0)}(z)=\bra{0}\sigma(z) \sigma(1) \ket{0}_{\mathbb{C}}^c$ and the cross ratio $z_{con},z_{A,B}$ is respectively given by $u_B/u_A$ and $\bar{u}_{A,B}/u_{A,B}$.

Since the holomorphic and antiholomorphic parts are factorized in the Virasoro block \cite{Fitzpatrick:2014vua}, the chiral identity block is just a square root of the usual identity block. The ratio is given by~\cite{Asplund:2014coa}
\begin{align}
    \frac{G_n(z)}{G_n^{(0)} (z)}&= \left(\frac{1}{|\ap_O|^2} |z|^{1-\ap_O} \left|\frac{1-z^{1-\ap_O}}{1-z}\right|^2 \right)^{-\Delta_n/2} 
    \nonumber\\
    &=\left| \frac{z^{\ap_O /2}-z^{-\ap_O /2}}{\ap_O (z^{1/2}-z^{-1/2})}\right|^{-\Delta_n},
    \label{eq:conf-block}
\end{align}
where $\ap_O=\sqrt{1-12\frac{\tilde{\Delta}}{c}}=\sqrt{1-24\frac{\Delta_O}{c}}$. Here we emphasize again that the conformal dimension in $\ap_O$ is twice as large as the one discussed in \cite{Asplund:2014coa,Hartman:2013mia} since the local operator in the HHLL correlator in concern is $\tilde{\Psi}$, whose conformal dimension is $\tilde{\Delta}=2\Delta_O$.

By rewriting $z=e^{i\omega}$ and $\bar{z}=e^{-i\bar{\omega}}$,\footnote{$\bar{\omega}$ is not the complex conjugate of $\omega$ but is defined from $\bar{z}$.} \eqref{eq:conf-block} reduces to
\begin{equation}
    \frac{G_n(z)}{G_n^{(0)} (z)}= \left( \frac{\sin \frac{\ap_O \omega}{2}}{\ap_O \sin \frac{\omega}{2}}
    \frac{\sin \frac{\ap_O \bar{\omega}}{2}}{\ap_O \sin \frac{\bar{\omega}}{2}} \right)^{-\Delta_n/2}.
    \label{eq:conf-block-sin}
\end{equation}

\subsubsection{Connected Entropy}
When $z=u_B/u_A\equiv z_{con}$ (connected phase), after the analytical continuation to the Lorentzian time,
we have
\begin{align}
    z_{con}&=\left(\frac{x_A-t+i\ap}{x_A-t-i\ap}\frac{x_B-t-i\ap}{x_B-t+i\ap}\right)^\kappa. 
\end{align}
%
%
Similarly, by replacing $t\rightarrow -t$,
\begin{align}
    \bar{z}_{con}&=\left(\frac{x_A+t+i\ap}{x_A+t-i\ap}\frac{x_B+t-i\ap}{x_B+t+i\ap}\right)^\kappa. 
\end{align}
Using the coordinate transformation \eqref{thrmap}, the cross ratios can be written in terms of the global coordinates:
\begin{equation}
    z_{con}=\left(\frac{e^{i(\theta_A-\theta_B)}}{e^{i(\tau_A-\tau_B)}}\right)^\kappa, \quad \bar{z}_{con}=\left(e^{i(\theta_A-\theta_B)}e^{i(\tau_A-\tau_B)}\right)^\kappa.
\end{equation}
We can immediately read out $\omega$ and $\bar{\omega}$ from this.
Assuming the trivial branch, we can show
\begin{equation}
    \frac{G_n(z_{con})}{G_n^{(0)} (z_{con})}= \left[\frac{\cos(\ap_O \kappa(\tau_A-\tau_B))-\cos(\ap_O\kappa (\theta_A-\theta_B))}{\ap_O^2 \left(\cos(\kappa(\tau_A-\tau_B))-\cos(\kappa (\theta_A-\theta_B))\right)}
    \right]^{-\Delta_n/2}.
    \label{eq:conn-EE-CFT}
\end{equation}
by some simple trigonometric algebra.

This analysis leads to the final expression of the growth of connected entanglement entropy:
\ba
\Delta S_{AB}^{con,{CFT}}=\frac{c}{6}\log \left[\frac{|u_1||u_2|}{|w_1-w_2|^2|u'_1||u'_2|\ap_O^2}
\cdot \left|\left(\frac{u_1}{u_2}\right)^{\f{\ap_O}{2}}-\left(\frac{u_1}{u_2}\right)^{-\frac{\ap_O}{2}}\right|^2\right]. \label{CFTconEE}
\ea

On the other hand, our previous gravity dual result \eqref{connected HEE} can be rewritten as follows:
\ba
\Delta S^{con,AdS}_{AB}=\frac{c}{6}\log\left[\frac{\left|\left(\frac{u_1}{u_2}\right)^{\frac{\chi}{2\kappa}}-\left(\frac{u_1}{u_2}\right)^{-\frac{\chi}{2\kappa}}\right|^2}{\chi^2\left|\left(\frac{u_1}{u_2}\right)^{\frac{1}{2\kappa}}-\left(\frac{u_1}{u_2}\right)^{-\frac{1}{2\kappa}}\right|^2}\right].
\label{AdSconEE}
\ea


The BCFT result (\ref{CFTconEE}) exactly agrees with the holographic one  by identifying the parameters as follows:
(\ref{AdSconEE}) 
\begin{equation}
    \ap_O=\frac{\chi}{\kappa}.   \label{Identifyap}
\end{equation}
Since $\kappa=\eta_0$, this is precisely what we have observed for $\chi$ under the rescaling by $\eta_0$ \eqref{eq:chi-transform}.
Thus, it implies $12\frac{\tilde{\Delta}}{c}=\frac{M^\prime}{R^2}$ when $\eta=\eta_0$. This is exactly equivalent to the relation we argued \eqref{eq:AdS EM tensor}. In this way, We have explicitly confirmed the perfect matching between the BCFT and the gravity dual by looking at entanglement entropy as well as the energy-momentum tensor.

To see the first law of entanglement, we need to subtract the vacuum contribution in the CFT given by
\begin{equation}
    S_{AB}^{{vac}}=\frac{c}{6}\log \frac{|x_B-x_A|^2}{\epsilon^2}
\end{equation}
from the connected entropy discussed above.\footnote{The vacuum contribution is equivalently given by the full contribution with $\Delta_O=0$ and $\kappa=1$.} What we computed in this subsection is $\Delta \left.S_{AB}\right|_\mathrm{SD}$, which is the full contribution $S_{AB}$ minus the contribution with the induced boundary but with no operator $S_{AB\ \mathrm{no\, op.}}$.
We can explicitly confirm that $\Delta S_{AB}
=\Delta_O \left.S_{AB}\right|_\mathrm{SD} + S_{AB\ \mathrm{no\, op.}} -S_{AB}^{{vac}}$ in the short distance limit satisfies the first law with the energy-momentum tensor \eqref{EMgencft} within the current CFT.
Via the first law (\ref{firstL}), \eqref{CFTconEE} leads to the energy density 
\ba
T_{tt}=\frac{c}{\pi}\left[-\frac{1}{6}(\kappa^2-1)+4\kappa^2\frac{\Delta_O}{c}\right]\cdot H(t,x),
\ea
which perfectly agrees with the energy-momentum tensor (\ref{EMgencft}), which was directly computed from the CFT.

\subsubsection{Disconnected Entropy}
When $z=\bar{u}/u\equiv z_{dis}$ (disconnected phase), after the analytical continuation to the Lorentzian time,
we have
\begin{align}
    z_{dis}=\bar{z}_{dis}^{-1}&=\left(\frac{x-t-i\ap}{x-t+i\ap}\frac{x+t-i\ap}{x+t+i\ap} e^{-2i\varphi} \right)^\kappa \nonumber\\
\end{align}

Using the coordinate transformation \eqref{thrmap}, the cross ratio can be written in terms of the global coordinates. In contrast to the connected entropy, we leave the branch choice from $w^\prime$ to $u$ and the cross ratio unspecified. Then,\footnote{Although $z$ and $\bar{z}$ can independently have a different branch, we consider $\omega=\bar{\omega}$ here, which turns out to be the one corresponding to the holographic entanglement entropy.}
\begin{equation}
    z_{dis}=e^{2\kappa i(\theta+\varphi-l\pi)} e^{-2\pi m} \Rightarrow \omega_{dis}=2\kappa \left(\theta+\varphi-\pi l -\frac{\pi}{\kappa} m \right) \quad (l,m\in\mathbb{Z}).
\end{equation}
It follows that
\begin{align}
    &\frac{G_n(z_{dis})}{G_n^{(0)} (z_{dis})} \nonumber\\
    =&\left( \frac{
    \displaystyle\sin \frac{\ap_O \omega_{dis}}{2}
    }{
    \displaystyle\ap_O \sin \frac{\omega_{dis}}{2}
    }
    \frac{
    \displaystyle\sin \frac{\ap_O \bar{\omega}_{dis}}{2}
    }{
    \displaystyle\ap_O \sin \frac{\bar{\omega}_{dis}}{2}
    } \right)^{-\Delta_n/2} 
    =\left(
    \frac{\sin \left(\ap_O \kappa \left(
    \displaystyle\theta -\pi l +\varphi - \frac{\pi}{\kappa} m  
    \right)\right)}{\ap_O \sin \left( 
    \displaystyle\kappa \left(\theta -\pi l +\varphi - \frac{\pi}{\kappa} m  \right)\right)}
    \right)^{-\Delta_n} 
    \label{eq:dis-first-br}\\
    =&\left( \frac{
    \displaystyle\sin \frac{\ap_O (-\omega_{dis})}{2}
    }{
    \displaystyle\ap_O \sin \frac{-\omega_{dis}}{2}
    }
    \frac{
    \displaystyle\sin \frac{\ap_O (-\bar{\omega}_{dis})}{2}
    }{
    \displaystyle\ap_O \sin \frac{-\bar{\omega}_{dis}}{2}
    } \right)^{-\Delta_n/2} 
    =\left(
    \frac{\sin \left(\ap_O \kappa \left(
    \displaystyle \pi l -\theta -\varphi +\frac{\pi}{\kappa} m
    \right)\right)}{
    \displaystyle \ap_O \sin \left(\kappa \left(\pi l -\theta -\varphi +\frac{\pi}{\kappa} m \right)\right)
    }
    \right)^{-\Delta_n}.
    \label{eq:dis-next-br}
\end{align}
As we have discussed for the connected entropy, the identification (\ref{Identifyap}) $\ap_O=\chi/\kappa$ reproduces the holographic result for the disconnected entropy as well by choosing $(l,m)=(-1,1)$ for \eqref{eq:dis-first-br} and $(l,m)=(1,0)$ for \eqref{eq:dis-next-br}; each choice of the branch corresponds to $\theta^{min}=\theta$ and $(2-1/\chi)\pi -\theta$, respectively. $l$ corresponds to choosing an appropriate branch so that $0\ge \arg u\ge \pi$ is satisfied; This is nothing but choosing an appropriate $\theta^{min}$. To justify the branch choices of $m$ within the CFT, we need a careful consideration of the monodromy, which we discuss in the Appendix \ref{app:branch}.


\subsection{Entanglement Entropy after Rescaling by $\eta$}
As it has been mentioned in the previous sections, entanglement entropy after an arbitrary rescaling by $\eta$ \eqref{rescaleex} is simply given by the replacement $\chi\rightarrow\chi/\eta$. Since the entanglement entropy is written in terms of $\ap_O \kappa$, it becomes $\ap_O \kappa/\eta$ and this is consistent with $\chi$ becoming $\chi/\eta$ in holographic entanglement entropy.




\section{Conclusions and Discussions}
In this paper we studied the local operator quench in two-dimensional holographic BCFTs from both the gravity dual analysis based on the AdS/BCFT and the field theoretic one. We eventually confirm that these two independent approaches agree with each other perfectly after our careful analysis summarized below.

In the gravity dual description, the local operator quench is described by putting a massive particle in a three-dimensional AdS. The boundary of the BCFT corresponds to  an end of the world brane (EOW brane) in this AdS geometry. The backreaction of the massive particle deforms the profile of the EOW brane in a nontrivial way. For example, if the mass of the particle is too heavy, the EOW brane gets folded and thus the gravity dual does not seem to make sense as in the right of Fig.\ref{deformationfig}. However, we noticed that the relation between the mass of the particle in the AdS and the conformal dimension of the dual operator in the BCFT, gets modified into (\ref{relasd}) because the second boundary emerges in the dual BCFT. This modified relation nicely avoids the folding problem. Moreover, we showed that an appropriate rescaling of the angular coordinate (by the parameter $\eta$) before mapping into the asymptotically Poincare AdS coordinate allows us to remove the extra boundary and leads to the BCFT on a half plane. 
Using this construction of the gravity dual, we computed the energy-momentum tensor and holographic entanglement entropy.

In the BCFT description of the local operator quench, 
we first constructed the conformal map which transforms the upper half plane into  a semi-disk, which becomes a strip like region with two time dependent boundaries via a Wick rotation. This allows us to analytically calculate the energy-momentum tensor and entanglement entropy, where we also applied the standard holographic CFT treatment to evaluate the correlation functions. These calculations perfectly reproduce the results obtained from the gravity dual.

Finally we would like to complete this paper with a few future directions. In this local quench, we assumed that the local operator is inserted along the boundary of the BCFT. Thus it will be an interesting future problem to extend our construction of the gravity dual to the case where the local operator is inserted in the bulk of a BCFT. Furthermore, our holographic calculation assumes a vanishing one-point function in the BCFT. It is interesting to take the matter contribution as well \cite{Suzuki:2022xwv}. The BCFT counterpart might be calculated by taking a certain limit such as the Regge limit \cite{Kusuki:2019gjs,Kusuki:2021gpt}, which is presumably related to $\ap\rightarrow 0$ limit. Another intriguing problem is to explore the brane-world interpretation of our holographic local quench in a BCFT, which allows us to interpret it as a two-dimensional gravity coupled to a CFT and which relates our entanglement entropy computation to the island formula.


\section*{Acknowledgements}
We are grateful to Hao Geng, Satoshi Iso, Yuya Kusuki, Tokiro Numasawa, Mark Van Raamsdonk and Kenta Suzuki for useful discussions and comments. This work is supported by MEXT-JSPS Grant-in-Aid for Transformative Research Areas (A) "Extreme Universe", No.\,21H05182, No.\,21H05184, and No.\,21H05187. T.\,M. thanks the Atoms program very much which helps him to stay in YITP, Kyoto U. for this collaboration 
and is supported by SOKENDAI and Atsumi International Foundation. T.\,T. is supported by JSPS Grant-in-Aid for Scientific Research (A) No.21H04469,  by the Simons Foundation through the ``It from Qubit'' collaboration, by Inamori Research Institute for Science and by World Premier International Research Center Initiative (WPI Initiative) from MEXT. T.U is supported by JSPS Grant-in-Aid for Young Scientists 19K14716.

\begin{appendices}
\section{Embedding the Global AdS to the Flat Space}\label{sec:Embedd}
The $(d+2)$-dimensional anti-de Sitter (AdS) space is obtained as embedding  to the flat space time $\mathbf{R}^{2,d+1}$ with  
 \begin{equation}
     \Vec{X}\cdot \Vec{X}= - (X_0)^2 - (X_{d+2})^2 + (X_1)^2 + \cdots + (X_{d+1})^2 = - R^2 \label{eq:hyper}
 \end{equation}
The problem of finding geodesics $\Vec{X} (s)$ in  AdS is thus a variational problem with the constraint \eqref{eq:hyper}. Let $\lambda (s)$ be a Lagrange multiplier associated with the constraint, then the action is given by 
 \begin{align}
     S = \int ds L,\quad L = \frac{1}{2} \dot{\Vec{X}}\cdot \dot{\Vec{X}} - \lambda(s) F(\Vec{X}),   \quad
     F(\Vec{X})=  \Vec{X}\cdot \Vec{X} + R^2.
 \end{align}
 
 Equations of the motion for $\Vec{X},\lambda$ are thus given by  
 \begin{align}
     \Ddot{\Vec{X}}+ 2 \lambda \Vec{X} = 0, \quad 
     \Vec{X}\cdot \Vec{X} + R^2 =0
 \end{align}
 Taking derivatives two times  of the constraint gives 
 \begin{equation}
     \Ddot{\Vec{X}} \cdot \Vec{X} = - \dot{\Vec{X}}\cdot \dot{\Vec{X}} = -1.
 \end{equation}
 The second equality holds because we focus on spacelike geodesics. This fixes the Lagrange multiplier to be,

 \begin{equation}
     \lambda(s)= -\frac{1}{2 R^2},
 \end{equation}
and 
 \begin{align}
 \Ddot{\Vec{X}}-\frac{1}{R^2}\Vec{X}=0 \;  \Rightarrow \; 
 \Vec{X}(s) = \Vec{p} \cosh{\frac{s}{R}} + \Vec{q} \sinh{\frac{s}{R}}.
 \end{align}
 The constraint implies $\Vec{p}$ and $\Vec{q}$ satisfy
 \begin{equation}
     \Vec{p}\cdot \Vec{p} =  -\Vec{q}\cdot \Vec{q}=- R^2, \quad  \Vec{p}\cdot \Vec{q} = 0.
 \end{equation}
    We can freely set $s=0$ at the one end of the geodesic (call it A). Thus 
    \begin{equation}
         \Vec{p}= \Vec{X}_A
    \end{equation}
    We call another end as B, and by taking the inner product 
    of $\Vec{X}_B= \Vec{p} \cosh{\frac{s_B}{R}} + \Vec{q} \sinh{\frac{s_B}{R}}$ and $\Vec{X}_A$
    \begin{equation}
        \cosh{\frac{s_B}{R}} = -\frac{\Vec{X}_A\cdot\Vec{X}_B}{R^2}
    \end{equation}
    and we obtain
    \begin{equation}
        \Vec{q} = \frac{1}{\sinh{\frac{s_B}{R}}}{\Vec{X}_B- \cosh{\frac{s_B}{R}}\Vec{X}_A}
    \end{equation}
The length of the geodesics between A and B $d(A,B)$ is just the difference of parameter 
\begin{equation}
     d(A,B)= \int_A^B \sqrt{d\Vec{X}\cdot d \Vec{X}}=s_B
\end{equation}

Let us use the above formalism to compute the length of the geodesic connecting two points in global $AdS_{3}$, where the metric takes the form \eqref{Gmet}. The relation between  global coordinates $(\tau, \theta, r)$  and embedding coordinates $\vec{X}$ is

\begin{align}
    X_{0}= \sqrt{R^2 +r^2}\cos{\tau}, \quad
    X_{1} = r \sin{\theta},\quad
    X_{2} = -r \cos{\theta},\quad
    X_{3}= \sqrt{R^2 +r^2}\sin{\tau}.
\end{align}

This relation allows us to write the relevant inner product $\Vec{X}_A\cdot \Vec{X}_B$ in terms of the global coordinates

\begin{align}\label{AppendixA geodsics}
    \Vec{X}_A\cdot \Vec{X}_B& =  -{X_A}_0 {X_B}_0 -{X_A}_3 {X_B}_3+{X_A}_1 {X_B}_1+{X_A}_2 {X_B}_2 \\
    &= -\sqrt{\left(R^2+r_A^2\right)\left(R^2+{r_B^2}\right)} \cos(\tau_A-\tau_B) + r_A r_B\cos(\theta_A-\theta_B).
\end{align}
Thus we conclude the geodesics length between two generic points is given by 
\begin{align}\label{AppendixA EE}
    d(A,B)=& R \cosh^{-1}{\left(\sqrt{\left(1+\frac{r_A^2}{R^2}\right)\left(1+\frac{r_B^2}{R^2}\right)} \cos(\tau_A-\tau_B) - \frac{r_A r_B}{R^2}\cos(\theta_A-\theta_B)\right)}
\end{align}
In the case $r_A,r_B \gg R$, and recall $\cosh^{-1}{x}= \log{(x+\sqrt{x^2-1})}$, we obtain
\begin{equation}
    d(A,B)= R\log{\left(\frac{2r_A r_B}{R^2}\left(\cos(\tau_A-\tau_B)-\cos(\theta_A-\theta_B)\right)\right)}
\end{equation}

\section{Details of the Gravity-inducing Boundary Curve}
\begin{figure}[h]
    \centering
    \includegraphics[width = 100mm]{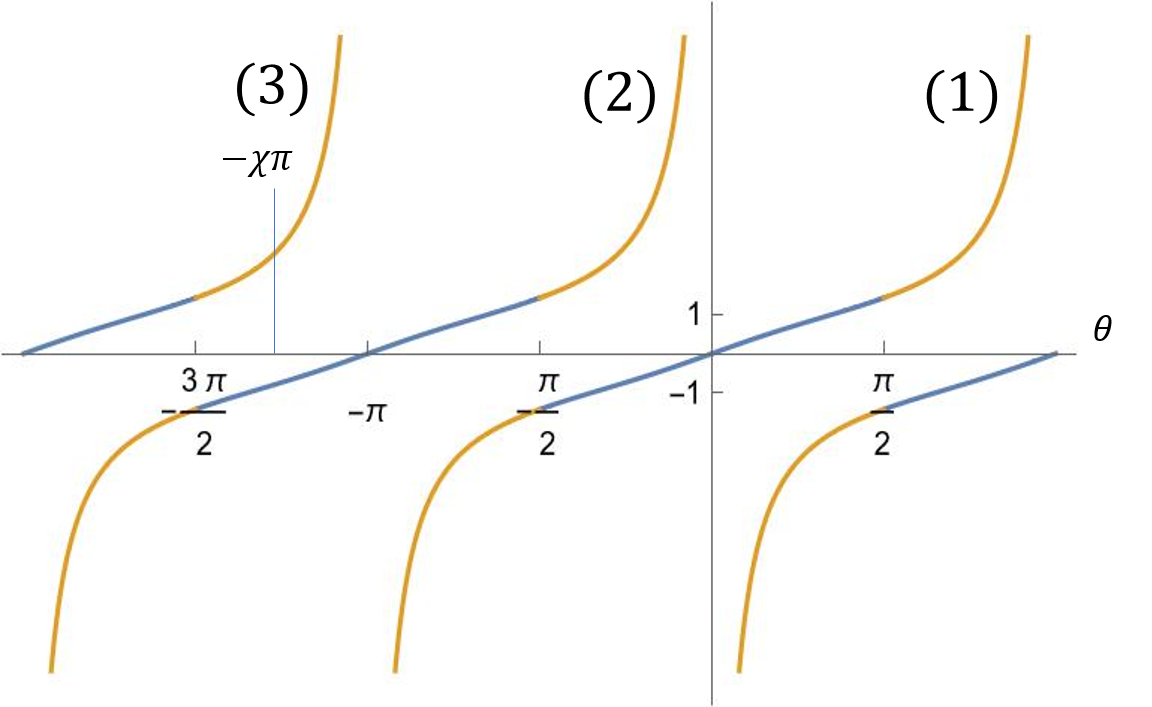}
    \caption{The blue curve is correspond to$+$ sign and the orange curve is correspond to $-$ of (\ref{theta to x}).We should take (1) and (3) branch for example.}
    \label{fig:theta to x}
\end{figure}

 From (\ref{mappo}), we can associate $x$ with $\theta$ for each fixed time $t$ at the conformal boundary. By solving the forth equation of (\ref{mappo}), which is just a quadratic equation for $x$, we can get 
 \begin{equation}\label{theta to x}
      x= \alpha \cot{\theta}\left(-1\pm \sqrt{1+ \tan^2{\theta}\left(1+\frac{t^2}{\alpha^2}\right)}\right).
 \end{equation}
The branch should be selected so that the map (\ref{theta to x}) is smooth for $\theta\in [-\pi,\pi)$. Thus for $-\frac{\pi}{2}< \theta < \frac{\pi}{2}$ we take $+$ sign and, for $-\pi < \theta < -\frac{\pi}{2}$and$\frac{\pi}{2}< \theta < \pi$ we take $-$ sign. In the other region, we take the sign so that the map has $2 \pi$ periodicity. Please see the Fig.\ref{fig:theta to x}.
 Especially we have the gravity-inducing boundary $Z(t)$  at $\theta = -\frac{\pi}{\chi}$, where
 \begin{align}
     Z(t) = \left \{
     \begin{array}{ll}
        \alpha \cot{(\frac{\pi}{\chi})} \left(1+ \sqrt{1+\tan^2{(\frac{\pi}{\chi})}(1+\frac{t^2}{\alpha^2})}\right), & \frac{2}{3} < \chi < 1  \\
        \alpha \cot{(\frac{\pi}{\chi})} \left(1 - \sqrt{1+\tan^2{(\frac{\pi}{\chi})}(1+\frac{t^2}{\alpha^2})}\right).   &  \frac{1}{2} \leq \chi < \frac{2}{3}
     \end{array}
     \right.
 \end{align}
 For $\chi \leq \frac{1}{2}$, $Z(t)\leq 0$ and it looks there is no physical region for the holographic CFT. 
 \\The critical point $\chi = \frac{1}{2}$  can be seen when we consider whether the map between the tilde global coordinate$(\tilde{r},\ti{\theta},\ti{\tau})$ and the tilde Poincare$(\ti{z},\ti{x},\ti{t})$ is well-defined.
By a careful inspection of coordinate transformation \eqref{mappo}, choosing a particular domain for $\theta$ (or $\tilde{\theta}$) corresponds to considering a brane profile with a fixed winding number.
Let us first discuss the $-\pi\le \theta <\pi$ case in detail and comment briefly about other branches at last.

In the original global AdS spacetime with the massive particle, we chose the domain of $\theta$ to be $[-\pi,\pi)$. This results in the domain of $\tilde{\theta}$ \eqref{tildtheta}.
Since we discuss the profile of the EOW brane by bringing the solution from the Poincare metric via \eqref{mappo} from the metric with $\tilde{\theta}$ \eqref{deficitg}, we are implicitly assuming the existence of the map between them. However, due to the deficit angle in $\tilde{\theta}$ and hence not covering the whole Poincare spacetime, such a single-valued map does not exist for some values of $M>0$ given a fixed domain of $\theta$.

\begin{figure}
  \centering
  \includegraphics[width=13cm]{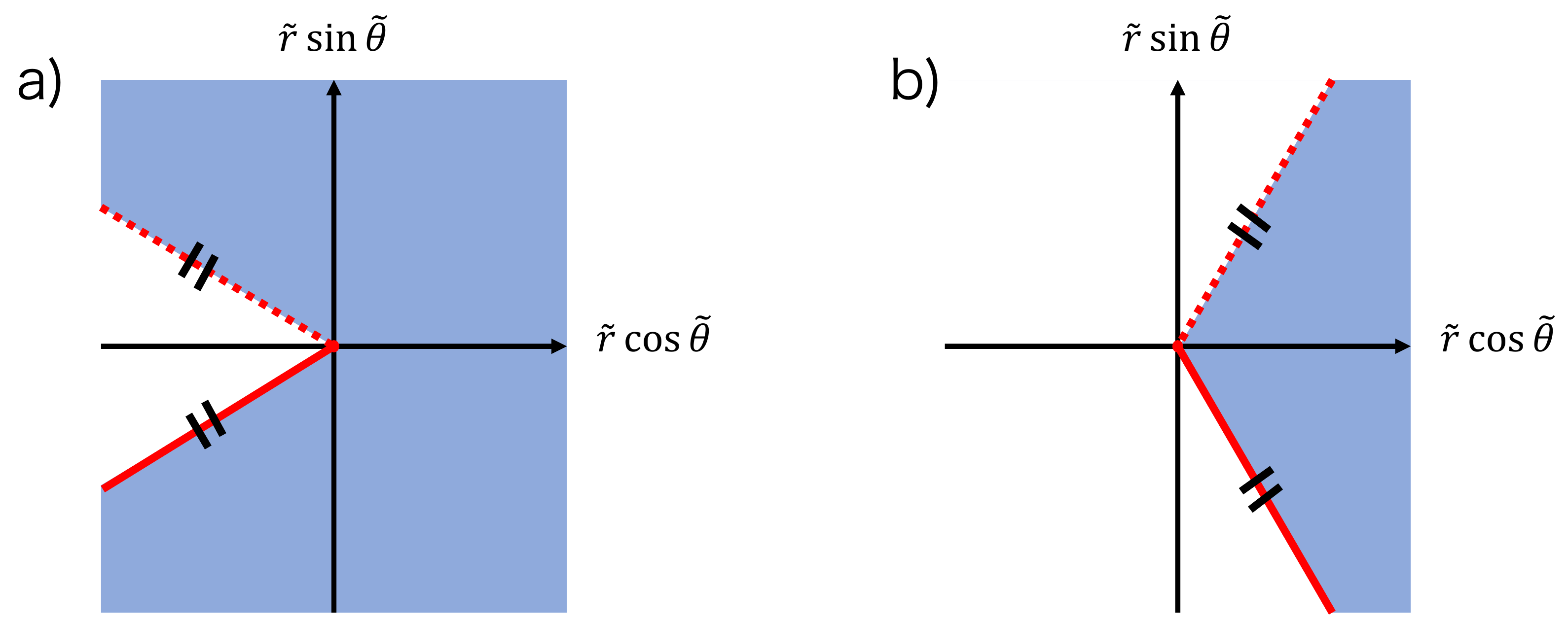}
  \caption{The allowed region for $-\pi \chi\le \tilde{\theta} < \pi \chi$ when a) $\frac{1}{2}\le\chi(<1)$ and b) $\chi\ge\frac{1}{2}$. The strike-though denotes an identification of two lines.}
\label{fig:deficit-geom}
\end{figure}

The deficit geometry in the rescaled coordinates $(\tilde{r},\tilde{\theta})$ is shown in Fig.\ref{fig:deficit-geom}.
From \eqref{mappo}, we expect the rescaled metric \eqref{deficitg} is mapped to the Poincare metric
\ba
ds^2=R^2\left(\frac{d\tilde{z}^2-d\tilde{t}^2+d\tilde{x}^2}{\tilde{z}^2}\right).
\ea
Let us focus on the relation
\ba
\tan\tilde{\theta}=\frac{\alpha^2 -\tilde{z}^2-\tilde{x}^2+\tilde{t}^2}{2\alpha\tilde{x}},
\label{tan}
\ea
By dividing into two cases according to the value of  ${\chi}<1$.

First, let's consider 
$\frac{1}{2}\le{\chi}$
as shown in a) in Fig.\ref{fig:deficit-geom}, we need some case analyses.
    
For $-\frac{\pi}{2}<\ti{\theta}<\frac{\pi}{2}$, $\cos\tilde{\theta}\ge 0 \Leftrightarrow \tilde{z}^2+\tilde{x}^2-\tilde{t}^2\le \alpha^2$, there is no constraint on $\tan \tilde{\theta}$ and the conformal boundary region  $-\frac{\pi}{2}<\ti{\theta}<\frac{\pi}{2}$ is mapped to the conformal boundary region $x^2< t^2 +\alpha^2$. On the other hand, for $\cos\tilde{\theta} < 0$, $\tan \tilde{\theta}$ is constrained due to \eqref{tildtheta} as follows.
\ba
\begin{cases}
\tan\tilde{\theta} < \tan\left(\pi \chi\right) <0 & \text{for}\ \tilde{x}>0 \quad (\tilde{\theta}>\pi/2),\\
\tan\tilde{\theta} \ge \tan\left(-\pi \chi\right) >0 & \text{for}\ \tilde{x}<0 \quad (\tilde{\theta}<-\pi/2).
\end{cases}
\label{cond-for-tan}
\ea
In our setup, we are interested in the region $x>0$, which corresponds to $0<\ti{\theta}<\chi \pi$ and only consider that region for a while. Rewriting \eqref{cond-for-tan} in terms of Poincare coordinates using \eqref{tan},
\ba
\tilde{x}\ge -\alpha \tan \left(\pi \chi \right)
+\sqrt{\alpha^2\tan^2 \left(\pi \chi\right)+\alpha^2+\tilde{t}^2}
\ea
and
\ba
\tilde{x}^2-\tilde{t}^2\ge \alpha^2
\ea
for $\tilde{x}\ge 0$ on the conformal boundary $\tilde{z}=0$. Actually there is a region which satisfies the both inequality and the region$ \frac{\pi}{2}\geq \frac{\pi}\chi$ is mapped to the region in the tilde Poincare coordinates satisfies above inequalities.
Next, we consider $\chi\le 1/2$
as shown in b) in Fig.\ref{fig:deficit-geom}, we need
\ba
\tan\left(-\pi \chi\right) \le \tan\tilde{\theta}< \tan\left(\pi \chi\right).
\ea
However, we have no solution satisfying this in Poincare coordinates.

From these analysis, we require
\ba
\frac{1}{2}<\chi = \sqrt{\frac{R^2-M}{R^2}} (\le 1) \quad \Longleftrightarrow \quad 0\le M <\frac{3}{4}R^2 (<R^2)
\label{restrictionM}
\ea
for the existence of the map from the rescaled coordinates $(\tilde{\tau}, \tilde{r},\tilde{\theta})$ to the Poincare coordinates $(\tilde{t},\tilde{z},\tilde{x})$. Since we focus on the brane profile obtained in this manner, we restrict to \eqref{restrictionM} for $M<R^2$.

Within this domain, the boundary of the BCFT is indeed given by $\tilde{\theta}=0$ and $2\pi-\pi \chi$ for $M>0$.\footnote{For $M=0$, the boundary is given by $\tilde{\theta}=0,-\pi$.}

The nontrivial endpoint of the EOW brane comes from $\tilde{\theta}=-\pi$. However, this is outside of the domain for $M>0$ as we chose $(-\pi<)-\pi \chi\le \tilde{\theta} < \pi \chi(<\pi)$. Within this domain, it seems that there is only one solution $\tilde{\theta}=0$. However, there should be two solutions from the viewpoint of the global patch. This paradox is resolved when you consider a proper origin of the angle such that the entire brane is within the fundamental domain.

Periodicity of $\tilde{\theta}$ given by $2\pi \chi$ is greater than $\pi/2$ regarding the previous analysis. This implies that by shifting the domain of $\tilde{\theta}$, we can always have two nodes in $\sin\tilde{\theta}$. It means the initial choice of $-\pi\le \theta<\pi$ may not appropriate; however, for $1/2<\chi$ all configurations are connected continuously and we will not cause a problem by evading using periodicity as we have discussed.

\section{Branch Choice in the Identity Conformal Block}\label{app:branch}
In addition to the branch of $w^\prime$ discussed above, we need to choose the correct branch to analytically continue to the Lorentzian correlator \cite{Asplund:2014coa}. To see the monodromy around $z,\bar{z}=1$, we expand $z$ and $\bar{z}$ around $1$ provided $\ap$ is sufficiently small compared to other quantities. Since $e^{-i\varphi}$ factor only shifts the imaginary part, it is irrelevant to the discussion here. The cross ratios are expanded as follows:
\begin{align}
    z_{dis}&\propto\left(\frac{x-t-i\ap}{x-t+i\ap}\frac{x+t-i\ap}{x+t+i\ap}  \right)^\kappa &=1-\frac{4i\kappa x}{(x-t)(x+t)} \ap + O(\ap^2) \label{eq:1}\\
    \bar{z}_{dis}&\propto\left(\frac{x-t-i\ap}{x-t+i\ap}\frac{x+t-i\ap}{x+t+i\ap}  \right)^{-\kappa} 
    &=1+\frac{4i\kappa x}{(x-t)(x+t)} \ap + O(\ap^2)\label{eq:2}\\
    z_{con}&\propto\left(\frac{x_1-t+i\ap}{x_1-t-i\ap}\frac{x_2-t-i\ap}{x_2-t+i\ap}\right)^\kappa 
    &=1+\frac{2i\kappa (x_2-x_1)}{(x_1-t)(x_2-t)} \ap + O(\ap^2) \label{eq:3}\\
    \bar{z}_{con}&\propto\left(\frac{x_1+t+i\ap}{x_1+t-i\ap}\frac{x_2+t-i\ap}{x_2+t+i\ap}\right)^\kappa
    &=1+\frac{2i\kappa (x_2-x_1)}{(x_1+t)(x_2+t)} \ap + O(\ap^2). \label{eq:4}
\end{align}

When the sign of the imaginary part changes, $z$ or $\bar{z}$ moves to another sheet. From \eqref{eq:1}, $z_{dis}\rightarrow e^{-2\pi i} z_{dis}$ for $t\gg x$. From \eqref{eq:2}, $\bar{z}_{dis}\rightarrow e^{2\pi i} \bar{z}_{dis}$ for $t\gg x$. These branches for the disconnected entropy correspond to $m=0$ to $m=1$ in \eqref{eq:dis-first-br} and \eqref{eq:dis-next-br}. The transition between two branches is determined so that the whole expression is continuous.

From \eqref{eq:3}, $z_{con}\rightarrow e^{2\pi i} z_{con}$ for $x_1\ll t\ll x_2$. Finally, from \eqref{eq:4}, there is no change in the branch as the sign of the imaginary part is always positive. For the connected entropy, by considering the Euclidean branch corresponding to the dominant OPEs due to the approximation of the conformal block. Then, it should cancels the additional phase during $x_1\ll t\ll x_2$ and leads to the trivial branch \eqref{eq:conn-EE-CFT}.

\end{appendices}


\bibliographystyle{JHEP}
\bibliography{Island}

\end{document}